\documentclass[twocolumn,amssymb,aps,pra,showpacs,showkeys,nofootinbib]{revtex4-1}

\usepackage{amsthm}
\usepackage{amsmath}
\usepackage{amssymb}
\usepackage{color}
\usepackage{adjustbox}
\usepackage{graphicx}

\begin{document}

\title[Quantum photonics at telecom wavelengths based on lithium niobate waveguides]{Quantum photonics at telecom wavelengths based on lithium niobate waveguides}

\author{Olivier Alibart, Virginia D'Auria, Marc De Micheli, Florent Doutre, Florian Kaiser, Laurent Labont\'e, Tommaso Lunghi, \'Eric Picholle, and S\'ebastien Tanzilli}

\affiliation{Universit\'e C\^ote d'Azur, CNRS, Laboratoire de Physique de la Mati\`ere Condens\'ee, France}
\email{sebastien.tanzilli@unice.fr}


\begin{abstract}
Integrated optical components on lithium niobate play a major role in standard high-speed communication systems. Over the last two decades, after the birth and positioning of quantum information science, lithium niobate waveguide architectures have emerged as one of the key platforms for enabling photonics quantum technologies. Due to mature technological processes for waveguide structure integration, as well as inherent and efficient properties for nonlinear optical effects, lithium niobate devices are nowadays at the heart of many photon-pair or triplet sources, single-photon detectors, coherent wavelength-conversion interfaces, and quantum memories. Consequently, they find applications in advanced and complex quantum communication systems, where compactness, stability, efficiency, and interconnectability with other guided-wave technologies are required. In this review paper, we first introduce the material aspects of lithium niobate, and subsequently discuss all of the above mentioned quantum components, ranging from standard photon-pair sources to more complex and advanced circuits.
\end{abstract}

\maketitle




\section{Introduction: integrated photonics, a paradigm shift for quantum engineering and technologies}
\label{Sec_Intro}

Over the last decades, quantum photonics has become a thriving field of research, promoting both fundamental investigation of quantum phenomena and a broad variety of disruptive quantum technologies~\cite{OBrien2009,Tanz_Genesis_2012}. From the fundamental side, entanglement has been exploited, among others, for more and more convincing nonlocality demonstrations~\cite{Brunner_BellRMP_2014,Hensen_15,Putz_LDLPRL_2016}, single-photon complementarity~\cite{Peruzzo_ScienceQDC_2012,Kaiser_QDC_2012}, and random walks~\cite{Peruzzo_2010,Crespi_2014}. From the applied side, we now find reliable quantum-enabled applications, among others, in secure communication, referred to as quantum cryptosystems~\cite{gisin_QKD_2002,Scarani_QKDRMP_2009,Aktas_DWDM_2016,Korzh_2014}, teleportation-based quantum communication~\cite{Martin_2012,Metcalf2014}, metrology~\cite{Dowling_Qmet_2008,Giovannetti2011}, as well as in fast computing algorithms~\cite{Crespi_2014,Deutsch85,DiVincenzo95,Nielsen00,Lanyon_2010}.

In this framework, integrated optics (IO) technologies allow realizing complex and scalable quantum circuits~\cite{OBrien2009,Tanz_Genesis_2012,Tillman_2013,Metcalf2013}, finding striking repercussions in all the above mentioned research areas, otherwise unreachable using bulk approaches. Notably, quantum random walks (QRW) and boson sampling are very good examples of new topics in quantum physics being enabled by integrated optics~\cite{kempe_03}. A QRW can be implemented via a constant splitting of photons into several possible waveguide. Experimentally, it amounts to chaining 50/50 beam-splitters, but the price to pay is the rapidly growing size of the setup. Until the 21-evanescently-coupled waveguides demonstrated in 2001 by the university of Bristol (UK)~\cite{Peruzzo_2010}, the largest realization using bulk optics used to be limited to a 5-item system. Yet, technology triggered a new pathway for quantum information processing as a 101-coupled-waveguides device was presented in 2014~\cite{Solntsev_2014}. This example clearly shows the potential of integrated optics (IO) when it comes to enabling new research fields.

The aim of this article is to establish that integrated photonics provides a powerful technological basis for a broad class of advanced quantum experiments. Most of the devices we shall discuss have taken advantage of guided-wave optics ability to dramatically enhance the efficiency of nonlinear interactions. This is because it permits maintaining high optical power densities over distances far exceeding those permitted by the diffraction limit in bulk devices. For passive devices, IO has provided a possibility of assembling many components on single chips, simplifying the realization and use of complex circuits.

However, IO faces one problem, that of coupling integrated devices to subsequent fiber systems. In seminal experiments~\cite{sanaka_new_2001,tanzilli_ppln_2002}, for instance, entangled photon pair collection efficiencies were limited, preventing from reaching the level of bulk optics approaches. Nowadays, monolithic integrated devices containing advanced photonics circuit powered by on-chip photon-pair sources are having a tremendous impact on the progress made in the field of quantum information science. More specifically, quantum photonics requires fully integrated devices for producing (multiple) pairs of entangled photons efficiently, and where photons can be coherently manipulated using frequency-conversion stages, dynamical routers and phase shifters, combined with external photons, and eventually detected. Those devices are therefore often considered as enabling quantum nodes where bits of quantum information (qubits) can be tailored at will and on-demand thanks to on-chip capabilities.

Among the multiple platforms available that are covered for example in Ref.~\cite{Tanz_Genesis_2012}, this review focuses on lithium niobate (LiNbO$_3$, in the following this material will be referred to as LN) only. The motivation lies in the fact that, compared to equivalent setups exploiting different platforms, such as silicon photonics~\cite{Bogaerts_LPRSilicon_2012,silverstone_2014} or silica~\cite{Meany_2015}, LN circuits enable highly efficient periodically poled waveguide (PPLN/W) photon-pair sources as well as ultra-fast electro-optic modulators. Those complementary aspects make them unique and very appealing for on-chip engineering of quantum light despite they feature a larger footprint and incompatibility with CMOS electronics.

In the following, we will start with briefly presenting LN properties from the material science perspective. Then, we will discuss in more details the applications of such a platform along five thematics: entangled photon-pair sources, heralded single photons, applications ranging from quantum interfaces to quantum memories, and the use of IO for linear quantum computing and bright squeezed state of light. We will finish with a conclusion and some perspectives related to LN-based integrated quantum photonic devices. We shall note that this review is not intended to be encyclopaedic. It rather intends to provide a honest overview on LN integrated optical technology as applied to various tasks, while referring to works of historical interest as well as current state-of-the-art.

\section{Lithium niobate, the material aspects} 
\label{Sec_LN}

\subsection{Generalities}

When considering the requirements for a quantum photonics platform, the material has obviously to be transparent in the wavelength ranges of interest. Moreover, when photon-pair generation is intended, high nonlinear coefficients are desired, as well as practical phase-matching solutions. Less critically, the damage threshold should be as high as possible, and the substrate should ideally be available in large scale, at low cost, with high quality and uniformity.
In this section, we will start with a quick comparison between several materials motivating why LN appears as a prime candidate for quantum photonics platforms. We will then discuss the properties of LN and the material structuration available for photon pair generation. We will finish with a quick overview of the post processing characterizations used to evaluate the performance of such devices for quantum applications. 

Generally speaking, three families of materials are currently used:
\begin{itemize}
\item non-centrosymmetric crystals (or $\chi^{(2)}$ crystals) are the most widely used, well-known, and exploited materials to obtain efficient nonlinear processes;
\item symmetric materials such as silica (or $\chi^{(3)}$ materials), have lower nonlinearities, however they also have lower losses, therefore less efficient frequency conversions can be compensated by much longer interaction lengths. Note that artificial $\chi^{(2)}$ nonlinearities can be induced via poling techniques~\cite{pruneri_1998}, despite reduced overall optical nonlinearity and the fact that poled gratings can usually not be engineered over more than a few cm lengths;
\item semiconductors such as silicon and AlGaAs are promising materials thanks to extremely controlled technological processes allowing to design very small footprint nanostructures potentially exploiting some of the highest nonlinear coefficients reported to date (\textit{e.g.}, in ZnSe $d\,\sim$\,54~pm.V$^{-1}$). However, practical phase-matching conditions or propagation losses are still issues at the present time~\cite{angell_1994}.
\end{itemize}
Tab.~\ref{materials} presents the most commonly used $\chi^{(2)}$ crystals benefiting from high enough fabrication quality as well as a compatibility with waveguide integration technologies. Although etching processes allow obtaining waveguide structures in beta-baryum borate (BBO) and lithium triborate (LBO), the associated propagation losses are usually very high $(\sim 7 \rm\,dB.cm^{-1})$~\cite{deglinnocenti_2008}, which is at least one order of magnitude higher compared to waveguide structures realized on LN, lithium tantalate (LT), and potassium titanyl phosphate (KTP).
\begin{table}[!htb]
\begin{center}
\begin{tabular}{|l|c|c|c|c|c|}
 \hline
Substrate & Highest NL coeff. & Waveguide integration \\
& (pm.V$^{-1}$) & \\
\hline
LiNbO$_3$ & 34~\cite{choy_1976} & titanium indiffusion \\
&  & proton exchange \\
\hline
LiTaO$_3$ & 15~\cite{shoji_1997} & proton exchange \\
\hline
KTP & 18.5~\cite{vanherzeele_1992} & ion exchange \\
\hline
BBO & 1.6 & He$^+$ implantation  \\
\cline{1-2}
LBO & 2.97 & + etching \\
\hline
\end{tabular}
\caption{Usual nonlinear crystals used for frequency conversion.\label{materials}}
\end{center}
\end{table}

The latter three materials are compatible with nonlinear coefficient engineering techniques which allow quasi-phase matching (QPM), and therefore precise engineering of photon-pair wavelengths over a broad range. In addition, these crystals exhibit non-zero electro-optics coefficient which means that one can modulate waveguide properties by applying electric fields, giving thus access to dynamic, and potentially high-speed, manipulation of light fields, and notably, single photons. 

In this paper, we will focus on the perspective offered by the LN platform in quantum photonics applications, thanks notably to unique optical-optical and electro-optical nonlinear capabilities, as well as in terms of mature technological processes required to tailor those properties.

\subsection{Lithium niobate properties}

\begin{figure}[h]
\begin{center}
\includegraphics[width=0.5\textwidth]{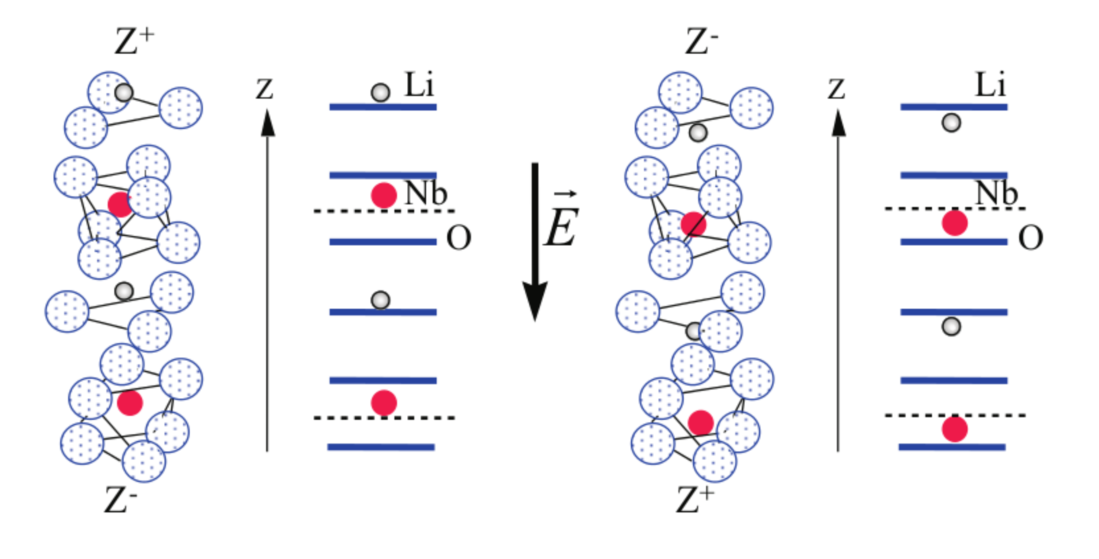}
\caption{Crystaline structure of lithium niobate. The small red and white circles are the niobium and lithium atoms, respectively. On the left picture, they are shifted above the oxygen planes inducing a positive spontaneous polarization field along the $z$ axis. Applying a negative electric field ($\vec{E}$ in the middle of the figure) during the fabrication permanently shifts the niobium and lithium atoms below the oxygen planes, therefore inverting the direction of the crystal's spontaneous polarization. \label{LNcell}}
\end{center}
\end{figure}

LN is a monocrystal grown in the form of a boule by Czochralski method~\cite{ballman_1965} with high reproducibility and homogeneity. If an electric field is applied while the crystal is cooled down below its Curie temperature ($\sim$1145$^\circ$\,C), the boule is in a ferroelectric phase~\cite{newnham_1975}, \textit{i.e.} it exhibits a spontaneous electric polarization due to a permanent shift of lithium and niobium ions relatively to oxygen planes (elementary cell shown in \figurename~\ref{LNcell}).
Also note that LN also has uniaxial negative birefringence, and is a photorefractive material. Note that, for quantum photonics applications, photorefractivity is mainly a detrimental effect, since pump light fields below 800\,nm dynamically changes the spatial-mode properties of the waveguides even at small pump powers. As will be shown later, this has repercussions on nonlinear efficiencies as soon as photon pair generation is considered since this effect randomly changes the phase matching condition (see section later) over time. Practically, this problem is circumvented by operating the crystal at a temperature above 50\,$^\circ$C to allow for fast charge recombination in the waveguide and therefore smoothing the waveguide properties over the measurement time.

\subsection{Nonlinear interactions using PPLN}
\label{Sec_NLI}

\begin{figure}[h]
\begin{center}
\includegraphics[width=0.5\textwidth]{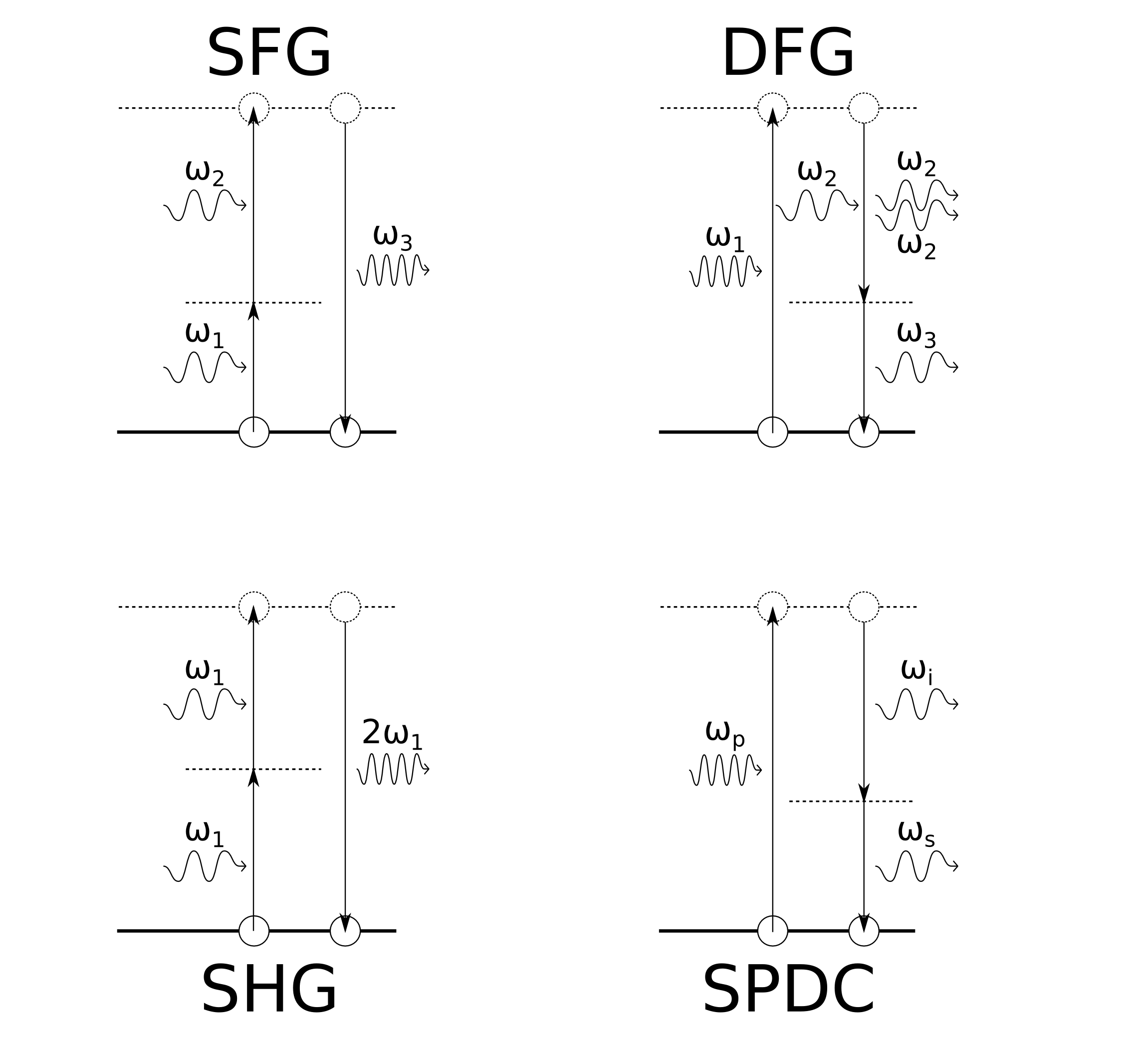}
\caption{Nonlinear three-wave mixing parametric processes: Sum Frequency Generation (SFG), Spontaneous Parametric Down-Conversion (SPDC), Second Harmonic Generation (SHG, a particular case of SFG in which $\omega_1=\omega_2$), Difference Frequency Generation (DFG, a SPDC process stimulated by injection of a wave at frequency $\omega_2$). Thick lines represent the ground energy level of the medium, dashed lines represent virtual states of excitation.\label{NLmixing}}
\end{center}
\end{figure}

Light propagating through a LN crystal induces an electric polarization in the material that can be approximated by a Taylor series as:
\begin{equation}
\vec{P}=\epsilon _0 \left( \chi ^{(1)}\vec{E} + \chi ^{(2)}\vec{E}\vec{E} +\chi^{(3)}\vec{E}\vec{E}\vec{E}+...\right) , 
\label{NLP}
\end{equation}
where $\epsilon_0$ is the vacuum permittivity, $\vec{E}$ the propagating electric field, and $\chi^{n}$ the $n^{\rm th}$ order electric susceptibility of the material\footnote{Even orders of $\chi$ exist only in non-centrosymmetric materials.}. Since the nonlinear coefficient decreases rapidly with the considered order ($\chi^{(n)}~\gg~\chi^{(n+1)}$) and LN is non-centrosymmetric, we limit our description to second-order nonlinearities, and more specifically to parametric effects, especially three-wave mixing~\cite{armstrong_1962}. In these processes, three electromagnetic fields interact in the dielectric medium, and energy is exchanged between the fields. This process obeys to energy conservation laws summarized in Tab.~\ref{Econservation}, and it
\begin{table}[h]
\begin{center}
\begin{tabular}{|l|c|c|c|}
  \hline
  Process & \multicolumn{2}{|c|}{Frequencies} & Conservation\\
  \cline{2-3} involved &  input & output & law\\
  \hline
  SFG & $\omega_1$, $\omega_2$ & $\omega_3$ & $\omega_3 =\omega_1+\omega_2$ \\
  SHG & $\omega_1$, $\omega_1$ & $\omega_3$ & $\omega_3=2\omega_1$ \\
  DFG & $\omega_1$, $\omega_2$ & $\omega_2$, $\omega_2$, $\omega_3$ & $\omega_3=\omega_1-\omega_2$ \\
  \hline
  SPDC & $\omega_p$ & $\omega_s$, $\omega_i$ & $\omega_p= \omega_s+\omega_i$ \\
  \hline
\end{tabular}
\caption{Energy conservation rules of nonlinear parametric processes.}
\label{Econservation}
\end{center}
\end{table}
can be seen as successive transitions between virtual energy levels. Following this representation, three kinds of processes can be identified, as depicted in \figurename~\ref{NLmixing}:
\begin{itemize}
\item \emph{Sum-frequency generation (SFG)}: here, two photons at frequency $\omega_1$ and $\omega_2$ are annihilated and one is created at a frequency $\omega_3$ corresponding to the sum frequency $\omega_1+\omega_2$. A particular case is the second-harmonic generation (SHG) where the two input photons have the same energy~\cite{armstrong_1962};

\item \emph{Difference frequency generation (DFG)}: here, two photons at frequencies $\omega_1$ and $\omega_2$ are injected in the crystal ($\omega_1\geq\omega_2$); one photon at $\omega_1$ is annihilated, while two photons are created, respectively at frequencies $\omega_2$ (via stimulated emission) and $\omega_3$, with $\omega_3=\omega_1-\omega_2$.
\item 

\emph{Spontaneous parametric down-conversion (SPDC)}: here, one input pump photon at frequency $\omega_p$ is annihilated in the medium and two photons, signal at $\omega_s$ and idler at $\omega_i$ are spontaneously created. Differently from previous processes, only the upper virtual level of energy is fixed by the input frequency. The intermediate levels are free, so that a continuum of transitions is possible.
\end{itemize}
It is interesting to note that those three processes are fully exploited in quantum photonics for various applications as outlined in the following sections.

As an example among possible nonlinear optical effects with second-order materials, the SPDC process is widely exploited for photon-pair generation (see Section~\ref{Sec_EPPS}). As for all other processes, the nonlinear coefficient of the material couples the three interacting fields, via a given nonlinear coefficient, depending on their polarization states. In the case of PPLN, among all possible nonlinear interactions, three are widely exploited in quantum photonics applications. Type-0 interaction couples, via the $d_{33}\simeq$27\,pm/V, pump, signal, and idler fields that are all polarized along the extraordinary ($e$) axis of the crystal. Type-I interaction couples, via the $d_{31}\simeq$5\,pm/V, an $e$-polarized pump field with ordinary ($o$) polarized signal and idler fields. Finally, type-II interaction couples, via the $d_{24}$ ($=d_{31}$ for symmetry reasons), an $o$ polarized pump field with orthogonally polarized signal and idler fields.

\begin{figure}[h]
\begin{center}
\includegraphics[width=0.5\textwidth]{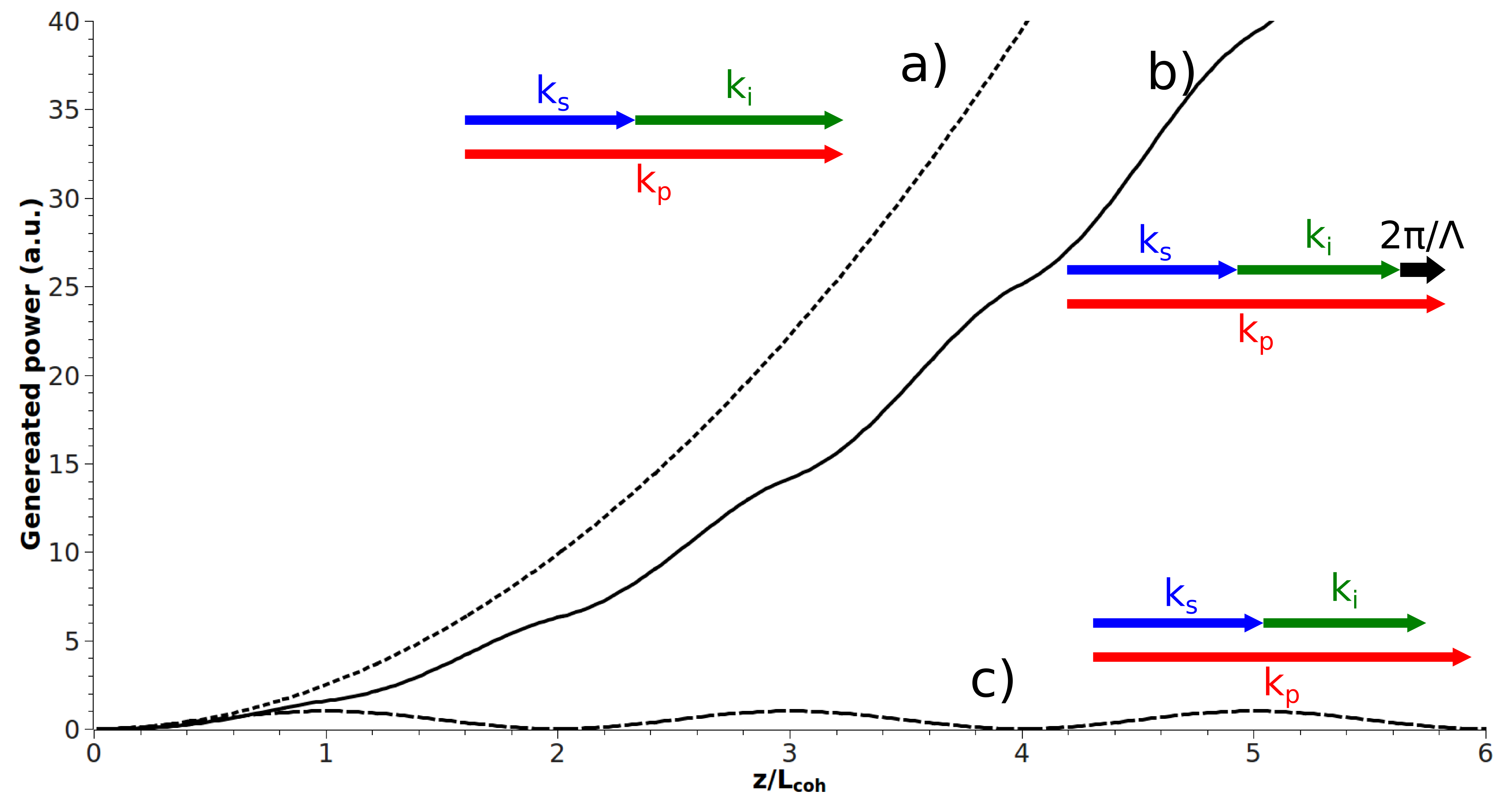}
\caption{Power generated as a function of the propagation length for different phase-matching conditions: a)~perfect phase matching (dotted), b)~quasi-phase-matching (solid), c)~phase mismatch (dashed). Inset drawings show the vectorial representation of the phase-matching condition, where the indices $\{i,j,k\}$ represent the three interacting fields.}
\label{QPM}
\end{center}
\end{figure}

However, nonlinear mixing must also obey phase matching condition (also referred to as momentum-conservation in the particle-like picture). For instance, for the SPDC process, this imposes $\vec{k}_p = \vec{k}_s+\vec{k}_i$, with $k_{p,s,i}$ the $k$-vectors of the three coupled fields. When this condition is fulfilled, constructive interference occurs all along the crystal and the efficiency of the nonlinear process increases quadratically with the interaction length (see \figurename~\ref{QPM}, curve a), as is the case for SFG, SHG, and DFG processes. Oppositely, when the $k$-vectors are not perfectly phase-matched, the interaction between the three beams shows constructive interference only over a length called the coherence length ($L_{\rm coh}$) after which the above mentioned interference becomes destructive. Consequently, constructive and destructive interference occur repeatedly, leading to a strongly reduced efficiency which oscillates with the propagation distance through the crystal as shown in \figurename~\ref{QPM} (curve c). 

Because of chromatic dispersion, perfect phase matching (PM) condition is not easily fulfilled for an arbitrary combination of wavelengths~\cite{miller_1965}. To achieve this, different techniques have been demonstrated:
\begin{itemize}
\item Birefringent PM: the exact dispersion compensation is obtained by exploiting the difference of refractive indices between different polarizations. Even if it can be tuned by temperature or angle variations, PM cannot always be fulfilled depending on the involved wavelengths and material. In addition, the temperature at which PM is reached might be incompatible with other practical constraints, for instance, LN is photorefractive and it is usually required to operate above room temperature;
\item Modal PM~\cite{fejer_1986}: it exploits the refractive index mismatch between different propagating spatial mode orders. Once again, this PM condition cannot be achieved for a large set of parameters;
\item Quasi-phase matching (QPM)~\cite{armstrong_1962} relies on the periodical poling of the crystal polarization and, consequently, the sign of the nonlinear coefficient. The fundamental advantage of this approach is that almost any phase mismatch can be compensated by appropriately engineering the poling period. Therefore, an extremely large set of wavelengths can be addressed. 
\end{itemize}

\begin{figure}[h]
\begin{center}
\includegraphics[width=0.5\textwidth]{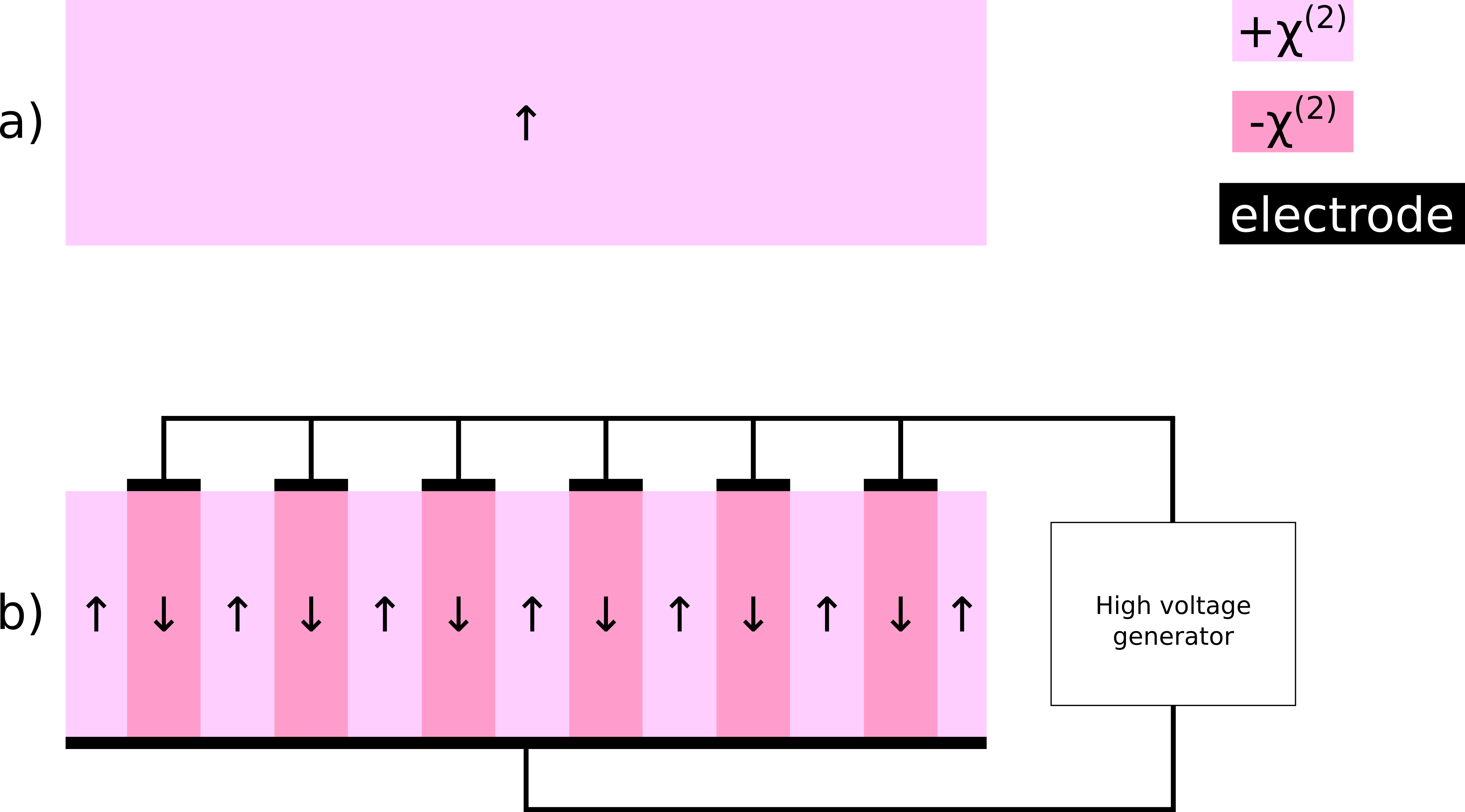}
\caption{a) Bulk LN crystal with its spontaneous polarization oriented in $+Z$ (crystallographic axis); b) The same crystal after a high voltage field has been applied through electrodes: crystal cells located between electrodes show a polarization inverted.}
\label{PPLN}
\end{center}
\end{figure}

Among these techniques, QPM is the best established one and is used for a wide range of frequency conversions. Although many techniques of micro- or nano-domain engineering have been demonstrated~\cite{shur_2015}, the most common and effective way to reverse the material spontaneous polarization consists in applying an intense electric field (typical voltage $\sim$~21\,kV.mm$^{-1}$ for LN) through the crystal as schematically represented in \figurename~\ref{PPLN}. The periodic poling is imprinted through standard photolithographic techniques by depositing an electrode pattern on the crystal, either using a metallic deposition~\cite{yamada_1993} or a electrolytic liquid~\cite{webjorn_1994}. Periodicity typically varies between 5 to 35\,$\mu$m. LN substrates prepared this way are referred to as periodically-poled lithium niobate (PPLN).

\subsection{Enhancing nonlinear effects with waveguides}
\label{subsec:NLOwg}

By definition, an integrated optical circuit must include structures that allow guiding of light between different locations of the chip. Waveguides are obtained by locally increasing the refractive index of the considered material. Typically, LN waveguides feature widths of a few microns and lengths of up to 10\,cm. Common increases of refractive index ($\delta$n) are reported in table~\ref{tab_wg}. Waveguide architecture provides the following advantages:
\begin{itemize}
\item higher confinement of the electric fields leading to increased nonlinear process efficiencies (several order of magnitudes compared to bulk-crystal approaches);
\item longer interaction lengths (centimeter scales) which also enhance nonlinear process efficiencies;
\item integration of on-chip linear-optic functions (tunable couplers, wavelength demultiplexers as well as phase shifters);
\item efficient interfacing with optical-fiber technologies.
\end{itemize}

Two prime techniques have emerged for waveguide fabrication on LN substrates: titanium doping by thermal indiffusion~(Ti-ind) and proton exchange~(PE). In addition, many research efforts are currently deployed to create new waveguide structures that will be discussed at the end of this section. 

Titanium indiffusion stands as the oldest technique~\cite{sugii_1978}. Although it allows to fabricate waveguides with lower optical confinement, this technique increases the refractive index of both ordinary and extraordinary axis. Therefore, Ti-ind structures guide both transverse electric (TE) and transverse magnetic (TM) polarization modes~\cite{sohler_1989}. In addition, Ti-ind process preserves the nonlinear properties of the LN~\cite{Hofmann_99}.
A thin titanium film (of typically 100\,nm thick) with the desired waveguide pattern is deposited on the substrate surface through standard lithographic processes. Ti-stripes are later diffused into the substrate for several hours at high temperature ($\sim$1000$^{\circ}$\,C) in controlled atmosphere. The periodic poling is made after the waveguide fabrication, since poling does not withstand such high temperatures~\cite{nouroozi10}.

All proton-exchange~(PE) fabrication techniques are based on the substitution of lithium ions by protons in the crystal. Compared to Ti-ind, PE leads to higher confinement but increases only the refractive index of the extraordinary axis (decreasing the refractive index of the ordinary axis), so only TM-polarized light can be guided. The fabrication process starts by periodically poling the crystal. Then, a buffer layer (usually silica) with the reciprocal pattern of desired waveguides is deposited using standard lithographic techniques. The samples are later immersed for several hours into a liquid source of protons (usually pure benzoic acid) at a temperature between 160 and 200\,$^{\circ}$\,C. During this process, Li ions out-diffuse from the crystal matrix and are replaced by protons that indiffuse into the medium. After this step, the concentration of protons accumulated at the surface is so high that the superficial crystalline structure layers are altered, leading to high optical losses and an almost suppressed nonlinear coefficient. To restore the optical properties of the LN material, PE is complemented with one (or two~\cite{Roussev06}) annealing step that diffuses protons deeper into the substrate and relaxes the stress in the crystalline structure, partially restoring the nonlinear properties of the material~\cite{bortz93}. This technique, called annealed-proton exchange~(APE), allows to fabricate waveguides with low transmission losses and high conversion efficiency (see Tab.~\ref{tab_wg}).

To improve this technique and exploit the full nonlinear coefficient of LN, a further diffusion step in a lithium-rich bath can be introduced. While protons are diffusing within the material, lithium ions take their place back, replacing protons closer to the surface. This improved version of APE is called reverse-proton exchange (RPE)~\cite{Roussev06, Lenzini15}. After RPE, the confinement region is buried below the surface where the nonlinear properties of the material remain intact. In addition, the refractive index profile has a circular symmetry such that the overlap between the three mixing modes is maximal, resulting in the most efficient nonlinear parametric process~\cite{Parameswaran02}. The circular symmetry also allows better coupling with gaussian-shaped modes, as those guided in single-mode fibres. However, the optimization of the two diffusion processes is challenging and so far electrode control of the buried waveguide properties have never been demonstrated.

A third PE variation technique referred to as soft-proton exchange (SPE) has been developed to keep the mode close to the surface without affecting the nonlinear coefficient~\cite{chanvillard_2000}. With this technique, the exchange is performed at higher temperatures ($\sim$\,300\,$^{\circ} $C) in a proton bath buffered with lithium ions. The less acid bath reduces the proton substitution rate while the higher temperature increases the proton diffusion speed, preventing their accumulation at the surface. This soft alteration of the crystalline structure preserves the LN optical properties. In addition, this technique allows higher refractive index jumps leading to a more efficient optical confinement with respect to the other techniques presented above~\cite{Korkishko_01}. 

\begin{table*}[!hbtp]
\begin{center}
\begin{tabular}{|c|c|c|c|c|c|}
\hline
{\bf Technique} & {\bf $\Delta$n @ 632 nm} & {\bf Polarization}&{\bf Transm. Loss} & {\bf Note}\\
\hline
Ti-ind & $\leq$0.01 & TM \& TE & $\leq$0.1 dB/cm & Nonlinearity preserved\\
\hline
PE & 0.1 & TM & $\geq$1 dB/cm & No nonlinearity\\
\hline
APE & 0.02 & TM & $\leq$0.1 dB/cm & Nonlinearity reduced\\
\hline
RPE & 0.015 & TM & $\leq$0.1 dB/cm & Buried waveguide, nonlinearity preserved\\
\hline
SPE & 0.03 & TM & $\leq$0.1 dB/cm & Surface waveguide, nonlinearity preserved\\
\hline
\end{tabular}
\end{center}
\caption{Summary of common waveguide integration techniques on LN associated with their respective properties.\label{tab_wg}}
\end{table*}

For all the above mentioned techniques, the fabrication parameters are chosen by engineering the effective index of the modes propagating in the waveguide. This depends on the geometrical characteristics of the waveguide (width, depth) and on the refractive-index profile~\cite{Roussev06, Korkishko_01} induced by the waveguide fabrication. The latter information is usually obtained using the so-called m-line technique~\cite{Tien_71, Ulrich_73}. Those techniques are currently very well mastered and are even used at the commercial scale. However, there are many research groups developing more advanced waveguide structures to address some particular issues, such as weak mode confinement, large mode asymmetry or propagation losses due to crystal structure modification. At the moment, those waveguide devices have mainly been used for classical functions (second harmonic generation or wavelength filtering), but it is reasonable to expect a direct transfer toward quantum photonics. However, those structures are hardly scalable for now and the coupling efficiency with single mode fibre remains an issue. Despite those problems, they surely display what the future of integrated quantum photonics will be made of.

Basically, they differ from proton exchanged or indiffusion techniques by a surface machining to create a ridge structure allowing much higher confinement~\cite{Geiss:10,Lu:12,Wang:14,Geiss:15} while avoiding substitution of lithium ions in the crystalline structure at the origin of excess losses. Unfortunately, the tolerance of such novel techniques induces rugosity along the ridge waveguide therefore introducing additional propagation losses. Moreover, the coupling efficiency to standard optical fiber has not been optimized yet and such structures report insertion losses over  6\,dB.

It is worth noting that the approach outlined in ref.~\cite{Chang:16} seems promising since it only relies on surface deposition of high index materials to create the guiding structure and avoid crystal structure modification or material machining.

\subsection{Characterization of LN chips}

Ideally, LN chips should feature low transmission losses and high nonlinear conversion efficiency in order to increase the success rate of quantum photonics experiment and drastically reduce the acquisition time. These quantities ultimately determine the final design of the chip and the optimization of the fabrication processes. Moreover, quantum states engineered through these photonic chips should be characterized in order to quantify their fidelity to the ideal state. In the following, we discuss how to characterize these quantities.

\subsubsection{Transmission losses} 
\begin{figure}[h!]
\begin{center}
\includegraphics[width=0.9\columnwidth]{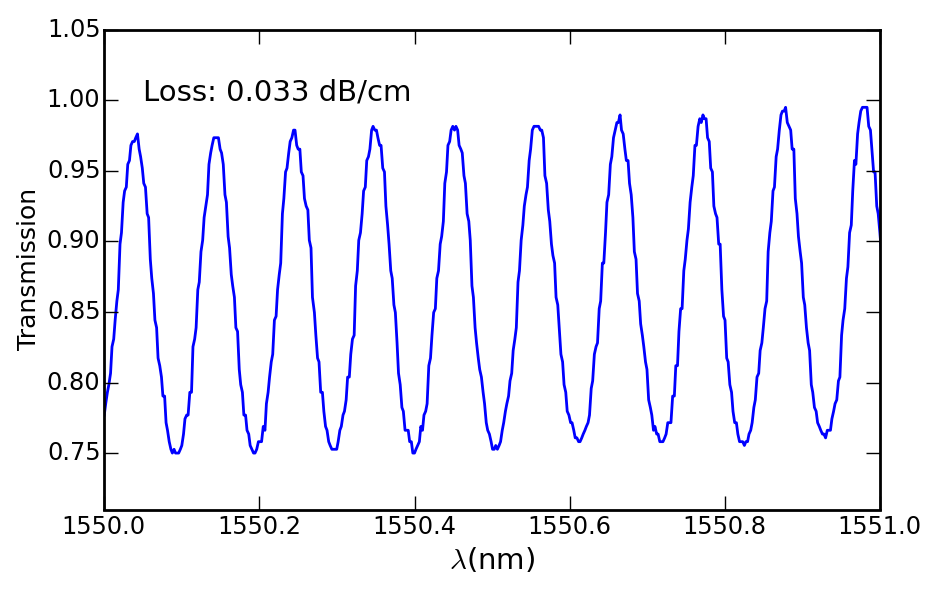}
\caption{Example of transmission curve obtained at the output of a home-made LN waveguide for losses estimation. The losses coefficient is inferred from Eq.~\ref{Eq_loss} taking into account the obtained contrast and reflectivity of the end factes.\label{Fig_loss}}
\end{center}
\end{figure}

Transmission losses ($\mu$) of a straight LN waveguide can be measured by exploiting the refractive index mismatch between LN ($\sim$2.15 at 1550\,nm) and air ($\sim$1). A low-finesse cavity is created by polishing a straight waveguide on both end-facets orthogonally to the direction of the propagating beam. 
\figurename~\ref{Fig_loss} gives an example of the transmission spectrum of the light exiting from the waveguide. The contrast~($C$) of the Fabry-Perot resonances is used to measure $\mu$~\cite{Regener85}, which reads 
\begin{equation}\label{Eq_loss}
\mu = 4.34\Big[{\rm ln}\frac{1-\sqrt{1-C^2}}{C}-{\rm ln}(R)\Big],
\end{equation}
where $R$ is the end-facet (Fresnel) reflection coefficient
\begin{equation}
R = \frac{(n_{eff}-1)^2}{(n_{eff}+1)^2},
\end{equation}
and $n_{eff}$ is the effective index of the waveguide.

While the sensitivity of this technique increases for smaller losses, it can only provide an upper bound on $\mu$. In practice, the contrast of the fringes is also  decreased by the quality of the polishing at the input/output facets.

\subsubsection{SPDC conversion efficiency and joint-spectral amplitude}

When the aim is to generate correlated photon pairs for a particular purpose, the energy correlation of the produced photons is an important figure of merit that needs to be estimated. We describe here the solutions that have emerged from the quantum community to provide such figures.
The photon quantum state is derived from a nonlinear Hamiltonian associated with SPDC process as outlined in~\cite{Migdall2013}. After an interaction length $L$, in the so-called low-pumping regime approximation, the bi-photon quantum state reads:
\begin{eqnarray}
|\psi\rangle_{SPDC} = |0\rangle+\\
\eta\int\int d\omega _s d\omega _i f(\omega _s, \omega _i)e^{-i\Delta k L /2}\hat{a}^\dagger (\omega _s) \hat{a}^\dagger (\omega _i) |0\rangle ,
\end{eqnarray}
where $\eta$ is a constant related to the probability of the spontaneous process, 
$\hat{a}^\dagger (\omega _{s, i})$ are the creation operators of signal and idler photons respectively, $\Delta k = k_p-k_s-k_i-\frac{2\pi}{\Lambda}$ is the QPM condition which relates the three mixing fields and the poling period $\Lambda$, and
${f(\omega_s, \omega_i) = \alpha(\omega_s + \omega_i) {\rm sinc}\Big[\Delta k\frac{L}{2}\Big]}$ is a function related to the pump spectral distribution $\alpha(\omega_s + \omega_i)$ and to the QPM conditions.

\begin{figure}[h!]
\begin{center}
\includegraphics[width=\columnwidth]{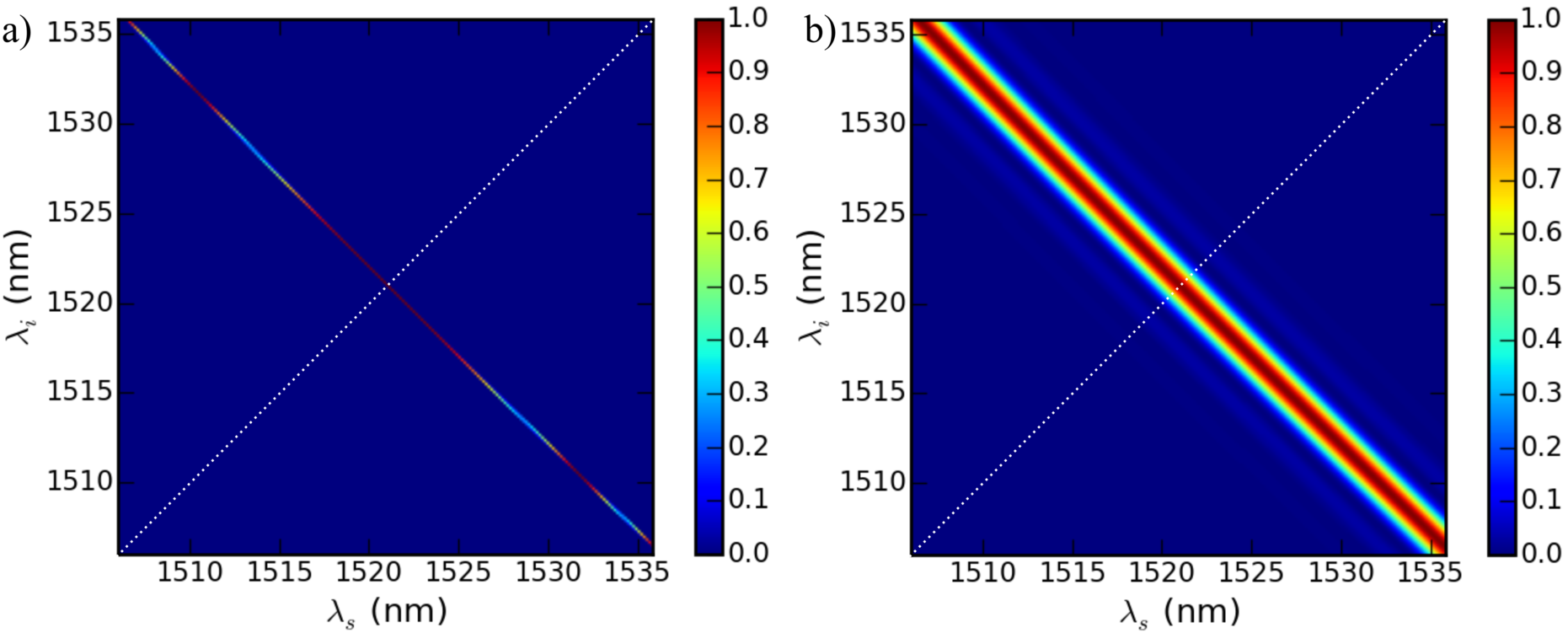}
\caption{Calculated JSA for a type-0 PPLN based SPDC source under (a) CW and (b) pulsed pumping regimes, respectively. The width of the joint spectrum is related to the spectral purity of the source. A broader line indicates a statistical mixture of different spectral modes. The curves are calculated following the model presented in Ref.~\cite{Kim_05}.\label{Fig_JSA}}
\end{center}
\end{figure}

$f(\omega _s, \omega _i)$ is called joint spectral amplitude (JSA) function and $\vert f(\omega_s, \omega_i)\vert^2$ is the joint spectral intensity (JSI) function~\cite{Brecht_13, Harder_13}. This figure not only quantifies the probability distribution of the bi-photon state, but also provides direct information about its purity~\cite{Kim_05}. Here, we only present an intuitive approach and we refer to~\cite{Brecht_13, Migdall2013} for a more rigorous discussion. Figure~\ref{Fig_JSA} shows two simulated JSA functions for a SPDC process pumped either with (a) a continuous wave (CW) pump or (b) an ultrafast pulsed pump. The JSA gives the degree of spectral correlations between the emitted two-photon state where narrower lines correspond to spectrally defined twins. Each photon of the pair corresponds to a unique spectral mode so that the output state is pure. Broadband pumps reduce the degree of spectral correlations between the twin photons which are emitted in a mixture of different spectral modes~\footnote{For sake of completeness let us mention that, differently from the JSA, JSI can provide only an upper bound for the purity of the quantum state~\cite{Harder_13, Jin_13}}.

To our knowledge, no direct measurement of the JSA have been reported so far, while the JSI is usually acquired trough spectrally resolved single-photon coincidence measurements~\cite{Bruno_14, Kim_05, Avenhaus_09}. In practice, this strategy depends on the brightness of the source and the maximal count rate of the single-photon detectors. These constraints impose a trade-off between the spectral resolution and the total integration time, resulting in hours-long measurements with modest spectral resolutions. Long integration times are discouraging, therefore this measurement is rarely implemented.

\begin{figure}[h!]
\begin{center}
\includegraphics[width=\columnwidth]{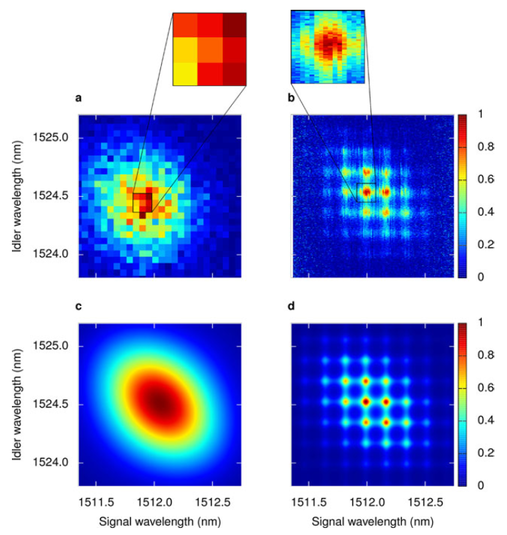}
\caption{(a)~Experimental JSI obtained with SPDC-based measurement with a sampling rate of 25$\times$25 pixels, and an integration time of 120~min. (b)~Experimental JSI obtained with DFG-based measurements with a sampling rate of 141$\times$501 pixels, an integration time of 45\,min. (c)~Theoretical prediction for the SPDC measurement. (d)~Theoretical prediction for the DFG measurement. Figures and data reproduced from Ref.~\cite{Eckstein_15} with permission of the auhthors.\label{Fig_JSI}}
\end{center}
\end{figure}

In Ref.~\cite{Liscidini_13}, an interesting alternative approach has been proposed to measure the JSI. The authors begin with an analogy between SPDC and DFG processes stimulated by vacuum fluctuations. This analogy is supported by the fact that DFG and SPDC share the same phase-matching configuration and pump spectrum, and thus the same spectral correlation function~\cite{Helt_12}. Standard DFG process can be seen as amplified-spontaneous-emission. In other words, by seeding a nonlinear source with a strong pump and an idler beam, the resulting difference-frequency beam is $N(\omega_i)$ times more powerful than the corresponding spontaneous signal. $N(\omega_i)$ corresponds to the number of photons in the idler beam. Within this framework, the spontaneous emission is amplified by several ($\sim$~8-9) orders of magnitudes above usual SPDC process. Therefore, single-photon detectors can be replaced by classical optical spectrum analyzer and the entire SPDC spectrum can be characterized by simply scanning the idler laser. This analogy has been experimentally verified in several works investigating different photonic platforms~\cite{Eckstein_15,Jizan_15,Fang_14} or different observables, \textit{e.g.} polarization~\cite{Rozema_15}. Figure~\ref{Fig_JSI} (a) and (b) show a direct comparison between standard and stimulated-DFG approaches. Thanks to the extremely high resolution offered by the DFG technique, measurements are able to resolve the Fabry-Perot interferences within the waveguide due to the high index mismatch at the waveguide facets. The results agree with the theoretical predictions (reported in \figurename~\ref{Fig_JSI} c and d for both approaches). In addition to the extremely higher-resolved measurement (100 times more points), DFG measurements takes only one third of the time.\newline

In summary, regarding current technology, LN is one of the most widespread substrate to develop integrated quantum circuits. It allows efficient generation of quantum states through nonlinear processes, because of its highest non-linear coefficient. For quantum photonics purposes, it is extremely robust and versatile thanks to the exploitation of the QPM technique. Its manufacturing is well established, enabling the integration of almost all possible elementary optical function. This not only increases the efficiency of the nonlinear process but it also expands the number of optical functions that could be integrated on chip. 

The characterization of the different properties of integrated quantum chips suffers from the limited counting capabilities of current single-photon detectors. Surprisingly, quantum phenomena can sometimes be partially mimicked using classical stimulated process. The resulting classical measurements allow for faster acquisitions and larger spectral resolutions. An example of this approach, the spectral characterization of the bi-photon state, has been given.

\section{Entangled photon pair sources}
\label{Sec_EPPS}

An entangled state of two particles exhibits remarkable (non-classical) properties~\cite{Brunner_BellRMP_2014,Hensen_15,Putz_LDLPRL_2016,Nielsen00,Hall_Quantum_Challenge}. It is as if these particles would exhibit a synchronized randomness, which is a purely quantum mechanical feature with no equivalent in classical physics, where all measurement outcomes are predictable (at least in principle).

Quantum information science has found ways to exploit these unique features to develop new powerful quantum protocols and applications without classical equivalents. Some of the most prominent examples emphasizing the interplay between fundamental and applied aspects of the field are given by teleportation~\cite{Bennett_Qtele_1993,Bouwmeester_Qtele_1997,Boschi_Qtele_1998,kim_Qtele_2001,Marcikic_Qtele_2003,Ursin_Tele_Danube_2004,Landry_Qtele_PlainPalais_2007,Kaiser_cw_tele_2015}, entanglement swapping~\cite{Pan_Swap_1998,Kaltenbaek_Inter_Indep_2009,Aboussouan_dipps_2010,McMillan_SR2photI_2013}, also referred to as quantum relays~\cite{Collins_QRelays_2005}, and repeaters~\cite{briegel_quantum_1998,duan_LDQcom_2001,simon_quantum_2007,Sangouard_DLCZRMP_2011}, which are all enabled by entanglement~\cite{Brunner_BellRMP_2014}. From the more applied side, entanglement allows for absolutely secure communication~\cite{gisin_QKD_2002,Scarani_QKDRMP_2009,BB84,Ekert_Crypto_1991}, quantum cryptosystems being now commercially available~\cite{QKD_commercial}. It also allows to perform metrological measurements with accuracies outperforming classical devices~\cite{Dowling_Qmet_2008,Giovannetti2011,Crespi_protein_concentration_2012} and has the potential to solve complex mathematical problems in significantly reduced computational times~\cite{Peruzzo_2010,Deutsch85,DiVincenzo95,Nielsen00,O'brien03,Pittman03,Okamoto_cNOT_2011,Politi08,Marshall09,Smith09,Politi09,Matthews09,Sansoni_EntangChip_2010,Laing10,Matthews10}.
In principle, the above mentioned experiments and applications can be performed with any entanglement carrier and observable. However, photonic systems have been identified as predominant, thanks to their weak sensitivity against decoherence effects when they propagate in optical fibres, straightforward generation and detection, the availability of high-performance (fibre) optics, and their high speed of travel. Most future quantum systems will therefore likely rely on photonics, which is why it is important to continue developing efficient, stable, and compact entangled photon pair sources. 

This section is outlined as follows. In subsection~\ref{Pioneering_ent_sources}, some of the pioneering energy-time, time-bin, and polarization entanglement sources based on PPLN/W are described, which are still used in today's research laboratories. Subsection~\ref{Hybrid_polar_ent_sources} gives details about the realization of some high-performance polarization entanglement sources based on hybrid approaches. The idea for these sources is to exploit, at the same time, the highly efficient generation of energy-time entangled photon pairs, and the straightforward detection of polarization entanglement. Subsections~\ref{Freq_bin_ent_sources} and \ref{Path_ent_sources} are devoted to frequency-bin and path entangled photon pair sources based on PPLN/Ws. These rather new concepts might have a great impact for future quantum information applications, such as entanglement-based secret key distribution, metrology, and lithography. Finally, some advanced sources are described (subsection~\ref{Adv_phot_ent_sources}), for entanglement generation in both ultra-narrowband and broadband operation regimes. Narrowband sources are usually intended for light-matter studies, such as the development of quantum memories. On the other hand, broadband sources can be exploited for high-rate quantum key distribution in wavelength division multiplexed networks.

\subsection{Pioneering entanglement sources \label{Pioneering_ent_sources}}

\begin{figure}[h!]
\includegraphics[width=0.48\textwidth]{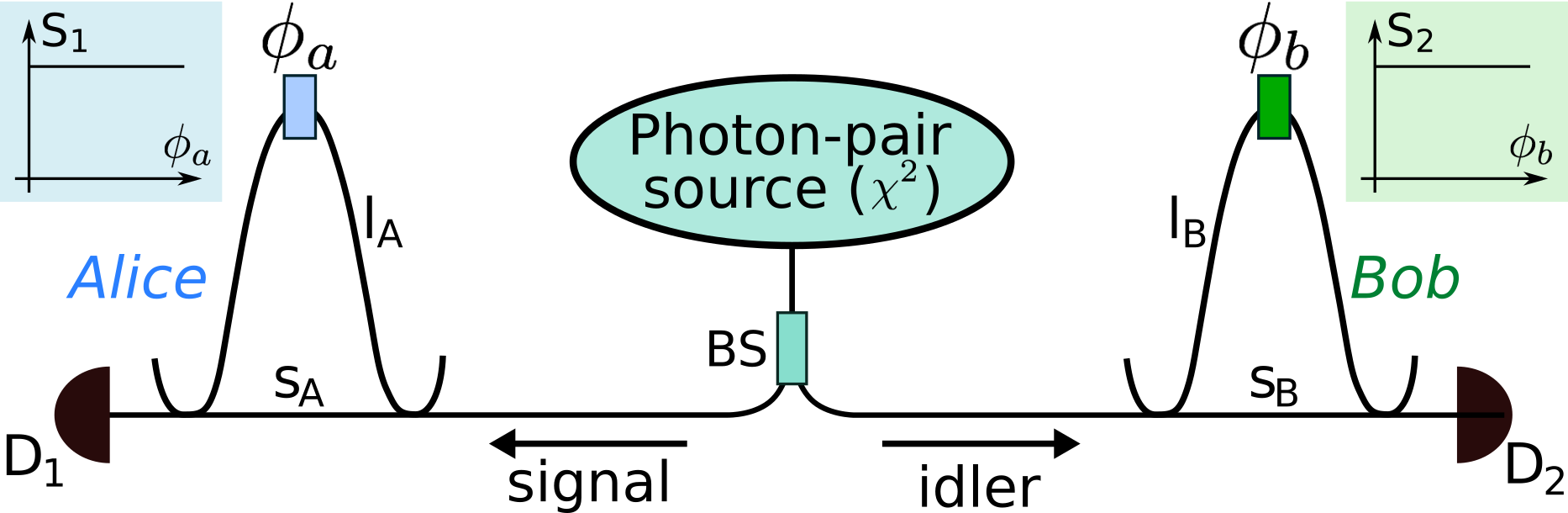}
\caption{Principle for generating and analysing energy-time entangled photons, as explained in Ref.~\cite{tanzilli_ppln_2002}. A source generates a flux of energy and time correlated photon pairs, which are sent, after separation at a beam-splitter (BS), to two unbalanced interferometers at Alice's and Bob's sites (Franson configuration~\cite{Franson_Bell_1989}). The individual photons can either travel along a short or a long path ($\rm s_{A,B}$ or $\rm l_{A,B}$). The phase difference between the two arms is given by $\phi_{\rm A,B}$, and the path length difference in both interferometers is set to be bigger than the coherence time of the individual photons to avoid single photon interference at each interferometer's output. In half of the cases, both photons take opposite paths. In the other half of the cases, both photons take the same path ($\rm s_A - s_B$ and $\rm l_A - l_B$), and those events can be post-selected from the others by time-tagging. If both photons exit the interferometer through the bottom output (towards detectors D$_1$ and D$_2$), then they are projected into the following state: $| \psi \rangle \propto |{\rm s_A} \rangle |{\rm s_B} \rangle + e^{i\,\left( \phi_{\rm A} + \phi_{\rm B}\right)} |{\rm l_A} \rangle |{\rm l_B} \rangle$. The contributions $|{\rm s_A} \rangle |{\rm s_B} \rangle$ and $|{\rm l_A} \rangle |{\rm l_B} \rangle$ become indistinguishable when the photon pair generation time is unknown to more than the travel time difference between short and long paths, which can be achieved by a continuous wave pump laser for which the spontaneous generation of photon pairs inside the PPLN/W occurs at totally random times (energy-time entanglement). Or, a pulsed pump may be used, for which photon pairs can be spontaneously generated at two different times, which match the travel time difference of the interferometers (time-bin entanglement). Both strategies make it impossible to know whether both photons took the short or long paths. In this case, entanglement is observed, which is expressed as a phase-dependent sinusoidal oscillation of the two-photon coincidence rate, $R_{\rm c}$, between the two detectors, $R_{\rm c} \propto 1 + V \cdot \sin \left( \phi_{\rm A} + \phi_{\rm B}\right)$. Here, $V$ is the visibility of the interference finges, which has to be above 71\% to demonstrate the presence of entanglement.\label{Fig_Energy-time_Tanz_2002}}
\end{figure}

In 2001 and 2002, two pioneering photonic entanglement sources were demonstrated~\cite{sanaka_new_2001,tanzilli_ppln_2002}, generating energy-time and time-bin entanglement by taking advantage of the highly efficient type-0 nonlinear interaction in a PPLN/W. The waveguide structure was fabricated using the soft proton-exchange technique. Through spontaneous parametric down conversion (SPDC), a vertically polarized pump photon, $| V \rangle_{\rm p}$, is converted to a pair of vertically polarized signal and idler photons, $|V \rangle_{\rm s} |V \rangle_{\rm i}$, which are wavelength correlated by energy-conservation. Entanglement may then be revealed using suitable unbalanced interferometers, referred to as Franson configuration~\cite{Franson_Bell_1989}, as explained in the caption of \figurename~\ref{Fig_Energy-time_Tanz_2002}.

In 2001, a group in Tokyo (Japan) generated photon pairs at 854\,nm and revealed energy-time entanglement with an unbalanced free-space Michelson interferometer, obtaining two-photon interference raw visibilities of $80 \pm 10\%$~\cite{sanaka_new_2001}. Conversely, a collaboration between Geneva university (Swiss) and Nice university (France) used a fully guided wave approach in which they generated photon pairs in the telecom range (1310\,nm)~\cite{tanzilli_ppln_2001}, distributed them in standard optical fibres, and attested for both energy-time and time-bin entanglement. Two-photon interference fringe visibilities of 92\% and 84\% were obtained, respectively with error bars of about 1\% only~\cite{tanzilli_ppln_2002}, representing a clear violation of the Bell inequalities ($>71\%$)~\cite{Clauser_consequences_1974}. Shortly after that, the Geneva/Nice collaboration also showed that entanglement is preserved even after distribution over 11\,km in a fibre link within the Swiss telecom network~\cite{Thew_nonmax_2002}, which opened new perspectives in the field of quantum communication. One of the limitations of energy-time and time-bin entanglement is that half of the photon pairs are intrinsically lost, as they are projected into non-entangled time-bins.

\begin{figure}[h!]
\includegraphics[width=0.48\textwidth]{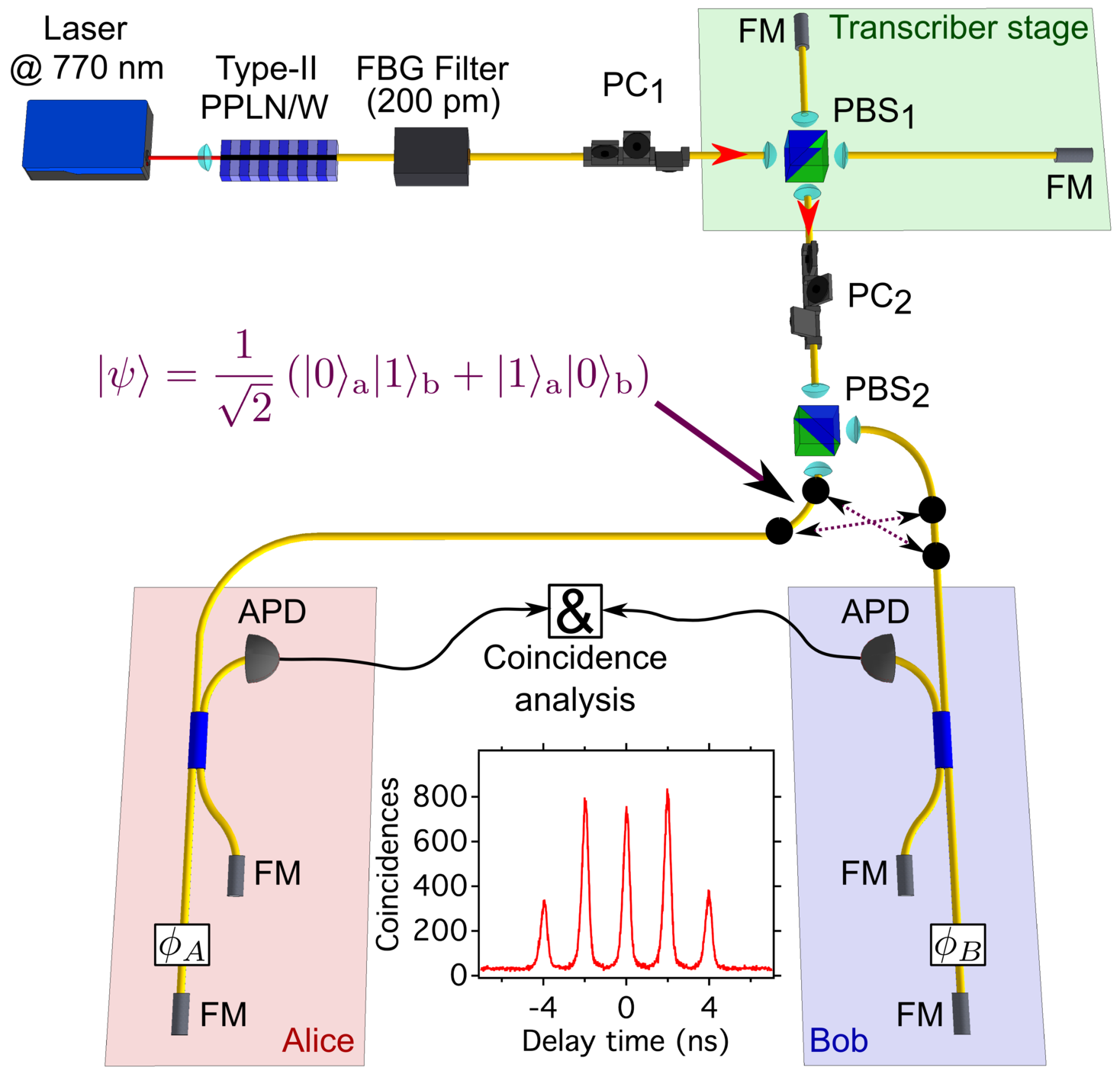}
\caption{Time-bin source setup in which only 25\% of the paired photons projected into non-entangled time-bins. A 770\,nm CW laser pumps a type-II PPLN/W in which cross-polarized photon pairs at 1540\,nm are generated ($| H \rangle | V \rangle$). A transcriber stage, composed of a polarizing beam-splitter (PBS) and two Faraday mirrors (FM) introduces a time-delay between the two photons. When the paired photons are separated on PBS$_2$, a cross-time bin state is generated, $|\psi \rangle = \frac{1}{\sqrt{2}} \left( |0 \rangle_{\rm A} |1 \rangle_{\rm B} + |1 \rangle_{\rm A} |0 \rangle_{\rm B} \right)$. Analysing this state with standard Franson-type unbalanced interferometers leads to a coincidence histogram exhibiting five coincidence peaks. Entanglement is measured in the central three peaks, corresponding to 75\% of the generated photons, as oposed to 50\% for standard energy-time experiments. Figure inspired from Ref.~\cite{Martin_TB_2013}.  \label{Fig_TimeBinNew}}
\end{figure}

Our group, in collaboration with teams in Marcoussis (France) and Singapore have recently demonstrated an advanced approach where only 25\% of the detected pairs are projected into non-entangled time slots. The corresponding experimental setup is shown in \figurename~\ref{Fig_TimeBinNew}. We used a PPLN/W, in which the waveguide was realized through titanium indiffusion, which allows guiding horizontal and vertical polarization modes, therefore making it possible to exploit the type-II SPDC interaction. Here, a horizontally polarized pump photon, $| H \rangle_{\rm p}$, gives birth to a pair of cross-polarized signal and idler photons, $|H \rangle_{\rm s} |V \rangle_{\rm i}$. Via a transcriber stage, we introduce a polarization dependent delay between the two photons, allowing to generate two photon states in which both photons are in opposite time bins, which can be used advantageously to reduce the above-mentioned projection losses~\cite{Martin_TB_2013}.

Another heavily exploited entanglement observable is polarization, which can be straightforwardly generated by exploiting type-II SPDC in a PPLN/W. By sending the cross-polarized photons, $|H \rangle_{\rm s} |V \rangle_{\rm i}$, to a non-polarizing beam-spitter, and considering only cases in which the pair has been split up, the following state is generated:
\begin{equation}
| \psi \rangle = \frac{1}{\sqrt{2}} \left( |H \rangle_{{\rm s},a} |V \rangle_{{\rm s},b} + |V \rangle_{{\rm s},a} |H \rangle_{{\rm s},b} \right),
\end{equation}
in which $a$ and $b$ denote the two different spatial modes after the beam-splitter.
By ensuring that signal and idler photons are indistinguishable over all observables, except their polarization modes, a maximally polarization entangled state is generated, $| \psi \rangle = \frac{1}{\sqrt{2}} \left( |H \rangle_a |V \rangle_b + |V \rangle_a |H \rangle_b \right)$. Indistinguishability is typically achieved by QPM engineering and filtering (spectral modes), a birefringent walk-off compensator (temporal modes), and choosing appropriate optical components (polarization dependent losses).
In this perspective, in 2007, a group in Osaka (Japan) exploited the type-II interaction in a PPLN/W for the first time to generate polarization entangled photon pairs at 1560\,nm. In a Bell inequality test, raw visibilities of $\sim 76\%$ were reported, despite a fairly low two-photon coincidence rate on the order of $\sim$ 1\,s$^{-1}$~\cite{suhara_generation_2007}. By optimizing the experimental setup using fibre optics components and reducing losses, our group demonstrated coincidence rates of $\sim$ 1\,kHz and raw visibilities of about 83\%~\cite{martin_polar_2010}.

\begin{figure}[h!]
\includegraphics[width=0.48\textwidth]{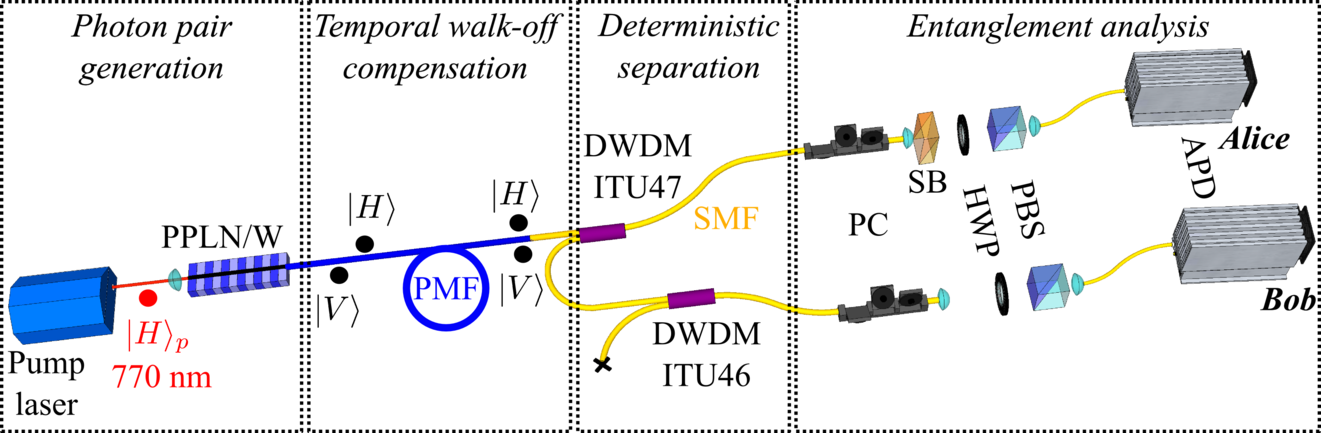}
\caption{The PPLN/W based polarization entangled photon pair source with deterministic pair separation is composed of the following components. A 770\,nm pump laser is coupled via a mirror (M) and lens (L) into a titanium indiffused PPLN/W and type-II SPDC is exploited for generation of cross-polarized photon pairs. The temporal walk-off is compensated using a standard polarization maintaining fibre (PMF) in which the sign of birefringence is inverted compared to the PPLN/W. Photon pairs are deterministically separated using two standard dense wavelength division multiplexers (DWDM) and entanglement is analysed using the standard setup, made of fibre polarization controllers (PC), half-wave plates (HWP), a Soleil-Babinet phase compensator (SB), polarizing beam-splitters (PBS), and avalanche photo detectors (APD) in the Geiger mode. A logic AND gate (\&) registers coincidences between the detectors. Figure inspired from Ref.~\cite{Kaiser_TypeII_2012}. \label{Fig_type-II_Kaiser_2012}}
\end{figure}

Note that for the above-mentioned experiments, photon pairs have been split up probabilistically, \textit{i.e.}, using a non-polarizing beam-splitter, which leads to a 50\% loss in the coincidence rate~\cite{suhara_generation_2007,martin_polar_2010}. 
In 2012, our team demonstrated that type-II-generated photon pairs can be deterministically separated into two standard telecommunication wavelength channels. As shown in \figurename~\ref{Fig_type-II_Kaiser_2012}, we used an all-guided wave approach made of off-the-shelves telecom components, \textit{i.e.}, a polarization maintaining fibre for temporal walk-off compensation, and standard dense wavelength-division multiplexers for deterministic photon pair separation. Using low-noise InGaAs single photon detectors, raw fringe visibilities exceeding 97\% have been obtained at coincidence rates up to $\rm$1\,kHz~\cite{Kaiser_TypeII_2012}. Another strategy based on an interlaced bi-poling grating structure on a single type-II PPLN/W has been proposed in 2009~\cite{Thya_type-II_proposal}, and demonstrated by two groups in Paderborn (Germany) and Osaka (Japan)~\cite{Herrmann_postselection_free_2013,Suhara_2QPM_for_polar_2009}. Here, two QPM conditions are obtained simultaneously, generating non-degenerate signal and idler photons with opposite polarization modes which can be conveniently separated using coarse wavelength splitters. Entanglement visibilities of up to 95\% have been reported, which underlines the potential of this approach.

For more details on the above-mentioned pioneering entanglement sources based on PPLN/Ws, we recommend also a very extensive review paper, written by T. Suhara in 2009~\cite{Suhara_PPLNW_review}.

\subsection{Hybrid polarization entanglement sources \label{Hybrid_polar_ent_sources}}

Despite the relatively narrow emission spectrum, which can be advantageous for some applications, type-II SPDC is unfortunately several orders of magnitude less efficient compared to the type-0 interaction. However, the latter interaction is not capable of generating polarization entanglement in a natural fashion.
In order to exploit the high efficiency, and the broadband emission of type-0 sources for polarization entanglement generation, several (interferometric) hybrid approaches have been developed.

\begin{figure}[h!]
\includegraphics[width=0.48\textwidth]{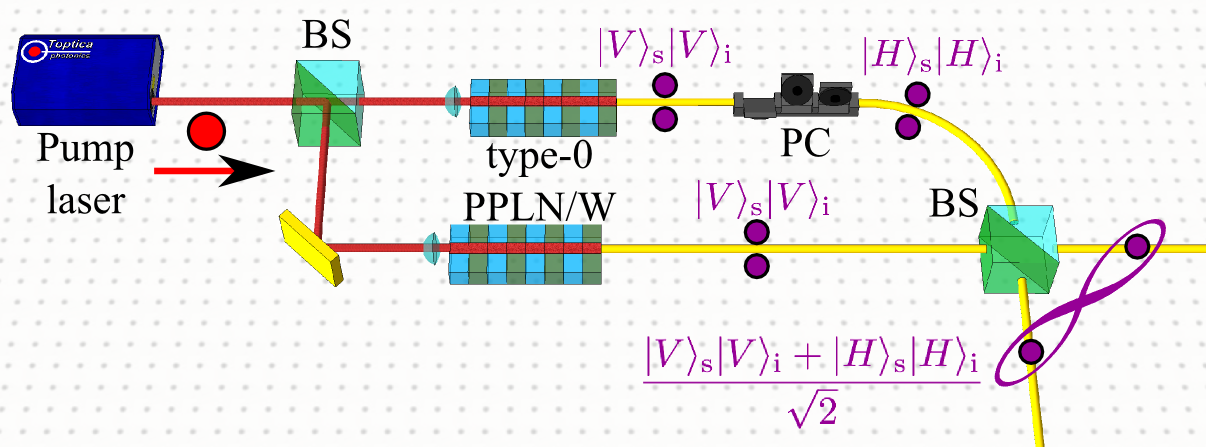}
\caption{Hybrid polarization entanglement source based on two type-0 PPLN waveguides in a Mach-Zehnder interferometer configuration. In each PPLN/W, vertically polarized photon pairs, $|V \rangle_{\rm s} |V \rangle_{\rm i}$, are generated. The pairs at the output of one PPLN/W are rotated by 90$^{\circ}$ in order to obtain $|H \rangle_{\rm s} |H \rangle_{\rm i}$. Both contributions are sent to a non-polarizing beam-splitter at the interferometer output. If a pair is separated, then it is projected into the polarization entangled state $|\psi \rangle = \frac{1}{\sqrt{2}} \left( |V \rangle_{\rm a} |V \rangle_{\rm b} + e^{i\,\phi} |H \rangle_{\rm a} |H \rangle_{\rm b}\right)$. Note that, in order to measure high quality entanglement, the phase, $\phi$, of the interferometer needs to be stabilized, which requires an active stabilization system and a phase shifter. Figure inspired from Ref.~\cite{yoshizawa_generation_2003}.\label{Fig_MZI_polar_2003}}
\end{figure}

For example, a group in Tsukuba (Japan) demonstrated, in 2003, a telecom wavelength, broadband, and highly efficient, polarization entangled photon pair source based on two simultaneously pumped type-0 PPLN/Ws in a Mach-Zehnder interferometer configuration~\cite{yoshizawa_generation_2003}, and similar realizations have been reported subsequently~\cite{Yoshi_polar_2004,Herbauts_active_routing_2013}. As shown in \figurename~\ref{Fig_MZI_polar_2003}, both PPLN/Ws generate vertically polarized signal and idler photons, $|V \rangle_{\rm s} |V \rangle_{\rm i}$, at the same rates and with identical emission spectra. The polarization of the pairs at the output of one PPLN/W is rotated by $90^{\circ}$, leading to $|H \rangle_{\rm s} |H \rangle_{\rm i}$. Both contributions are then sent to a non-polarizing beam-splitter, and the separation of a photon pair on this device leads to a polarization entangled two-photon state $|\psi \rangle = \frac{1}{\sqrt{2}} \left( |V \rangle_{\rm a} |V \rangle_{\rm b} + e^{i\,\phi} |H \rangle_{\rm a} |H \rangle_{\rm b}\right)$. The phase factor $\phi$ is related to the interferometer's path length difference, which necessitates an active phase stabilization system for high quality entanglement analysis.

\begin{figure}[h!]
\includegraphics[width=0.48\textwidth]{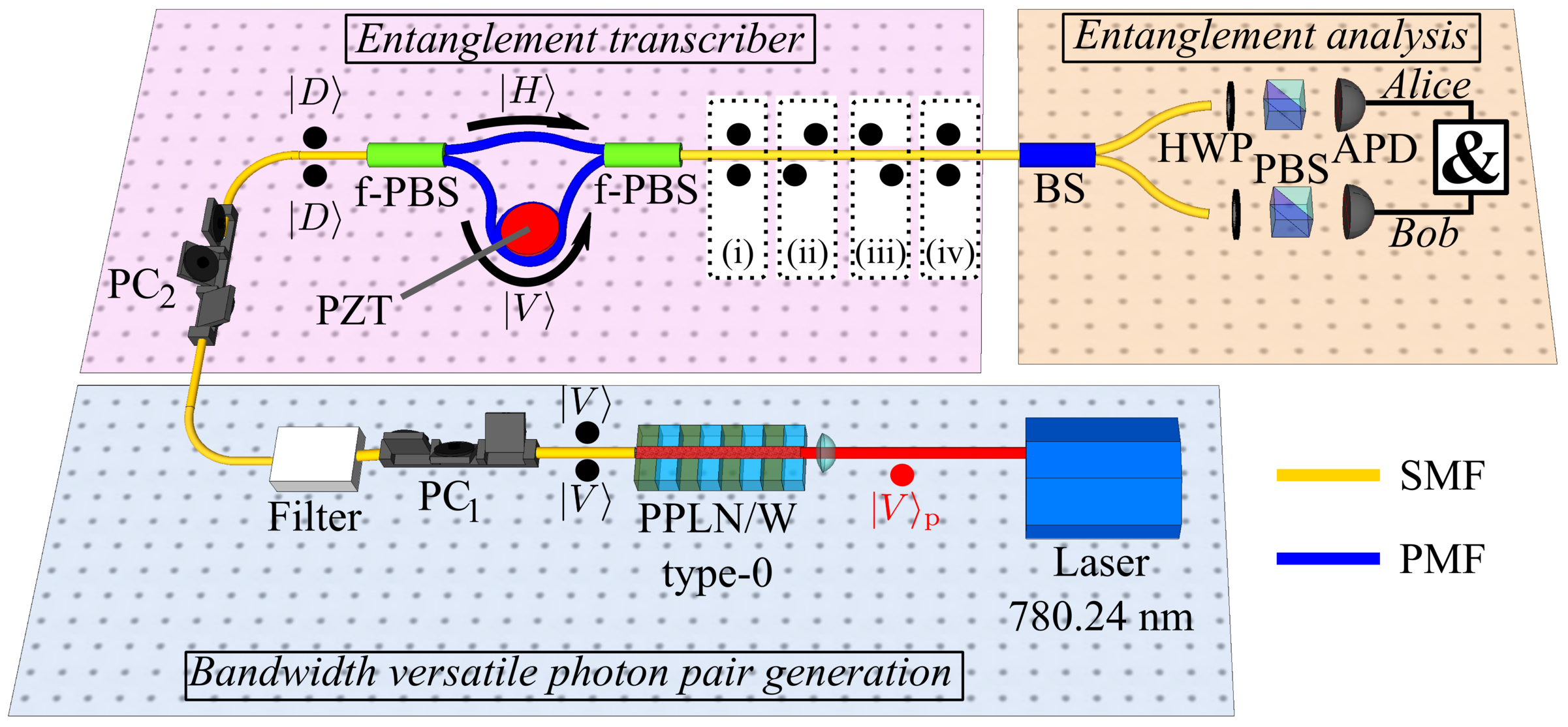}
\caption{Polarization entanglement generation from a standard high-efficiency energy-time entanglement source. (a) A 780\,nm laser pumps a type-0 PPLN/W in which vertically polarized energy-time entangled photon pairs are generated and filtered down to the desired spectral bandwidth. (b) Then, the paired photons are rotated to the diagonal state $| D \rangle | D \rangle$ and sent to the transcriber stage, which is essentially a polarization dependent delay line made of fibre polarizing beam-splitters (f-PBS) and polarization maintaining fibres (PMF) of different lengths. A piezo-electric fibre stretcher (PZT) allows to stabilize the phase of the transcriber. At the output of the transcriber stage, four contributions are observed, (i) to (iv). A polarization entangled state is generated when both photons exit the transcriber simultaneously (contributions (i) and (iv)), and the photon pair generation time remains unknown. Post-selection of these events is done using time-tagging electronics. (c) After the paired photons are separated on a beam-splitter (BS), a standard polarization Bell state analysis setup is employed to verify entanglement.
Figure inspired from Ref.~\cite{Kaiser_polar_transc_2013}.\label{Fig_source_polar}}
\end{figure}

Some of the issues of a Mach-Zehnder interferometer based source can be circumvented by using a Sagnac-loop configuration, as has been demonstrated by a collaboration of several Japanese groups~\cite{Lim_polar1_2008,Lim_polar2_2008,Lim_polar3_2008,Lim_polar4_2008,Lim_polar_2010}. Here, the PPLN/W is placed in the center of a fibre loop and simultaneously pumped in both directions, which ensures identical emission spectra. 
Because the generated photon pairs are collected in the same fibre loop, one obtains automatically a stable phase between the two contributions, which is at the very nature of a Sagnac loop configuration. Finally, both pair contributions are combined on a polarizing beam-splitter to generate the entangled state $| \psi \rangle$. For this configuration, a dual-wavelength polarizing beam-splitter is required, and some attention has to be paid to the multimode behaviour of the short wavelength pump light inside the fibre loop.\\
These issues may be overcome by exploiting SHG and SPDC processes on the same chip. As demonstrated by two Japanese teams in Saitama/Tokyo and Ibaraki, the PPLN/W chip can be pumped by a telecom wavelength laser which is frequency doubled (SHG) inside the device. The doubled light gives then birth to the desired photon pairs via SPDC in the same or a second PPLN/W~\cite{Arahira_ridge_2011,jiang_generation_2007}. It should be mentioned that, in this configuration, only non-degenerate photon pairs can be exploited, as the pump laser light needs to be rejected to avoid detector saturation.\\
Finally, a group in Kanagawa (Japan), and our team demonstrated that energy-time entanglement can be converted to polarization entanglement by a properly engineered polarization dependent delay line~\cite{takesue_generation_2005,Kaiser_polar_transc_2013,Kaiser_SourceLong_2014}. As shown in \figurename~\ref{Fig_source_polar}, the idea here is to generate energy-time entangled photon pairs in a type-0 PPLN/W and send them to an unbalanced interferometer in which photons in the longer (shorter) arm exit horizontally (vertically) polarized. Via time-tagging, only the events, in which both photons take the same path (\textit{i.e.}, $|V \rangle_{\rm s} |V \rangle_{\rm i}$ and $|H \rangle_{\rm s} |H \rangle_{\rm i}$ contributions) are post-selected.
If the photon pair creation time is unknown (for example when they are generated using a CW pump laser), then a polarization entangled state is obtained. Provided that the interferometer is phase-stabilized, such sources offer great long-term stability and entanglement quality.

\subsection{Frequency-bin entanglement sources \label{Freq_bin_ent_sources}}

Analysis of energy-time and time-bin entanglement requires actively stabilized unbalanced interferometers, and for long distance polarization entanglement distribution, active stabilization systems need to be employed to compensate dynamic polarization mode drifts. In this perspective, frequency-bin entanglement is an interesting candidate, as it necessitates only stabilizing the phase difference between two optical phase modulators in the microwave domain. The underlying concept is similar to energy-time entanglement, however, instead of performing entanglement analysis in the time domain, the frequency domain is exploited. A typical experiment exploits type-0 SPDC in a PPLN/W to generate a flux of (time-frequency) correlated photon pairs. After pair separation, standard telecom high-speed electro-optical phase modulators (typically based on LN waveguides) are used to create superposition of different frequency modes, and dense wavelength division multiplexers, followed by single photon detectors, are used for the quantum state projection~\cite{Olislager_frequencybin_2010,Olislager_frequencybin_2014}. So far, violation of the Bell inequalities has been demonstrated, and further performance improvements will make this type of entanglement interesting for quantum communication applications, such as entanglement swapping.

\subsection{Path entangled photon pair sources \label{Path_ent_sources}}

\begin{figure}[h!]
\begin{center}
\includegraphics[width=0.45\textwidth]{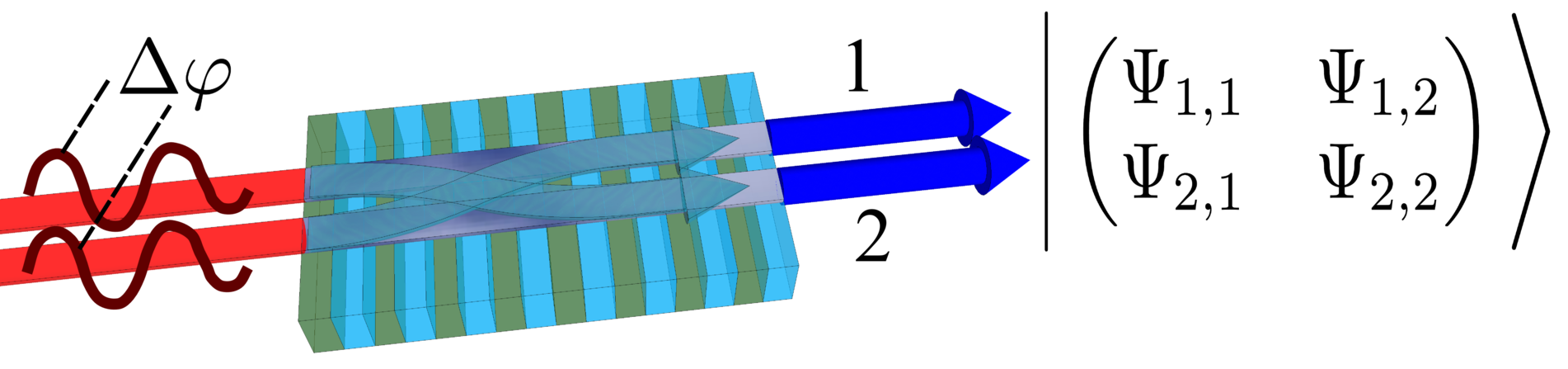}
\caption{Sketch of the proposed scheme for tunable generation of biphoton quantum states, with two pump beams (red) exciting a directional coupler and generating photon pairs (blue) via SPDC. The output state is described by a wavefunction in the spatial mode basis. Figure inspired from Ref.~\cite{Setzpfand_nonlinear_coupler_2016}.\label{NOON}}
\end{center}
\end{figure}

Path-entangled photonic states have recently attracted a lot of interest, as they might lead to interesting applications in both quantum metrology~\cite{Dowling_Qmet_2008} and lithography~\cite{Boyd_Litho_2011}. One particularly interesting class of states are the so-called $N00N$-states, which represent a coherent superposition of having $N$ photons in path $a$ and zero in path $b$, and conversely, $| \psi \rangle_{N00N} = \frac{1}{\sqrt{N}} \left( |N \rangle_a |0 \rangle_b + |0 \rangle_a |N \rangle_b \right)$.
Such states show an $N$ times higher phase sensitivity compared to classical light, which can be used to measure refractive index changes with superior performance~\cite{Crespi_protein_concentration_2012}.
Generation of the most elementary two-photon $N00N$-states directly on a PPLN/W chip has been reported recently by a Chinese collaboration~\cite{Jin_2014}, a group in Germany~\cite{Kruse_dualpath_2015}, and in Australia~\cite{Setzpfand_nonlinear_coupler_2016}.\\
The Chinese collaboration basically implemented the full source experimental setup of refs.~\cite{yoshizawa_generation_2003,Yoshi_polar_2004,Herbauts_active_routing_2013} on a monolithic device, which is an impressive achievement~\cite{Jin_2014}, which will be discussed in more detail in Section~\ref{Sec_ADV}.\\
For now, it is rather interesting to focus on the German and Australian devices that exploit guided-wave photonics to produce naturally path-entangled photon pairs. The concept lies on an ingenious solution involving two evanescent-field-coupled PPLN/Ws on the same chip as shown in \figurename~\ref{NOON}.

The sample is pumped in a single input port of the chip. The pump (at $\sim$775\,nm) remains localized in that particular waveguide whereas the created paired photons are delocalized over the coupled mode of the structure. Thanks to a proper engineering of the coupling constant, the pairs are naturally produced in one of the eigenmode of the coupled structures which are no longer the individual nonlinear waveguides but actually a coherent superposition of them. When those waveguides are decoupled, it happens that the phase matching naturally creates an entangled two-photon state $\left(|2,0\rangle+|0,2\rangle\right)/\sqrt{2}$. As no interferometer is involved, this device shows excellent stability and is largely independent of PPLN/W fabrication parameters and imperfections~\cite{Kruse_dualpath_2015}.\\
The Australian group has pushed the concept further, by showing that if both nonlinear waveguides are pumped coherently with a controllable phase relation (see also \figurename~\ref{NOON}), it is possible to tune the produced entangled state~\cite{Setzpfand_nonlinear_coupler_2016}. In all considered experiments, proper device engineering allowed to obtain quantum state fidelities exceeding 80\%, which underlines the relevance of these approaches and opens new perspectives in quantum metrology and lithography.

\subsection{Advanced photonic entanglement sources \label{Adv_phot_ent_sources}}

In the following, a few advanced photonic entanglement sources are described. They are intended to achieve the highest performances in two rapidly growing fields in quantum optics, namely quantum memories for the storage of single photons, and quantum key distribution (QKD) in wavelength-division-multiplexed telecom networks.

\begin{figure}[h!]
\includegraphics[width=0.48\textwidth]{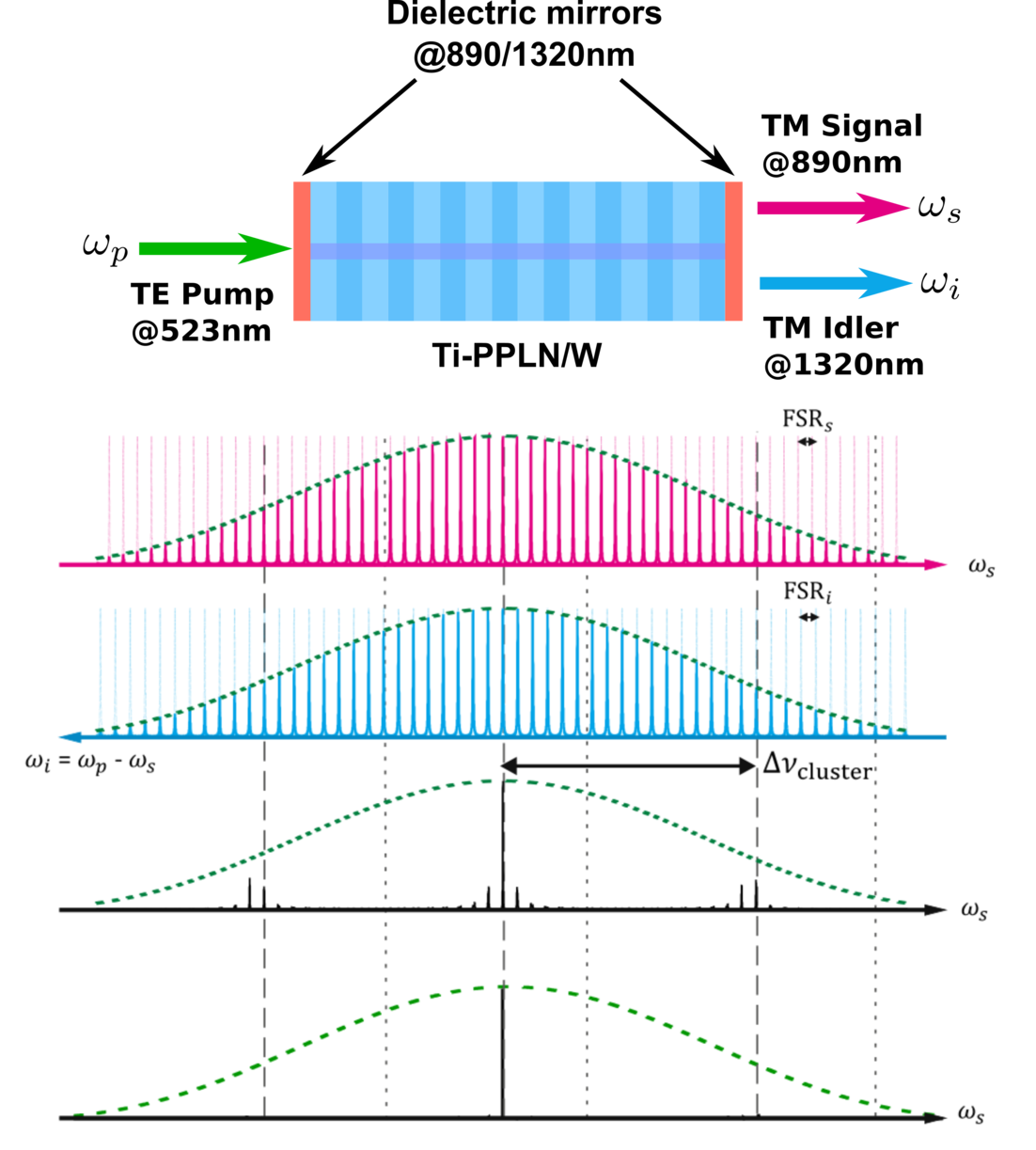}
\caption{Working principle of the PPLN/W OPO of the Paderborn group. (a) Schematics of the setup. Highly reflective coatings at 1320\,nm and 890\,nm are deposited on a titanium indiffused type-II PPLN/W, pumped by a 532\,nm laser. (b) PPLN/W emission (dashed lines) and cavity modes. The first (second) curve represents the cavity modes for the 890\,nm (1320\,nm) photons. Note that waveguide birefringence induces a different cavity mode spacing at the two wavelengths. Because of conservation of the energy ($\omega_{\rm i} = \omega_{\rm p} - \omega_{\rm s}$), SPDC can only occur when the cavity modes overlap. The third and fourth curves show the overlaps for a low and high finesse OPO, respectively. If the finesse is chosen to be high enough, then only one mode can be generated, which results in two-color ultra-narrowband photon pair source without the need of additional filters. Figure inspired from Ref.~\cite{Luo_type-II_OPO_2015}.\label{Fig_Paderborn_OPO}}
\end{figure}

Quantum memory devices have attracted high interest, as they allow to store a single photon excitation in a stationary device, and re-emit it on demand. This makes it possible to obtain photons in a deterministic fashion, which could tremendously boost the performance of many quantum information protocols and applications. One limitation of current quantum memories is their rather narrowband acceptance bandwidth, typically on the order of  $\rm 0.01-2\,pm$, which is much narrower than the emission of typical photon pair sources based on SPDC ($\rm 1-100\,nm$). In order to adapt the photon pair source to the memory, two ways are generally considered.
The broad spectrum may be filtered at the source itself in an optical parametric oscillator (OPO) configuration. To this end, highly reflective coatings are deposited on both facets of the PPLN/W, creating a waveguide optical cavity.
The emission inside the cavities narrow frequency modes is thus enhanced, and all other frequency modes are suppressed. This scheme has been successfully implemented by the Geneva group using a titanium indiffused type-0 PPLN/W OPO with 0.91\,pm spectral bandwidth at 1560\,nm~\cite{pomarico_waveguide_2009}. As degenerate photon pairs were exploited, an additional narrowband filter (10\,pm) was required to suppress the emission at non-degenerate cavity modes. They also performed energy-time entanglement analysis and demonstrated 81\% visibility in the standard Bell inequality test.
The group in Paderborn (Germany) used a titanium indiffused type-II PPLN/W OPO to obtain non-degenerate cross-polarized photon pairs at 890\,nm and 1320\,nm with a spectral bandwidth of 0.17\,pm at 890\,nm~\cite{Luo_type-II_OPO_2015}. As shown in \figurename~\ref{Fig_Paderborn_OPO}, thanks to the non-degenerate configuration, waveguide birefringence is exploited to allow the OPO to emit photons at one pair of cavity frequency modes only, which means that no additional filter is required. However, they did not yet perform entanglement analysis.
In both experiments, it was sufficient to stabilize the length of the optical cavity via temperature control, which is much less complex than the typical high-level stabilization systems required for bulk-optics based OPOs.

Another strategy towards obtaining a narrowband emission spectrum is to filter the photon pairs after they have been emitted by the PPLN/W.  Especially in the telecom range of wavelengths $\rm (1530\,nm - 1565\,nm)$, this can be conveniently achieved using ultra narrowband phase-shifted fibre Bragg gratings, which are commercially available for bandwidths ranging from 0.2\,pm to 10\,pm. Note that this filtering strategy is essentially reserved to waveguide-based photon pair sources only, which show high enough efficiencies to obtain a decent flux of photon pairs, even after filtering the spectrum from several 10s of nm down to the pm level. In this perspective, the Geneva group demonstrated an energy-time entangled photon pair source based on a type-0 PPLN/W with a spectral bandwidth of 10\,pm at 1560\,nm~\cite{halder_high_2008} which was subsequently used in an entanglement-swapping experiment without clock-synchronization~\cite{Halder_Ent_Indep_2007}. Our group took also advantage of the high SPDC efficiency of type-0 PPLN/Ws to demonstrate a hybrid polarization entanglement source achieving bandwidths of 0.2\,pm and 4\,pm at 1560\,nm. Near perfect visibilities of 99\% and 97\%, respectively, have been achieved in the standard Bell inequality test~\cite{Kaiser_polar_transc_2013,Kaiser_SourceLong_2014}. The 0.2\,pm source was further used to perform a continuous wave teleportation experiment~\cite{Kaiser_cw_tele_2015}, and the 4\,pm source was an essential building block for testing the foundations of quantum physics in a quantum delayed-choice experiment showing wave-particle complementarity~\cite{Kaiser_QDC_2012}.\\
It is expected that, in the near future, these sources will be eventually used for quantum-memory based experiments. For this, the wavelengths of the photon pair sources and quantum memories need to be matched, which will be described in more details in Section~\ref{Sec_QIM}.

Another very prominent field in quantum optics and quantum information is QKD, which allows to establish absolutely secure communication links~\cite{BB84,Ekert_Crypto_1991,gisin_QKD_2002,QKD_commercial}. In this perspective, entangled photon pairs are highly interesting as those links can be established in a device-independent manner, \textit{i.e.}, without the necessity to trust the photon pair source~\cite{Brunner_Bell_2014,Hensen_15,Putz_LDLPRL_2016,Acin_From_Bell_2006,Acin_DIQ_2007,Pironio_random_numbers_2010,Pironio_DIQ_2009,Bancal_DIQ_2011,Barreiro_DIQ_2013}. Today, researchers seek to improve the speed of QKD links in order to make them compatible with real world applications.
Similarly to standard telecommunication networks, the speed of these systems can be increased using wavelength division multiplexing techniques, \textit{i.e.}, by multiplexing (and demultiplexing) several independent signals into (and out of) the same fibre communication link.

\begin{figure}[h!]
\includegraphics[width=0.48\textwidth]{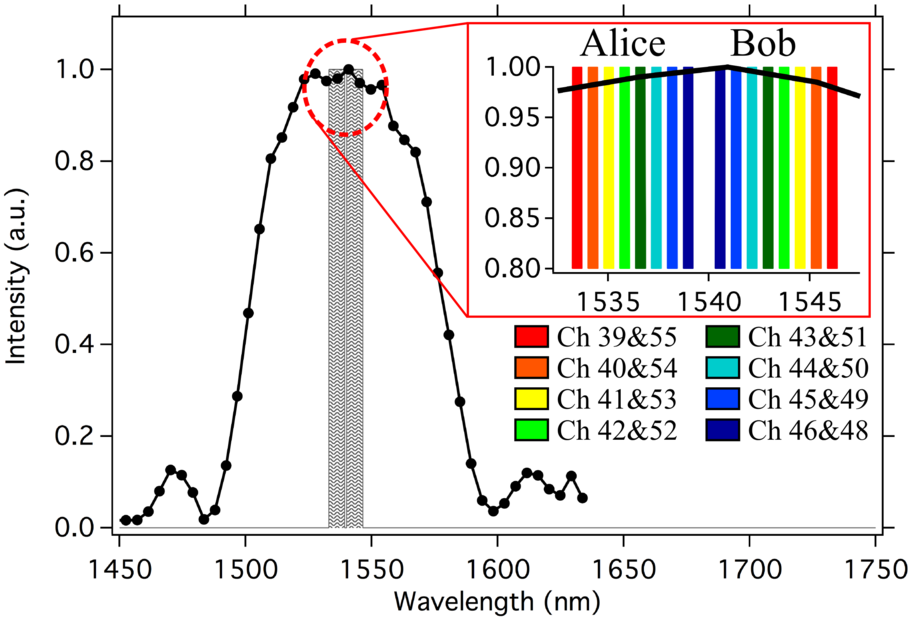}
\caption{Emission spectrum of an SPDC source, as can be obtained with the setup of \figurename~\ref{Fig_Djeylan_DWDM_setup}, see also Ref.~\cite{Aktas_DWDM_2016}. Thanks to the conservation of the energy, it can be decomposed in many correlated wavelength channel pairs. The black dots, connected by lines, represent the typical nondegenerate emission spectrum of a type-0 PPLN/W with 4\,cm length pumped by a 770\,nm continuous wave laser. Additionally, several telecommunication wavelength channels are shown, here in the heavily exploited standard 100\,GHz wavelength grid, as defined by the International Telecommunication Union. Wavelength correlated photon pairs are found in channels labeled with identical colors, which are located pairwise symmetrically around the degenerate wavelength of 1540\,nm. This allows using one source of entangled photon pairs to supply, at the same time, either multiple users with entanglement or to increase the coincidence rate in a two-use scanario by a factor equal to the number of exploited channel pairs.\label{Fig_Djeylan_DWDM_spectrum}}
\end{figure}

As shown in \figurename~\ref{Fig_Djeylan_DWDM_spectrum}, thanks to the conservation of the energy $(\omega_{\rm p} = \omega_{\rm s} + \omega_{\rm i})$, wavelength-correlated signal and idler photon pairs are found across the full emission spectrum of the SPDC source. If the spectrum is then demultiplexed into $2 \times N$ narrow wavelength channels, $N$ correlated, \textit{i.e.}, entangled, channel pairs are found. This allows one to analyse entanglement in $N$ channel pairs simultaneously, consequently enhancing the total coincidence rate by a factor $N$. It shall be noted that this requires a broadband photon pair source with an emission spectrum covering all the $N$ channel pairs. In addition, the source must be very efficient to obtain a high flux of photon pairs in each of the channel pairs.
First experiments in this perspective have been carried out by a Japanese group in 2008. They used a telecom wavelength broadband hybrid polarization entangled photon pair source based on a type-0 PPLN/W in a Sagnac loop configuration. Strong correlations across the full emission spectrum of the source were found, which demonstrate the potential for an $N$ times speed-up of coincidence rates~\cite{Lim_polar2_2008,Lim_polar3_2008,Lim_polar4_2008,Lim_polar_2010}. In 2013, a similar experiment was reported by a collaboration between Stockholm (Sweden), Vienna (Austria), and Waterloo (Canada), using a hybrid polarization entanglement source, this time based on two type-0 PPLN/W in a Mach-Zehnder configuration. In addition, they demonstrated active routing of entanglement between different users~\cite{Herbauts_active_routing_2013}. In the perspective of long-distance QKD using polarization entangled photon pairs, a major limitation is polarization mode dispersion in optical fibres. This causes a wavelength dependent phase shift, requiring adapted entanglement analysers for each channel pair~\cite{Lim_polar_2010}, and/or active compensation of birefringence fluctuation using integrated electro-optical devices based on lithium niobate~\cite{Xavier:09,Bonneau:12,Zhang:14}.

\begin{figure}[h!]
\includegraphics[width=0.48\textwidth]{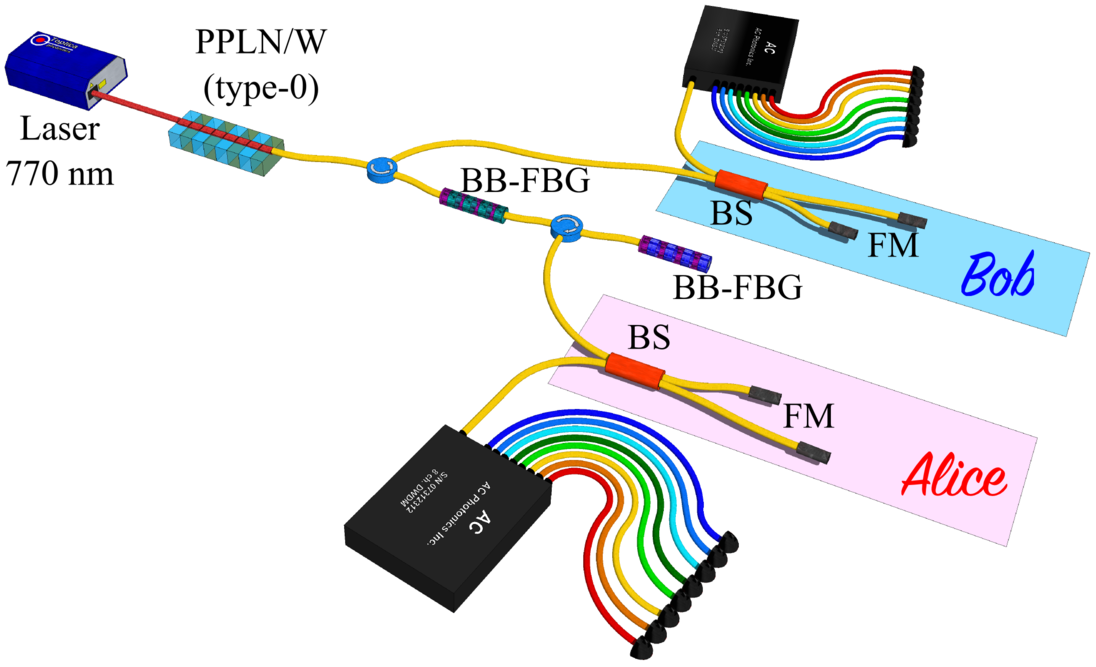}
\caption{Experimental setup for entanglement distribution in a dense wavelength division multiplexing environment as exploited in Ref.~\cite{Aktas_DWDM_2016}. A 770\,nm pump laser generates entangled photon pairs over a broad bandwidth in a type-0 PPLN/W. Thanks to two broadband fibre Bragg gratings (BB-FBG), Alice and Bob are supplied with photons below and above degeneracy, respectively. In order to reveal energy-time entanglement, photons pass unbalanced Michelson interferometers, made of a beam-splitter (BS) and Faraday mirrors (FM). At the interferometer out, the photons are demultiplexed into eigth standard telecom 100\,GHz wavelength channels, as shown in \figurename~\ref{Fig_Djeylan_DWDM_spectrum}. This allows to obtain an eight-fold increased coincidence rate. The strategy could be further used to demultiplex the spectrum into almost 100 channel pairs in the 50\,GHz grid.\label{Fig_Djeylan_DWDM_setup}}
\end{figure}

It has recently been demonstrated that this issue can be conveniently overcome by distribution energy-time entanglement and corresponding Michelson fiber interferometers for analysis purpose. As shown in \figurename~\ref{Fig_Djeylan_DWDM_setup}, our group used a type-0 PPLN/W whose emission is deterministically split into long and short wavelength parts by two broadband fibre Bragg gratings, and supplies two users with broadband energy-time entangled photon pairs.
At the end of the fibre link, spanning up to 150\,km, each user needed to employ only one unbalanced fibre interferometer, followed by a standard 8-channel telecom wavelength division multiplexer, to analyse entanglement with an 8 times increase coincidence rate~\cite{Aktas_DWDM_2016}. Note that by properly aligning these two interferometers, entanglement analysis can be further performed in almost 100 standard 50\,GHz telecom channel pairs simultaneously without the need to adapt the analyzers as a function of the wavelength. This corresponds to the typical emission bandwidth of a type-0 PPLN/W with a few centimetres of length ($\rm \leftrightarrow 80\,nm\,\,at\,\,1540\,nm$), and especially covers the commonly used telecom C-band of wavelengths $(1530 - 1565\rm\,nm)$~\cite{Kaiser_ultra_broadband_2016}.

To conclude, the development of entangled photon pair sources has benefited from both fundamental and applied repercussions which have stimulated the community to demonstrate highly specific photon pair sources. It has therefore produced a very large variety of different waveguide-based entanglement sources spanning over all possible observable and exploiting all at maximum the potential of each. This basically means that there is almost always one design which perfectly suits a targeted application.

\section{Heralded single photon sources}
\label{Sec_HSPS}

\begin{figure}[h]
\begin{center}
\includegraphics[width=0.5\textwidth]{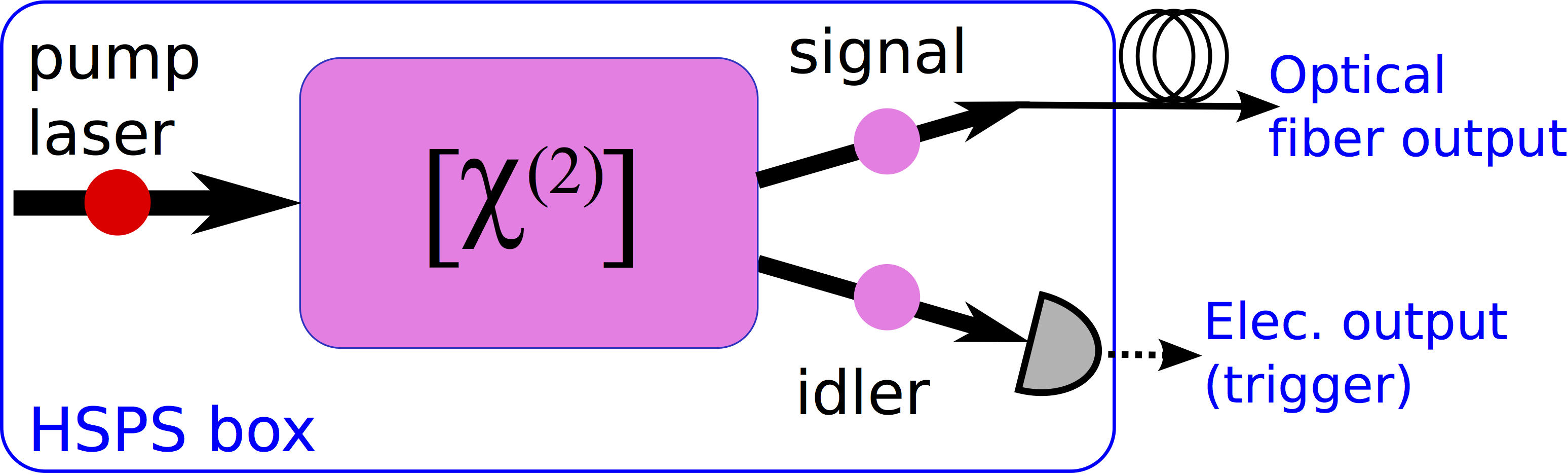}
\caption{Schematic representation of a heralded single photon source based on SPDC in a second-order nonlinear crystal ($\chi^{(2)}$). At the device output, simultaneous photons are spatially separated and the detection of the heralding photon is used to announce the presence of the heralded one. An HBT setup (see also \figurename~\ref{Fig_HSPS_Nicesetup} for this additional part) is generally triggered by the heralding detection events.\label{setupHSPS}}
\end{center}
\end{figure}

Beyond its fundamental interest~\cite{Peruzzo_ScienceQDC_2012,Kaiser_QDC_2012}, reliable generation of single photons is at the heart of many quantum optical technologies, including quantum information science, quantum metrology and detector calibration~\cite{Polyakov_2011_ReviewSinglePhoton}. Ideal sources should be able to emit, on demand, perfect single photon states at conveniently high rates and no multi-photon contributions. In anticipation to such archetypal cases, a valuable and pertinent alternative is represented by asynchronous heralded single photon sources (HSPS), as pertinently reported by the Geneva group in 2004~\cite{Fasel_HSPS_2004}. These schemes strongly rely on the production of photon pairs, as is the case with SPDC. As depicted in \figurename~\ref{setupHSPS}, after their generation, the photons are spatially separated. Because of the spontaneous character of the SPDC process, the emission time of a pair cannot be, in principle, precisely known. However, for each pair, the detection of one photon can be used to herald the emission time of its twin~\cite{Tanz_Genesis_2012}. Depending on the characteristics of the initial photon pair state~\cite{Walmsley_2008_HSPS} and on the heralding conditions~\cite{Laurat_2011_EPJD}, this configuration allows to obtain high quality heralded single photons. The rate of produced photons is proportional to that of the detected heralding ones, $R_H$, as well as to the heralding efficiency, $P_1$, representing the probability of observing one heralded photon per heralding event. In experiments, both figures increase with the mean number of generated photon pairs, $\langle n\rangle$, and decrease with optical losses on the heralded and the heralding photon paths~\cite{alibart_HSPS_2005}. Multi-photon contributions in HSPS are typically evaluated in terms of the second order autocorrelation function, $g^{(2)}(0)$. This quantity is proportional to the probability of accidentally heralding two photons instead of a single one. For perfect single photon Fock states, $g^{(2)}(0)=0$, while for heralded single photon states, it increases with $\langle n\rangle$. Experimentally, both $P_1$ and $g^{(2)}(0)$ can be obtained by characterizing the heralded photon statistics using either a bulk of fibre optics Hanbury-Brown-Twiss setup (HBT)~\cite{alibart_HSPS_2005}.

In the following, we will briefly present some realizations of HSPS relying on PPLN/Ws as enabling components. In particular, we will focus on three particular implementations. The first one reports the miniaturization of a source producing heralded photons at a telecom wavelength. In particular, it exploits LN-waveguide properties for the realization of a non-linear optical stage followed by integrated beam combination/splitting structures~\cite{Krapick_HSPStwocolor_2013}. The second and the third realizations show two different and complementary approaches to HSPS multiplexing~\cite{Ngah_HSPS_2014, Meany_HybridQcircuits_2014}. By doing so, both experiments aim at realizing ultra-fast sources able to emit single photons at telecom wavelength with high emission rate and low spurious event probabilities. In this sense, they demonstrate the capability of PPLN/W-based systems for future applications to fast quantum communication.

\begin{figure}[h]
\begin{tabular}{c}
\includegraphics[width=0.45\textwidth]{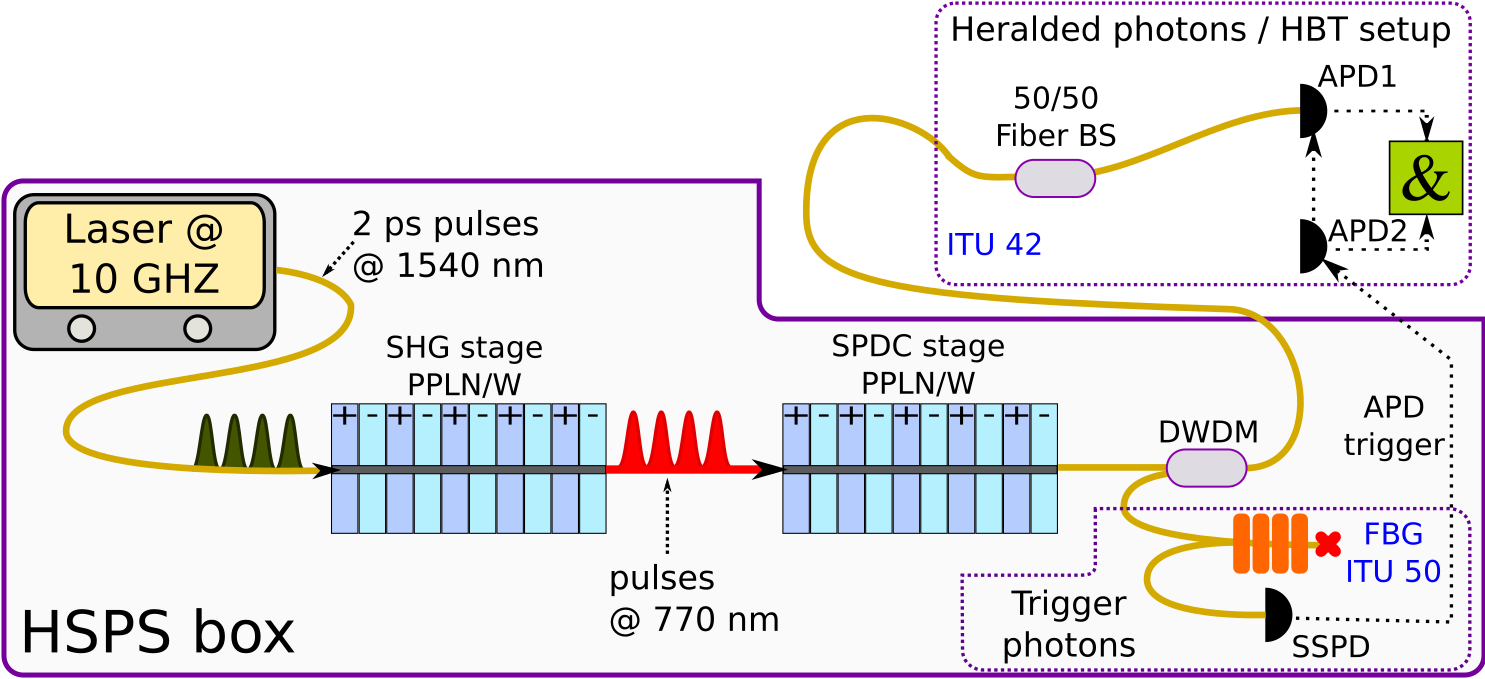}\\
\includegraphics[width=0.4\textwidth]{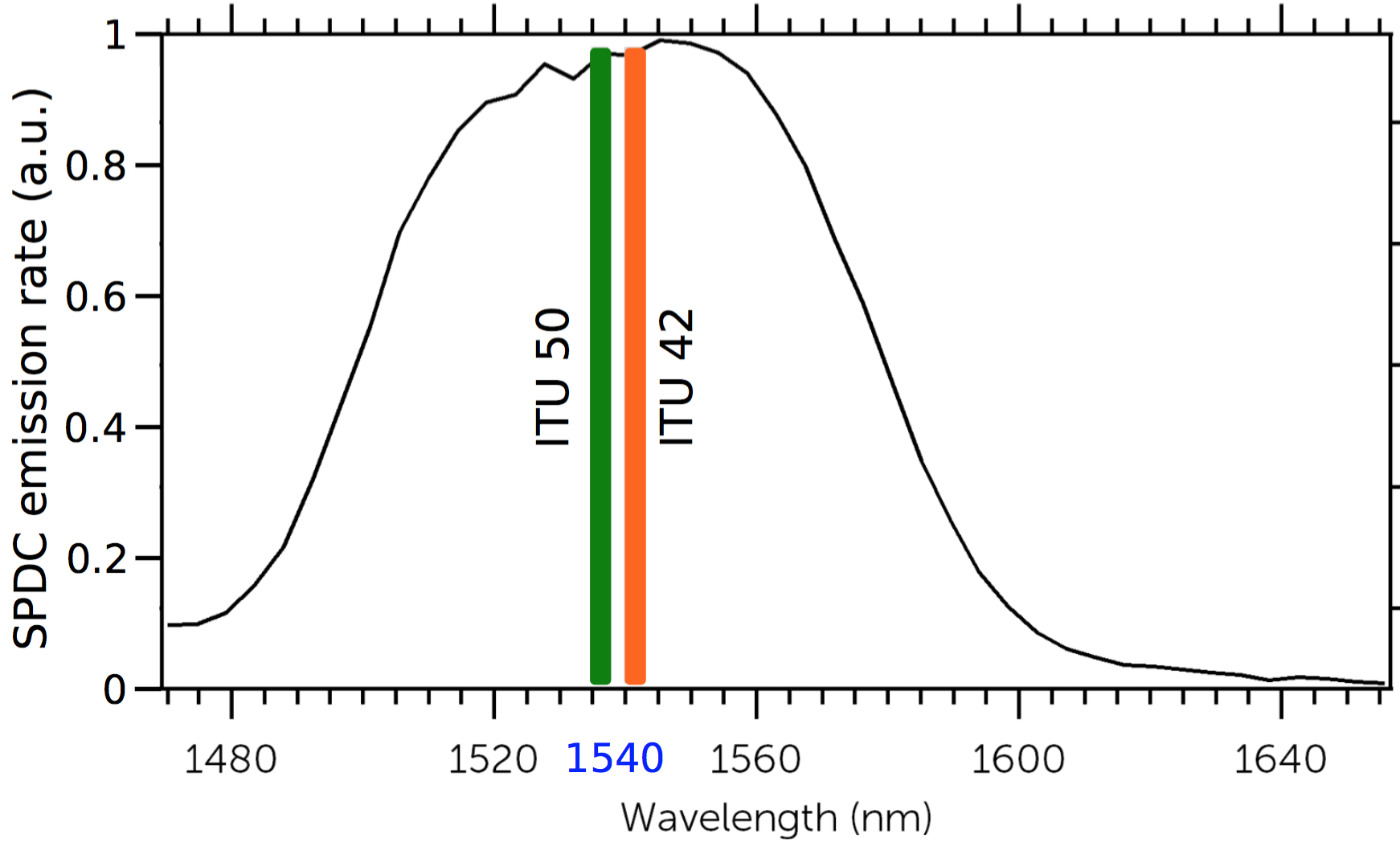}\\
\end{tabular}\caption{(Up) - Schematic of experimental setup for ultra-fast generation of heralded single photons. The telecom laser, operated at a repetition rate of 10 GHz, is first frequency doubled (SHG) and then used to pump the SPDC stage. Both processes take place in home-made PPLN/Ws. At the output, signal photons at 1537.40 (telecom channel 50) and idler photons at 1543.73 (channel 42) are obtained thanks to the combination DWDMs and a narrow-band fibre Bragg grating (not shown). Signal photons are directed towards the heralding path and detected by a superconducting single-photon detector (SSPD). Heralded idler photons are analyzed with a Hanbury Brown-Twiss (HBT) setup made of a beam splitter (BS) and two InGaAs avalanche photodiodes (APD1 and APD2). The detection in the HBT setup is triggered by the output signal of the SSPD. (Down) - SPDC spectrum out of the second PPLN/W. As shown, photon pairs are deterministically split up as a function of their wavelengths using fibre optics DWDM components into standard channels (50 and 42) belonging to the telecommunication grid. Figures inspired from Ref.~\cite{Ngah_HSPS_2014}.\label{Fig_HSPS_Nicesetup}}
\end{figure}

Straightforwardly, generation of entangled photon pairs via SPDC in PPLN/Ws can be advantageously exploited for HSPS (see also \figurename~\ref{Fig_HSPS_Nicesetup}). As already discussed, in particular, an adequate choice of QPM condition allows to produce heralded photons at telecom wavelength as demanded for long distance propagation in optical fibre. A first experiment is demonstrated in 2005 at the university of Nice, based on a SPE:PPLN/W pumped in CW regime and emitting non-degenerate photons at 1310\,nm (heralding) and 1550\,nm (heralded). Both photon pair separation and HBT setup were made using standard fibre components~\cite{alibart_HSPS_2005}. A $P_1\approx0.37$ was measured with a $g^{(2)}(0)=0.08$ for a heralding rate on the order of 100\,kHz. An interesting review of heralded single photon sources for quantum communication technology is available in~\cite{Castelletto_2008_ReviewHSPS} for an exhaustive list of realizations and it is worth mentioning an original work from a collaboration between US and Italian groups for developing an extremely low-noise heralded single-photon source~\cite{Brida_2012_LowNoiseHSPS} exploiting a PPLN/W.

\begin{figure}[h]
\begin{center}
\includegraphics[width=0.5\textwidth]{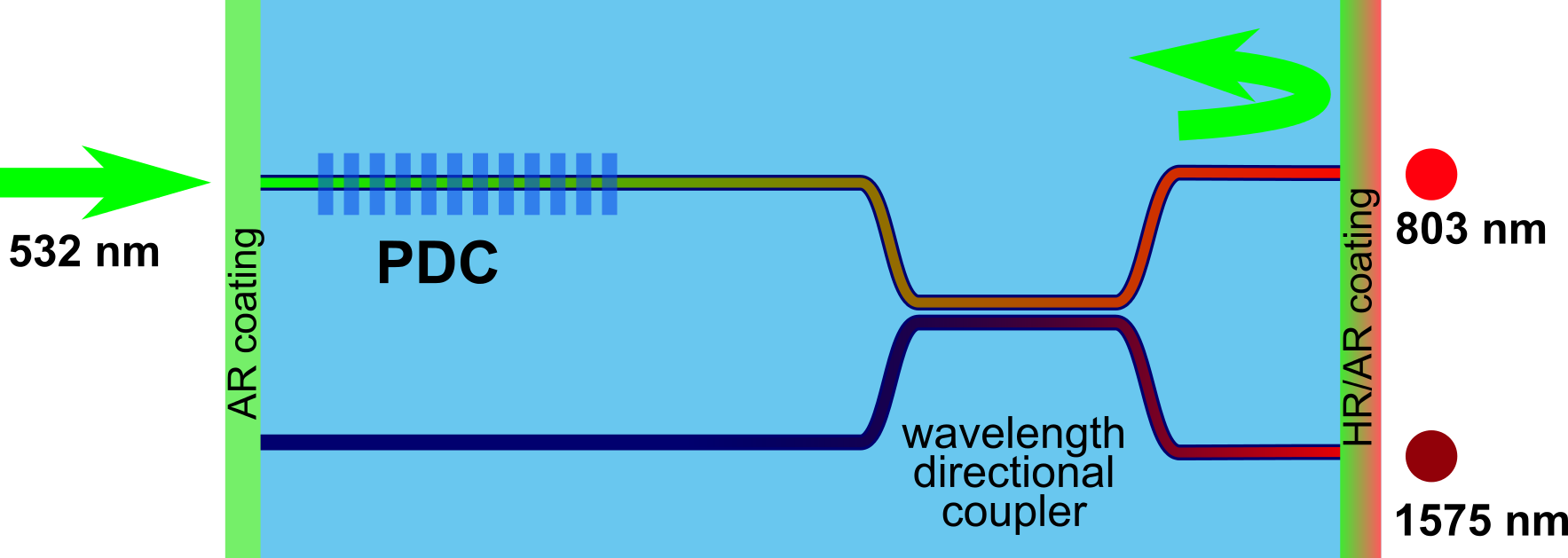}
\caption{Schematic representation of the Paderborn source made of titanium indiffuse waveguides. The device integrates, on a single substrate, a photon-pair generator (PPLN/W section) and an S-bend-type wavelength division demultiplexer (WDM coupler), that subsequently allows the spatial separation of heralding and heralded photons as a function of their wavelengths. Figure inspired from Ref.~\cite{Krapick_HSPStwocolor_2013}.\label{ChipHSPS}}
\end{center}
\end{figure}

Among the most recent experiments, a particularly interesting HSPS has been demonstrated in the pulsed regime in 2012 at the university of Paderborn~\cite{Krapick_HSPStwocolor_2013}. Here, the stages of photon pair generation and deterministic separation are both integrated on a single LN component (see Fig.~\ref{ChipHSPS}). The work clearly shows the potentiality of LN-chips as building blocks in complex and miniaturized quantum network realizations. In more details, a Ti:PPLN/W is used to produce, via type-0 SPDC\footnote{Based on the convention of this review paper, the related process is a type-0 process since all light field have the same polarization, but the authors have reported a type-I process in their paper.}, heralded single photons at the telecom wavelength of 1575\,nm paired with heralding photons at 803\,nm. At the output of the PPLN section, the generated twin photons are spatially separated thanks to an integrated passive wavelength division demultiplexer (96.5\% of efficiency). Estimated propagation losses inside the chip are as low as 0.07\,dB/cm. Anti-reflection coatings at the chip input and output facets allow to reduce in- and out-coupling losses. Downstream the chip, heralding photons are directed towards a single-photon detector (silicon avalanche photodiode) whose output triggers the detectors of an HBT setup measuring the heralded photons. A high heralding efficiency of $P_1\approx0.60$ is obtained. The authors investigate the produced photon quality as a function of the power of the SPDC pump beam. For low pump power (below 1\,$\mu$W), a heralding rate on the order of a few kHz is observed, associated with a measured $g^{(2)}(0)=3.8\cdot10^{-3}$. This result is consistent with negligible multi-photon events. For higher pump power ($\approx 100$\,$\mu$W), the heralding rate goes up to 105 kHz with a measured $g^{(2)}(0)$ close to 1. As discussed by the authors, this corresponds to a bright regime with a mean number of generated pairs per pump pulse of $\langle n \rangle$=0.24.
In the perspective of quantum communication, ultra-fast photon sources are mandatory to speed up data exchange. As shown by the Paderborn group, a simple way to reinforce the photon emission rate is to pump the SPDC at high pump power so as to increase $\langle n \rangle$. However, as discussed above, this comes at the price of non-negligible multi-photon contributions. In this context, HSPS multiplexing offers the advantage of high photon production rate combined with low SPDC pump power as demanded for small $\langle n \rangle$ values~\cite{Yamamoto_2007_10GHz}.

In 2015, our group has demonstrated the realization of an ultra-fast source emitting heralded single photons in the telecom C-band of wavelengths~\cite{Ngah_HSPS_2014}. This strategy exploits state-of-the-art telecom technology and nonlinear optical stages to time-multiplex the pump beam of a single SPDC process.
As depicted in \figurename~\ref{Fig_HSPS_Nicesetup}, a pump telecom laser emitting pulses at the wavelength of 1540\,nm and at the ultra-fast repetition rate of 10\,GHz is first frequency doubled via a second harmonic generation stage (SHG, consisting of an SPE:PPLN/W). This conversion stage fulfils the QPM condition for producing frequency degenerate photon pairs in another type-0 SPE:PPLN/W via SPDC. In other words, the wavelength of the emitted photons is centred around 1540\,nm. As was discussed in Section~\ref{Sec_EPPS} and more particularly \figurename~\ref{Fig_Djeylan_DWDM_spectrum}, energy conservation set by the SPDC process makes it possible to use a DWDM filter combination in order to pick up, within the emitted spectrum, twin photons slightly apart from perfect degeneracy (channels symmetrically spaced around the central wavelength) and to separate them deterministically. The entire setup downstream the SPDC stage, including the HBT arrangement for the heralded photon characterization, is made of standard telecom components only. The measured heralding efficiency is $P_1\approx42\%$. As discussed in~\cite{Sasaki_2014_SciRep}, the high repetition rate of the laser allows producing photon pairs at a high rate while, per each pump pulse, $\langle n \rangle$ is maintained as low as required for negligible two-photon events. Thanks to this strategy, single photons can be heralded at rates reaching values as high as 2.1\,MHz, with a constant $g^{(2)}(0)$ as low as 0.023. This $g^{(2)}(0)$ value not only proves negligible multi-photon contributions but also represents the best measurement reported to date in the literature for $R_H$ in the MHz regime~\cite{Ngah_HSPS_2014}. Since $R_H$ is directly proportional to the detection efficiency, it could be straightforwardly increased up to 10\,MHz by replacing the heralding detector with a more efficient one without affecting the $g^{(2)}(0)$ value.

\begin{figure}[h]
\begin{center}
\includegraphics[width=0.45\textwidth]{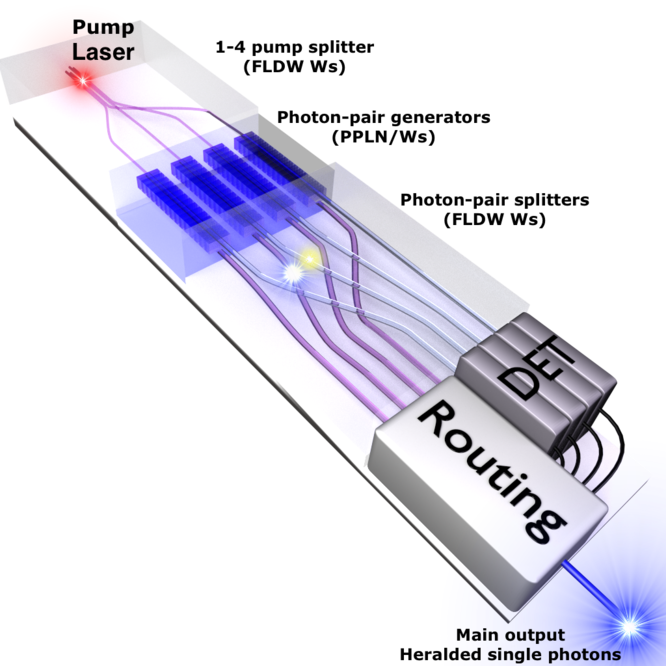}
\caption{Art-design schematics of a spatially multiplexed source of heralded single photons. The 1-4 splitter uses four evanescently coupled waveguides, in a 3D interaction area, to separate the input pump field into 4. It is then butt-coupled to the input facet of the nonlinear chip to pump four identical photon-pair generators, in this case PPLN/Ws, simultaneously. The QPM condition permits to produce photon pairs, by SPDC, at 1310\,nm and 1550\,nm. At the output, the photon pairs are collected using four FLDW integrated WDMs, operating at 1310/1550\,nm. The heralding photons at 1310\,nm are detected by one, out of 4, InGaAs detectors (DET stage). Upon detection of a photon, OR-gates located at the output of these detectors enable to know from which nonlinear waveguide the photon-pair originates, and to trigger fast optical switches (inside the routing stage) thanks to the corresponding heralding signal. Those switches then permit to route the comlementatry photon at 1550\,nm towards a single fibre optics output. Figure inspired from Ref.~\cite{Meany_HybridQcircuits_2014}.\label{Fig_hybrid}}
\end{center}
\end{figure}

Eventually, as proposed by A. Migdall and co-workers in 2002~\cite{Migdall_HSPSPRA_2002}, speeding up the rate at which heralded photons are made available can also be achieved by spatial multiplexing of several simultaneous SPDC stages. As shown in \figurename~\ref{Fig_hybrid}, a recent demonstration of this concept has been realized in 2014 thanks to a collaboration between our group and Macquarie and Sydney universities (Australia)~\cite{Meany_HybridQcircuits_2014}. In this work, thanks to a hybrid device incoporating the best of two state-of-the-art integrated photonics technologies, four identical sources of heralded single photons are spatially multiplexed using waveguide structures only. More specifically, hybridization means here that four PPLN/Ws integrated on a single substrate are interfaced in-between passive routers fabricated by femtosecond-laser direct writing (FLDW) on glass substrates. Additional fast fibre optical switches are placed at the output of the main device for enabling the routing of the heralded single photons towards a single fibre output. The first FLDW chip allows splitting a pulsed laser into to pump four identical type-0 SPE:PPLN/Ws. The chosen QPM condition permits generating, by SPDC, non-degenerate pairs of photons whose wavelengths are centred at 1310 and 1550\,nm. Note that the mean pump power is kept low enough in order to statistically create much less than one photon-pair at a time in either one of those four PPLN/Ws. Wherever they come from, the photon-pairs are then coupled to another FLDW routing chip that contains four individual integrated WDMs for separating the paired photons as a function of their wavelengths. After on-chip separation, the heralding photons at 1310\,nm are directed towards single-photon detectors (InGaAs avalanche photodiode), and the heralded photons at 1550\,nm are directed to a routing stage based on optical switches. Note that external wavelength fibre filters (not represented in \figurename~\ref{Fig_hybrid}) are employed for augmenting the extinction ratio and therefore the overall SNR. Depending on the detection results, \textit{i.e.}, which detector fires among the four, RF electronics enables controlling the switches for routing the 1550\,nm photons towards a single main fibre output. The setup performances are outlined in terms of coincidences-to-accidentals ratio (CAR) among heralded and heralding photons~\cite{Krapick_HSPStwocolor_2013}. This parameter characterizes the SNR of a photon source against background noise and provides, indirectly, information about two-photon-pair contributions by assuming a poissonian photon-pair statistical distribution. The data obtained with the setup depicted in \figurename~\ref{Fig_hybrid} clearly show how spatial multiplexing ensures an increase in photon rates while maintaining a fixed SNR. More precisely, an increase by a factor 3 was demonstrated, instead of 4 as expected, as one of the four nonlinear waveguides showed much higher losses compared to the three others. The experiment also showed, to date, the largest number of identical on-chip photon-pair sources having identical spectral properties.

In conclusion, the concept of heralded single photons based on photon-pair generation has proven to be very practical. For a long time, the probabilistic emission of paired photons was argued to prevent this kind of source to be scalable. This point has indeed been compensated by the highest emission rate reported to date and the single character of the heralded photons. In this context, PPLN waveguides are excellent candidates for real-world implementations of HSPSs for quantum communication. From one side, as demonstrated by many experiments, they allow to reliably and efficiently produce heralded photons at telecom wavelengths as demanded for long distance operation in optical fibres. At the same time, PPLN/Ws - and more generally speaking, LN-chips - are highly compatible with fibre-based telecom technology, thus opening the route towards a full exploitation of existing fibre networks and classical telecom systems.

\section{Quantum interfaces and memories}
\label{Sec_QIM}

In this Section, we address waveguide-based systems for which the very properties of LN structures have permitted demonstrations of both quantum interfaces and memories. The former corresponds to coherent wavelength transposition based on nonlinear optics of a photonic qubit prepared at one wavelength to another, while the latter permits coherent storage and retrieval of photonic excitation in and out of a matter system.
Quantum storage of photonic excitations often relies on energy transitions prepared in a well isolated ensemble of atoms, such as rubidium or caesium, or in a collection of rare-earth ions, such as thulium or praseodymium, embedded in a crystalline matrix. However, these energy transitions often belong to the visible range of wavelengths, preventing from storing directly telecom photonic qubits. This is why coherent quantum interfaces appear to be interesting to make flying and stationary qubits compatible, and possibly all designed in an entire waveguide approach~\cite{duan_LDQcom_2001,simon_quantum_2007,Sangouard_DLCZRMP_2011}.

\subsection{Quantum interfaces for coherent wavelength transposition}

Nowadays entanglement appears as essential for most of quantum technologies. In the particular case of quantum communications, the manipulation of entanglement has to be possible, for it to be of any practical value. Accordingly, a growing community of experimental physicists are working with entanglement transposition, especially in view of the connection of different users on a quantum network consisting of quantum channels along which photonic qubits travel~\cite{weihs_photonic_2001} and atomic ensembles are used to store and process the qubits~\cite{Julsgard_QMem_2004,Langer_LongLivedQM_2005}. It is thus timely to develop quantum information interfaces that allow transferring the qubits from one type of carrier to another, \textit{i.e.}, in the case of photonic carriers, able to provide back and forth a wavelength adaptation to atomic levels while preserving quantum coherence of the initial state~\cite{Giorgi_FrequencyHopping_2003,Tanz_Interface_2005,curtz_CFDC_2010,takesue_SPFDC_2010}. As a consequence, we will discuss in this section two type of interfaces: the ones for increasing the energy of the photonic carriers toward the visible region and those decreasing the energy toward the telecom range.

Chronologically, the first demonstration, in 2005, addressed an up-conversion transfer (increasing the frequency) for energy-time coded entangled photons. The inherently noise-free sum-frequency generation process was exploited to coherently transfer qubit carried by a photon at the telecom wavelength of 1312\,nm to another photon at 712\,nm, \textit{i.e.}, close to that of alkaline atomic transitions~\cite{Julsgard_QMem_2004,Langer_LongLivedQM_2005,Chaneliere_RemoteAtomEnt_06,Pan_QMRb_2008,Pan_QMRb_2009}. This operation has been shown to preserve entanglement in the following way: an initial photon at 1312\,nm, energy-time entangled with another photon at 1555\,nm (see also Sec.~\ref{Pioneering_ent_sources} and~\ref{Hybrid_polar_ent_sources}), was up-converted to 712\,nm. Provided the coherence length of the pump laser is longer than the size of the unbalanced analysis interferometers~\cite{Franson_Bell_1989} (see also caption of \figurename~\ref{Fig_Energy-time_Tanz_2002} for more details), the authors verified that the transfer actually preserves the quantum coherence by testing the entanglement between the produced photon at 712\,nm and the untouched photon at 1555\,nm. Although those two photons weren't originally entangled, the transferred entanglement was found to be almost perfect~\cite{Tanz_Interface_2005}. In other words, despite no energy conservation, up-conversion at the single photon level does not reduce the amount of entanglement.
\begin{figure}[h!]
\includegraphics[width=0.5\textwidth]{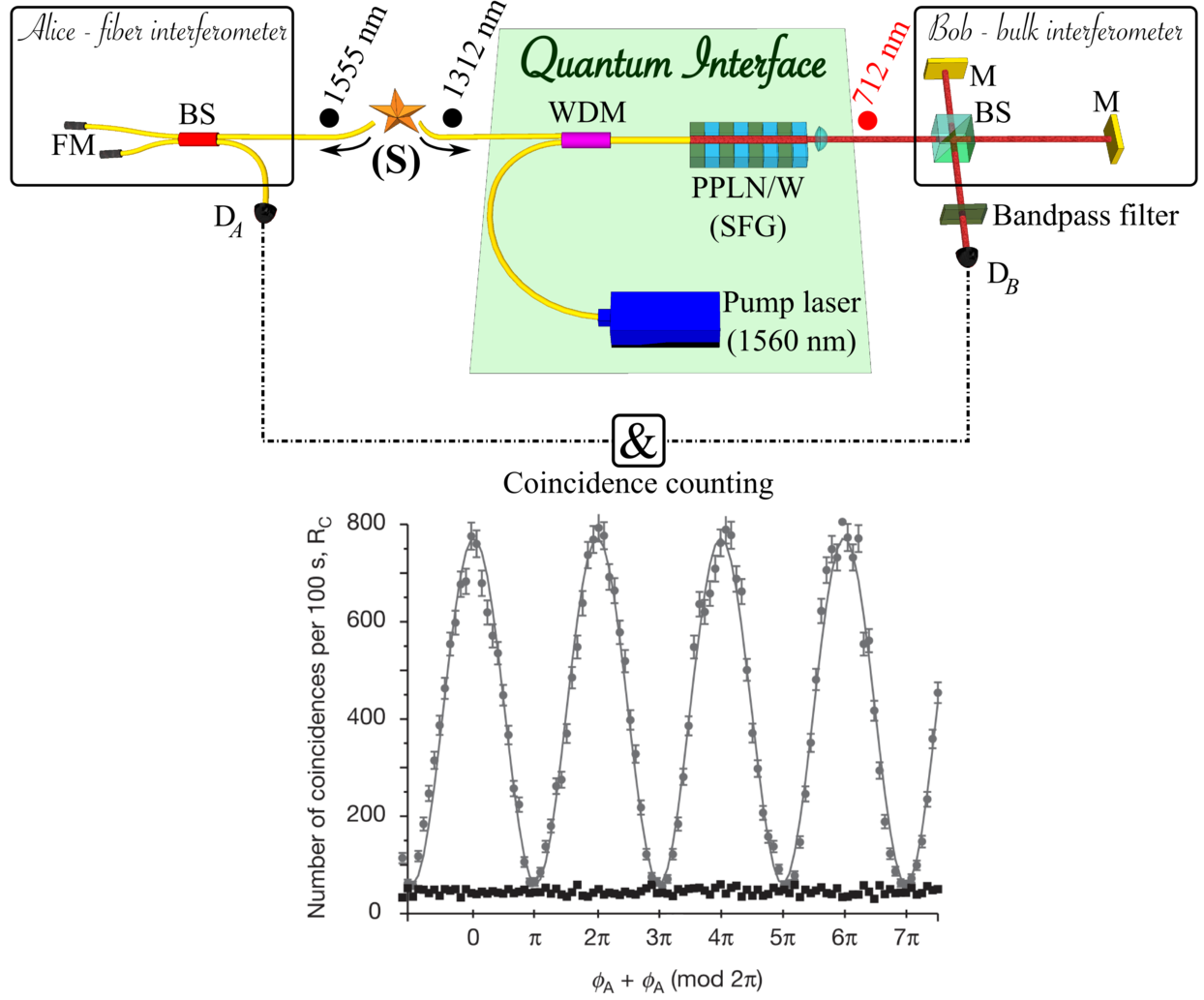}
\caption{\label{Fig_Interface} A photonic quantum interface enabling the coherent transfer of single qubits from photons at 1310\,nm, initially entangled with photons at 1555\,nm, to photons at 712\,nm. In this realization, it is shown that entanglement is preserved through the up-conversion process. A source (S) produces, by SPDC in a SPE:PPLN waveguide (not shown), pairs of energy-time entangled photons whose wavelengths are centered at 1555 and 1312\,nm. These photons are sent to Alice and Bob, respectively, via optical fiber links. The qubit transfer is then performed at Bob's location from a photon at 1312\,nm to a photon at 712\,nm using sum frequency generation (SFG) in another SPE:PPLN waveguide. This nonlinear waveguide is pumped by a CW, 700\,mW, and high coherence laser operating at 1560\,nm. The success of the transfer is tested by measuring the quality of the final entanglement between the 712\,nm photon and 1555\,nm photon using two unbalanced Michelson interferometers and a coincidence counter between detectors D$_A$ and D$_B$. The interference pattern obtained for the coincidence rate as a function of the sum of the phases acquired in the interferometers ($\phi_A + \phi_B$) shows a visibility greater than 97\%. Compared to the 98\% obtained with the initial entanglement (curve not represented), the fidelity of the transfer is therefore greater than 99\%. Figure inspired from Ref.~\cite{Tanz_Interface_2005}.}
\end{figure}

More specifically, the single photon up-conversion transfer is based on mixing the initial 1312\,nm single photon with a highly coherent pump laser at 1560\,nm in a SPE:PPLN waveguide, as depicted in \figurename~\ref{Fig_Interface}. The internal up-conversion efficiency of 80\% of the chip correspond to a single-photon up-conversion probability of $\sim$5\%, including the phase-matching filtering losses. The resulting photon at 712\,nm, filtered out of the huge flow of pump photons, and the remaining 1555\,nm photon were then analyzed using two unbalanced Michelson interferometers in the Franson configuration~\cite{Franson_Bell_1989} (\textit{i.e.}, the usual setup for testing energy-time or time-bin entanglement). A comparison of the degrees of entanglement before and after the up-conversion stage indicates no noticeable degradation and leads to a quantum information transfer fidelity higher than 99\%.

More recently, reverse process quantum interfaces, \textit{i.e.} decreasing the qubit-carrier photon frequency by DFG (see also section~\ref{Sec_NLI}, have been investigated. Similarly to SFG, this stimulated single photon down-conversion process is also a noise-free process and requires strong requirement on the coherence of the pump laser. Demonstrations of such quantum interfaces have been reported almost simultaneously in 2010 using single photonic qubits by the Geneva group~\cite{curtz_CFDC_2010} and NTT Japanese research groups~\cite{takesue_SPFDC_2010}, and in 2011 using entangled photon pairs at the University of Osaka~\cite{Ikuta_InterfaceNatCom_2011}.
The idea here is to proceed to the exact inverse operation to that discussed just above, \textit{i.e.}, transferring coherently, back to the telecom range, qubits initially coded on photons in the visible range. Both works, dealing with quite the same experimental procedures, propose a solution towards re-encoding qubits onto telecom photons after the read-out of a quantum memory based on alkaline atomic ensembles~\cite{Julsgard_QMem_2004,Chaneliere_RemoteAtomEnt_06,Pan_QMRb_2008,Pan_QMRb_2009}.
\begin{figure}[h!]
\includegraphics[width=0.5\textwidth]{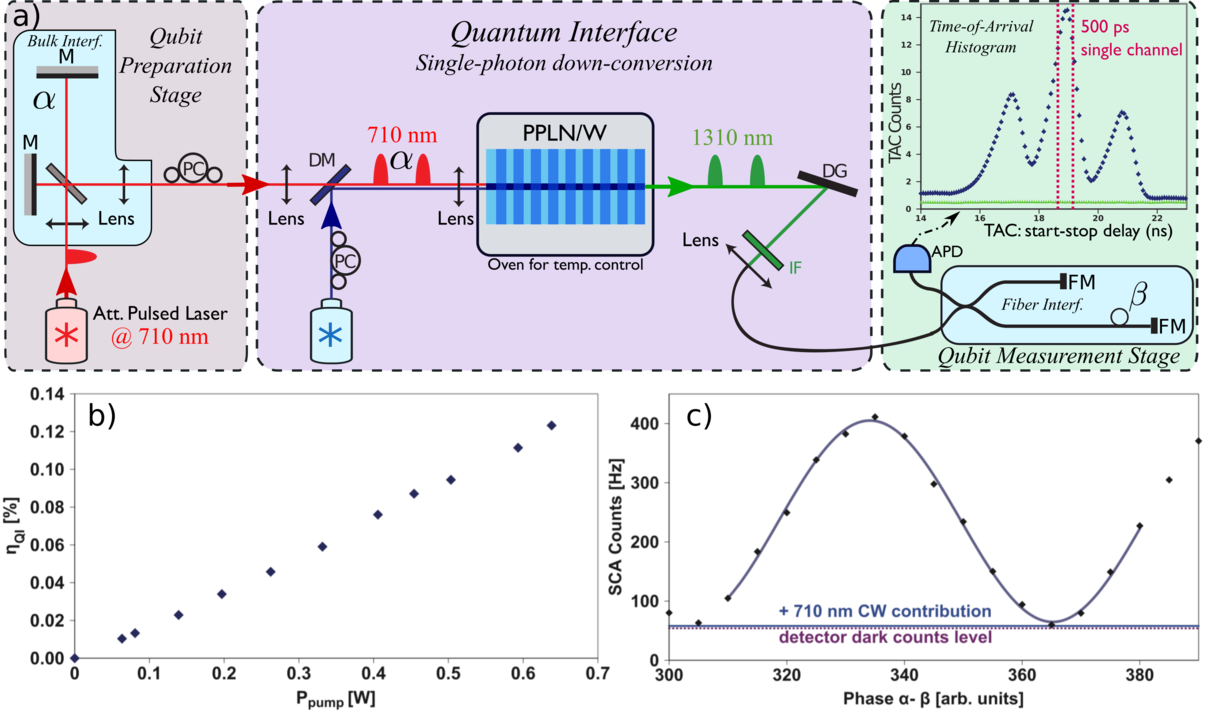}
\caption{\label{Fig_Interface_New}A photonic quantum interface dedicated to coherently transferring single qubits from photons at 710\,nm to photons at 1310\,nm by means of DFG. (a) Experimental set-up for the characterization of the coherence of the wavelength transfer. The setup comprises an attenuated pulsed laser mimicking a single photon source at 1310\,nm, a preparation (bulk) interferometer to code time-bin qubits on the 710\,nm photons (phase $\alpha$), a pump laser at 1550\,nm, a PE:PPLN/W for single photon down-conversion (or DFG), a fiber interferometer for time-bin qubit analysis (phase $\beta$), and a dedicated electronics for recording the difference in arrival times between the laser trigger and the detection events (inset, top right). (b)  Efficiency of the quantum interface as a function of the pump power. (c) Count rates for interfering events detected by a suitable AND-gate for an average number of prepared single photons reaching unity. As the phase $\beta$ of the fiber interferometer is scanned, a net (raw) interference visibility of 96\% (84\%) is reported. Figure inspired from Ref.~\cite{curtz_CFDC_2010} and data reproduced with permission of the authors.}
\end{figure}

Both single-qubit experiments exploit time-bin coded single photons out of an attenuated pulsed laser. Those qubits are then sent to single photon down-conversion stages based on PE:PPLN waveguides, and, eventually, entanglement fidelity is measured using either a fiber interferometer (see \figurename~\ref{Fig_Interface_New}(a) and Ref.~\cite{curtz_CFDC_2010}) or a PLC interferometer (see Ref.~\cite{takesue_SPFDC_2010}). The obtained results, in terms of quality of entanglement, are of the same order, \textit{i.e.}, showing a net visibility of almost 96\% for the Geneva group and of 94\% for the NTT group.

It is important to note that the quantum interfaces discussed above are not limited to the specific wavelengths reported in the corresponding papers, even if for some practical issues, the triplet 710\,nm, 1310\,nm and 1550\,nm was an optimal choice. Those are proof-of-principle experiments and suitable modifications of phase-matching conditions and pump wavelengths would enable engineering the result of the frequency-conversion processes to any desired wavelength triplet. More precisely, waveguide structures, as described in Sec.~\ref{Sec_LN}, provide the experimentalists with a broad range of accessible wavelengths by changing the poling period and the pump wavelength accordingly~\cite{tanzilli_ppln_2002}. For instance, having an original qubit carried by a photon at 1550\,nm and a pump laser at 1570\,nm would lead to a transferred qubit carried by a photon at the wavelength of 780\,nm, \textit{i.e.}, that of rubidium atomic transitions~\cite{Chaneliere_RemoteAtomEnt_06,Pan_QMRb_2008,Pan_QMRb_2009}.
In addition, the very high frequency-conversion efficiencies of such IO devices permit using modest pump powers to achieve reasonable qubit transfer probabilities. This is of great interest for applications requiring very narrow photon bandwidths, for instance when transferring photonic qubits to atoms~\cite{Shapiro_LongDistTele_2001,Julsgard_QMem_2004,Langer_LongLivedQM_2005,Chaneliere_RemoteAtomEnt_06,Pan_QMRb_2008,Pan_QMRb_2009}. In other words, ultra-bright frequency-converters make very narrow spectral filtering possible while maintaining a high photon-pair flux, with reasonable pump powers as described in Sec.~\ref{Sec_EPPS} and in Refs.~\cite{Aboussouan_dipps_2010,Halder_Ent_Indep_2007}.

To conclude this section, we must emphasize that interfacing photonic and atomic qubit is of great interest as it would facilitate the building of quantum networks and computers~\cite{duan_LDQcom_2001,simon_quantum_2007,simon_quantum_2010,Sangouard_DLCZRMP_2011}. In this perspective, recent frequency conversion experiments of photons correlated with a rubidium quantum memory have been performed
at Osaka University~\cite{arXiv:1607.07314} and at ICFO-Barcelona (Spain)~\cite{arXiv:1607.01350}.

Finally note that up-conversion, or hybrid, detectors based on PPLN/Ws have been widely employed to convert photons from the telecom range of wavelengths to wavelengths around or below 800\,nm, without the need to preserve the quantum state coherence, finding essentially applications in high-speed quantum cryptography~\cite{Diamanti_PRA_05,Thew_GHzQKD_2006}. A good review on those systems can be found in~\cite{Tanz_Genesis_2012}.

\subsection{Integrated quantum-memory devices}

Quantum memories are key elements for long distance quantum communication, more specifically in view of building real quantum networks. For instance, such devices can be employed for realizing deterministic single photon-sources~\cite{Matsukevich_DSPS_2006,Chen_DSPS_2006}, for turning quantum relays to much more efficient quantum repeaters~\cite{Simon_QR_2007,Simon_multiQR_2010}, and for further enabling quantum computation~\cite{Shor_DecoQC_1996,Knill_QECC_1997,Kok_QC_2007,knill_QC_2010}.

The performances of quantum memories can be quantified by four figures of merit : the storage time, the storage efficiency, the fidelity and the spectral acceptance bandwidth~\cite{tittel_PEQMSS_2009}. The storage efficiency represents the probability of absorption/re-emission in the right temporal and spatial modes. The fidelity corresponds to the overlap between the input and output qubit state. It also allows to define the storage time, reported as the duration over which the output qubit fidelity remains above a certain threshold. Eventually, a large memory spectral acceptance bandwidth is of great interest since it is directly connected to the performance of quantum communication links in terms of rate and multimodal operations.

In the quest towards optimal quantum memory realizations, various quantum storage protocols have been investigated and associated with different storage devices. Among others, were reported solutions based on alkaline cold atomic ensembles associated with electromagnetically induced transparency~\cite{chaneliere_storage_2005}, atomic vapours associated with off-resonant two-photon transitions~\cite{reim_towards_2010}, ion-doped crystals associated with photon echo or atomic frequency combs~\cite{de_riedmatten_ss_2008,chaneliere_light_2010}, \textit{etc.} Very pertinent review papers can be found in Refs.~\cite{lvovsky_optical_2009,tittel_PEQMSS_2009,Hammerer_interface_10}. 

The above mentioned combinations offer various interaction wavelengths and linewidths, ranging from the visible to the telecom band, and from a few MHz to almost 5 GHz, respectively. Naturally, the solutions based on ion-doped crystals, working at cryogenic temperatures (a few $^\circ$K), have been investigated using IO configurations. The first implementation was proposed by Hastings-Simon and collaborators, where they studied the Stark effect in Er$^{3+}$-doped Ti:LiNbO$_3$ waveguides~\cite{staudt_interference_2007,staudt_fidelity_2007}, and demonstrated the suitability of the scheme for protocols based on controlled reversible inhomogeneous broadening (CRIB) in solid-state devices~\cite{Kraus_CRIB_2006,Gisin_PEQM_2007,Lauritzen_SSM_2010}. This protocol is an extension of the photon echo technique showing better efficiency and a higher maximum fidelity. Storage times of 1 to $2\,\mu$s were demonstrated and a possibility of extending this to $6\,\mu$s discussed while the acceptance bandwidth is maintained at 125\,MHz. One of the attractive feature of this solution lies in the operating wavelength of 1550\,nm, being directly mapped to the transition ($^{4}I_{15/2} \rightarrow\,^{4}I_{13/2}$), making this memory fully compatible with standard telecom networks without any wavelength interface.

\begin{figure}[h!]
\includegraphics[width=0.5\textwidth]{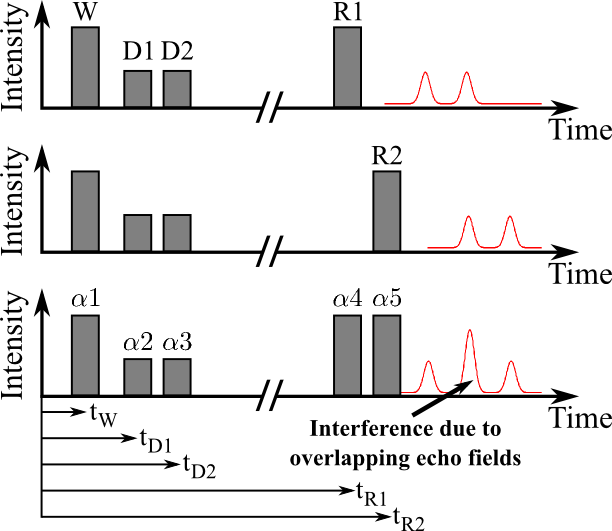}
\caption{\label{Fig_Memoire_pulse_sequence}Illustration of the pulse sequence used for time-bin pulses interference in a Ti:Er:LiNbO$_3$ waveguide memory. Figure inspired from Ref.~\cite{staudt_fidelity_2007}}
\end{figure}

More recently papers reported the characterization of Ti:Er:LiNbO$_3$ waveguides operated with stimulated photon-echo protocols, suitable for implementation of solid-state quantum memories. Details of the photon-echo operation principle can be found in~\cite{tittel_PEQMSS_2009,mitsunaga_time-domain_1992}. The first experiment showed the possibility of storing the amplitude and phase of two subsequent coherent pulses is depicted in \figurename~\ref{Fig_Memoire_pulse_sequence}~\cite{staudt_fidelity_2007}.

Two pulses are sent to the  memory after a bright "write" pulse. Then a second bright "read" pulse is used to retrieve the stored information, the relative times between the read pulse and the echo being the same as the relative times between the write and data pulses. To observe interference, two read pulses are sent to the memory with a time difference equivalent to the delay between the two data pulses. This pulse sequence configuration gives three echoes, but the middle one corresponds to the coherent superposition of two overlapping contributions, \textit{i.e.}, the second data pulse emitted by the first read pulse and the first data pulse triggered by the second read pulse. Consequently, an interference pattern can be observed thanks to the modulation of the phase carried by either the data or the read pulse. In this case, the authors obtained a near-perfect interference pattern visibility. This experiment therefore stands as a proof of the phase and amplitude preservation of two subsequent coherent pulses and the possibility of adjusting the phase.
Similarly to the previously described experiment, dealing with temporal interference, a second experiment has been dealing with spatial interference. It is based on a similar technology, and the aim was to demonstrate the coherent storage of the relative phase between two pulses in two independent quantum memories, located placed in an arm of a fiber interferometer~\cite{staudt_interference_2007}. The phase in the interferometer could be modulated using a fibre stretcher, leading to an interference pattern with a visibility of about 90\%. Therefore the phase was maintained over the storage time, and indistinguishability of photon echoes was proven.

\begin{figure}[h!]
\includegraphics[width=0.5\textwidth]{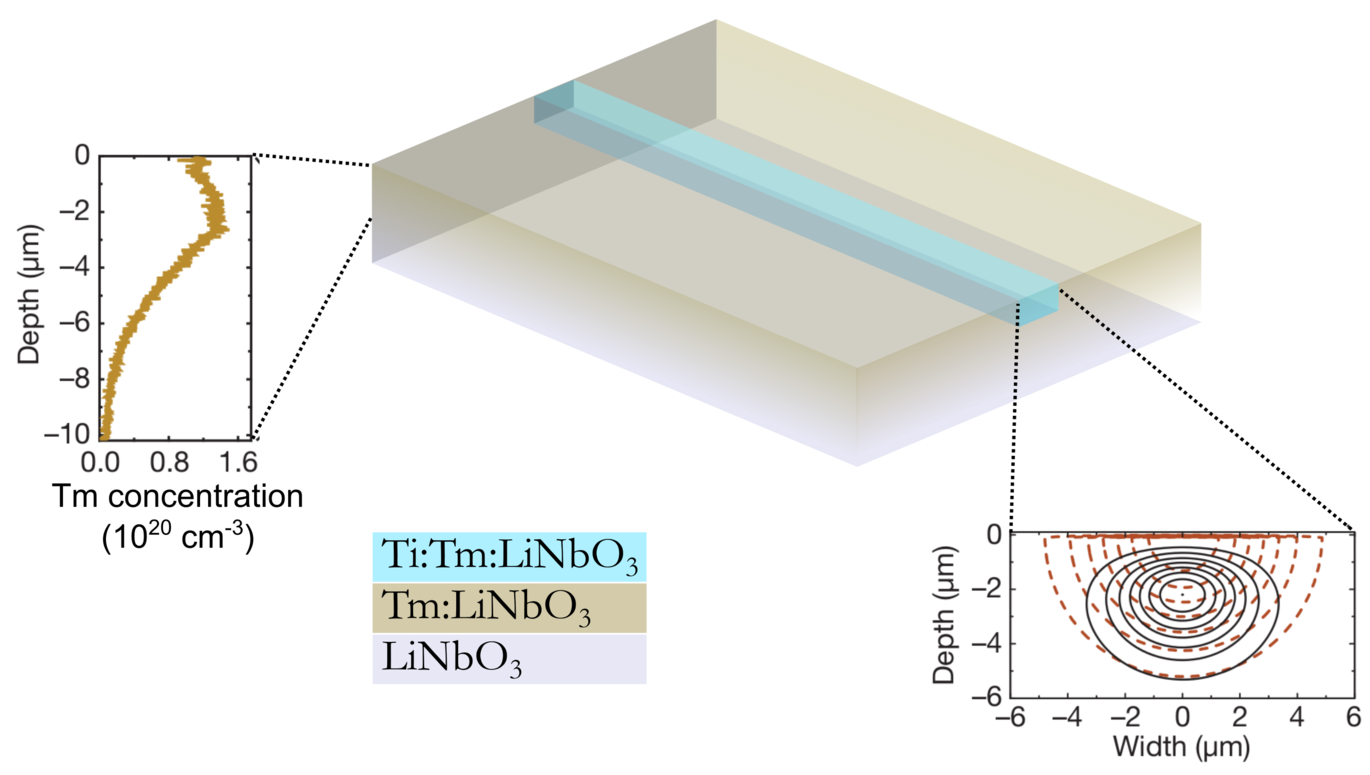}
\caption{\label{Fig_QM_waveguide}Illustration of the waveguide geometry and material layers for implementing a solid-state quantum memory. The Tm concentration profile is given on the left and the calculated intensity distribution of the fundamental TM-mode at 795\,nm is shown on the right. From the waveguide center, iso-intensity lines represent 100\%, 87.5\%, 75\%, \textit{etc.}, respectively, of the maximum intensity. Figure inspired from Ref.~\cite{Saglamyurek_BroadbandWQM_2011}.}
\end{figure}

Three years later, a thulium-doped LiNbO$_3$ waveguide has been investigated as an alternative to the Ti:Er:LiNbO$_3$ thanks to a collaboration between Calgary and Paderborn universities~\cite{sinclair_spectroscopic_2010,Saglamyurek_BroadbandWQM_2011}. With such a device, shown in \figurename~\ref{Fig_QM_waveguide}, the thulium (Tm$^{3+}$) transition of interest is at 795\,nm and the authors showed that the related transition has an optical coherence of 1.6\,$\mu$s when the sample is cooled at a temperature of 3\,K. A storage time of up to 7\,ns was demonstrated. The authors showed the coherent storage of time-bin entanglement using a photon-echo protocol. Thanks to this Tm-doped waveguide, they showed a broadband acceptance photon bandwidth of 5\,GHz, which corresponds to an increase by a factor 50 when compared to earlier published results for solid-state quantum memory devices. This is directly comparable with well-established systems based on atomic vapours, such as the ones demonstrated by the Oxford group~\cite{Walmsley_BroadMapping_2007,reim_towards_2010}.
Eventually, they obtained an input/output entanglement fidelity or more than 95\%. This realization highlights why IO technology could play a leading role in development of solid-state quantum memory systems.

\section{Advanced functions}
\label{Sec_ADV}

Together with the rise of ultra-bright integrated photon pair sources (see Section~\ref{Sec_EPPS}), integrated photonic circuits turned to be remarkable alternative to optical circuits based on bulk optical elements bolted to room-sized optical tables~\cite{Politi08}. Soon after, a new generation of integrated-photonics-enabled experiments was demonstrated~\cite{Peruzzo_2010,Tillman_2013,Bentivegnae1400255}. In this section, we aim at describing some of the more pertinent advanced circuits. It covers LN chips merging more than two elementary functions whose global performances have shown a significant step over bulk optics. This section will be divided in three parts: the first one encompasses demonstrators whose dense integration allows realizing otherwise impossible operations using bulk optics. We will start with a quantum relay chip applicable to long-distance quantum communication~\cite{Martin_IQR_2012} and also discuss a state-of-the-art chip allowing to generate and manipulate entangled photons~\cite{Jin_2014}. The second subsection will focus on realizations where dense integration enables in itself a new range of experiments, as for instance the generation of photon pairs in nonlinear waveguide arrays~\cite{Solntsev_2014}. Finally, the last subsection will deal with enhanced nonlinear interaction in integrated structures being exploited to demonstrate sources of advanced quantum states. The discussed experiments are generation of entangled photon triplets~\cite{Shalm_2013,Hubel_2010,Hamel_2014} and single photon interaction~\cite{Guerreiro_2014}.

\subsection{Integration of various optical functions}
As presented in Section~\ref{Sec_LN}, LN exhibits both optical-optical and electro-optical non-linear properties. It is therefore natural to exploit spontaneous parametric down-conversion to produce entangled photon pairs but also to integrate many channels that can be used to switch, recombine or make photons interfere at specific locations by use of electro-optically controllable couplers. Two favourable aspects of the IO implementations are: (a) the compactness and the monolithic construction of the chip ensure negligible phase drift between the channels during propagation; (b) most of the crucial functions for quantum processing are based on linear optical elements, including variable beam splitters and phase shifters~\cite{Knill01}, commonly achieved with current technologies on LN.

As a first example of dense integration, our group recently implemented a quantum relay chip for quantum communication, in a compact, stable, efficient, and user-friendly fashion.
\begin{figure}[h!]
\includegraphics[width=0.5\textwidth]{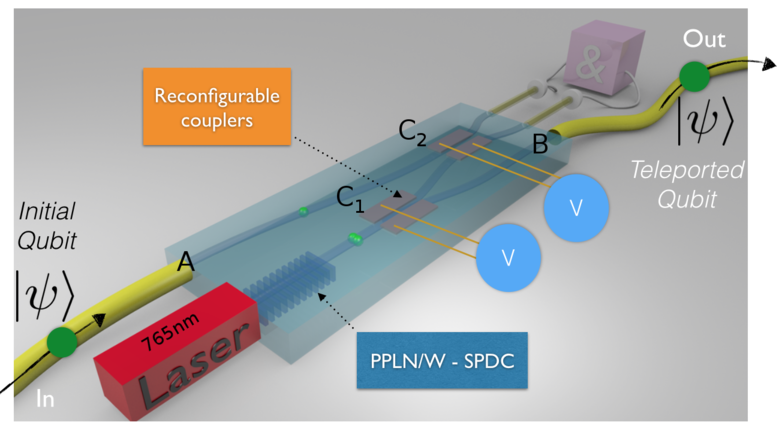}
\caption{3-D representation of a quantum relay chip made for extending one-way quantum cryptography systems. This chip gathers a PPLN section for photon-pair generation and two electro-optically controllable couplers $C_1$ et $C_2$. $D_1$ and $D_2$ are two detectors responsible for the Bell state measurement. At the end of the quantum channel, Bob's detector is triggered by the AND-gate (\&) placed after these two detectors. Figure inspired from Ref.~\cite{Martin_IQR_2012}.\label{relay_Nice}}
\end{figure}
\figurename~\ref{relay_Nice} presents a schematic of a chip on which all the necessary optical functions are merged for implementing on-chip teleportation~\cite{Martin_IQR_2012}. 
The basic operation is as follow: Let us suppose there is an initial qubit $|\psi\rangle$, encoded on a photon at 1530\,nm and travelling along a fiber quantum channel connected to port A to the relay chip. At the same time, a laser pulse at 766\,nm, synchronized with the arrival time of the photon, enters through the nonlinear zone of the chip which consists of a PPLN waveguide. Here, it generates by SPDC an entangled pair of photons (2 and 3) whose wavelengths are respectively centred at 1530\,nm and 1534\,nm. Then, the first 50/50 directional coupler ($C_1$) is used to separate the created entangled photons in such a way that photons 1 (sent by Alice) and 2 arrive at the same time at the second 50/50 coupler ($C_2$). If conditions on the polarization states, central wavelengths, and coherence times are met, photons 1 and 2 can be projected onto one of the four entangled ``Bell states''. The resulting state is identified by coincidence analysis (\&-gate) on detectors $D_1$ and $D_2$ placed at the output of the chip. As a result of this measurement, the quantum state $|\psi\rangle$ (initially carried by photon 1) can be teleported onto photon 3 that exits the chip at port B, ideally without any degradation of its quantum properties. This is made possible by photon 3 being initially entangled with photon 2. As a consequence the resulting electrical trigger is not only the signature of the presence of the initial carrier at the relay chip location but also of the departure of a new one coded with the same qubit state that remains unknown. 
From a practical point of view, when this teleportation process is repeated on all the qubits travelling along the quantum channel, we obtain, at the output of the chip, ``relayed'' photonic qubits coded on telecom photons. This relay function is ensured by the fact that each photon arrives synchronized with an electrical signal given by the AND-gate which allows triggering Bob's detectors at the end of the channel and therefore increasing both the SNR and the communication distance.
On the technological side (see \figurename~\ref{relay_Nice}), the chip features a photon-pair creation zone and two tunable couplers (which will be operated in 50/50 regime). The waveguiding structures are obtained using soft proton exchange on a Z-cut crystal. A 10\,mm long PPLN zone of poling period of 16.6\,$\mu$m allows the generation of twin photon at 1535\,nm at the temperature of $80^{\circ}$\,C. The directional couplers $C_1$ and $C_2$ consist of two waveguides integrated close to each other over mm-length. If the spacing is sufficiently small, energy is exchanged between the two guides. An electro-optical control of the coupling ratio is made possible using deposited electrodes enabling switching from 0 to 100\% coupling ratios and reconfigurable operation. To correctly separate photons 2 and 3 and entangle 1 and 2, the two couplers are set to the 50/50 ratio.

Once manufactured, the chip has been characterized using a weak coherent source in the single photon regime using the so-called HOM two-photon interference effect, which is seen as the major preliminary step towards achieving teleportation. We obtained a quantum interference featuring 79\%$\pm$25\% net visibility using two independent single photons, one external and one from the on-chip created pairs. The result is in good agreement with the theoretical expectation of 75\% accounting for the single photon statistics. The large error bars are essentially due to a non-optimal detection scheme and a lossy device (9\,dB) leading to a very low three-fold coincidence rate. However, it has been shown that the achievable quantum communication distance could be possibly augmented by a factor of 1.4 in that particular case.


Another excellent example of high integration is given by the chip from Nanjing University (China)~\cite{Jin_2014}. It gathers on a single substrate six different optical functions made of annealed proton exchanged waveguides integrated on a Z-cut PPLN crystal. Chronologically, the photons experience a Y-branch junction, transition tapers, a phase shifter, a nonlinear zone, and finally 50/50 couplers as well as wavelength filters. 

\begin{figure}[h!]
\includegraphics[width=0.5\textwidth]{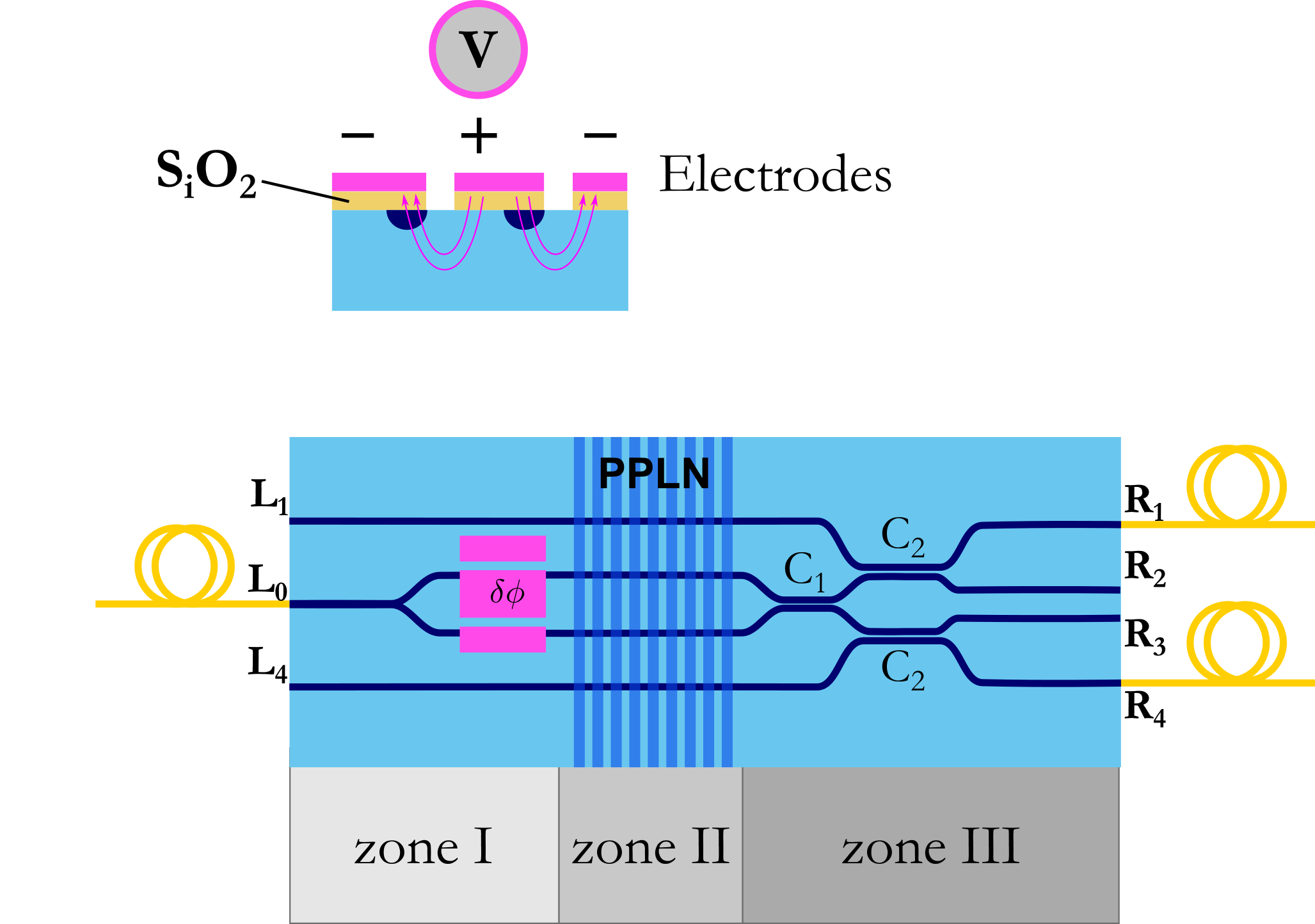}
\caption{The Nanjing University chip exhibits dimensions as small as 50\,mm$\times$5\,mm$\times$0.5\,mm. The widths of the single mode waveguides are 2\,$\mu$m for the pump (zone I) and 6\,$\mu$m for the entangled photons (zones II and III). The PPLN section is 10\,mm long with a period of 15.32\,$\mu$m. The interaction lengths of $C_1$ and $C_2$ are 650\,$\mu$m and 1300\,$\mu$m, respectively, with a separation of 4\,$\mu$m between the coupled waveguides. The electrodes are 8.35\,mm long with a separation of 6\,$\mu$m. The buffer layer (SiO$_2$) is etched along the gap between the electrodes to suppress DC drift. Figure inspired from Ref.~\cite{Jin_2014}.\label{chinese_chip}}
\end{figure}

The chip comprises three sections. In Section I, 780\,nm pump pulses are coupled into the central waveguide (L$_0$ on \figurename~\ref{chinese_chip}) and coherently split by a Y-branch single mode beam splitter toward two separated arms. Then, electrodes above each path are used to control their relative phase shift ($\delta\phi$). Eventually, transition tapers ensure efficient coupling of 780\,nm light into single mode waveguides at telecom wavelength (1560\,nm). Section II is dedicated to entanglement generation. It consists in a 10\,mm-long PPLN region, in which degenerate photon pairs at 1560\,nm are produced indistinguishably from either one of the two PPLN waveguides. Statistically, if much less than one pair is created at a time in the two waveguides, hence the photonic state is path-entangled, \textit{i.e.}, $|\psi\rangle=\left(|2,0\rangle+e^{i\delta\phi}|0,2\rangle\right)/\sqrt{2}$. Section III is designed to engineer entanglement. There, quantum interference is achieved using a 50/50 coupler ($C_1$). Depending on the applied phase, the output state remains path-entangled or is converted to the product state $|11\rangle$. Continuous morphing from a two-photon separated state to a bunched state is further demonstrated by on-chip fast control of electro-optic phase shift. The photon pairs are separated from the pump by on-chip wavelength filters ($C_2$) since the  photons at 1560\,nm experience a 100\% transfer in the $C_2$ couplers, toward waveguides $R_1$ and $R_4$ by evanescent coupling, whereas the pump at 780\,nm remains in $R_2$ and $R_3$.

From the technical side, a poling period of 15.32\,$\mu$m at a temperature of $25.5^{\circ}$\,C allows to generate pairs of photon with an associated brightness of $1.4\times 10^7$\,pairs.s$^{-1}$.nm$^{-1}$.mW$^{-1}$. It therefore requires a few tens of $\mu$W to operate in the single pair emission regime. The 780\,nm light suppression has been measured to be 29.2\,dB for $R_1$ and 31.4\,dB for $R_4$, accounted for the integrated filters and the propagation losses.

Single photon interference shows a visibility of 98.9\% and the period of the fringes is 7.1\,V with an offset voltage of 2.3\,V to compensate for a slight optical path difference between the two PPLN waveguides. The coupling efficiency from the chip to the output fibres provides an additional 1\,dB loss and the propagation losses for the entangled photons in waveguides $R_1$ and $R_4$ are 6.5\,dB and 8.6\,dB, respectively. The characterization of the emitted two-photon states has been performed thanks to a reverse HOM interference using an external beam-splitter. A visibility of 92\% has been reported.

\subsection{Advanced photonic state generation}

Quantum random walk and boson sampling are two iconic research trends since they belong to a class of mathematical problems extremely hard to simulate using classical computers when the number of walks becomes large~\cite{Dowling_book}. This is one peculiar situation where integrated quantum photonics turns out to be interesting by  allowing to work with more than 20 cascaded beam-splitters whose relative phases are inherently stable~\cite{Crespi_2014}. The chip demonstrated in Ref.~\cite{Solntsev_2014} through a collaboration between Australian and German groups adds an extra degree of complexity to random walk by investigating nonlinear random walk in a waveguide array to generate non-classical photon-pair states with a reconfigurable degree of path-entanglement.

\begin{figure}[h!]
\includegraphics[width=0.5\textwidth]{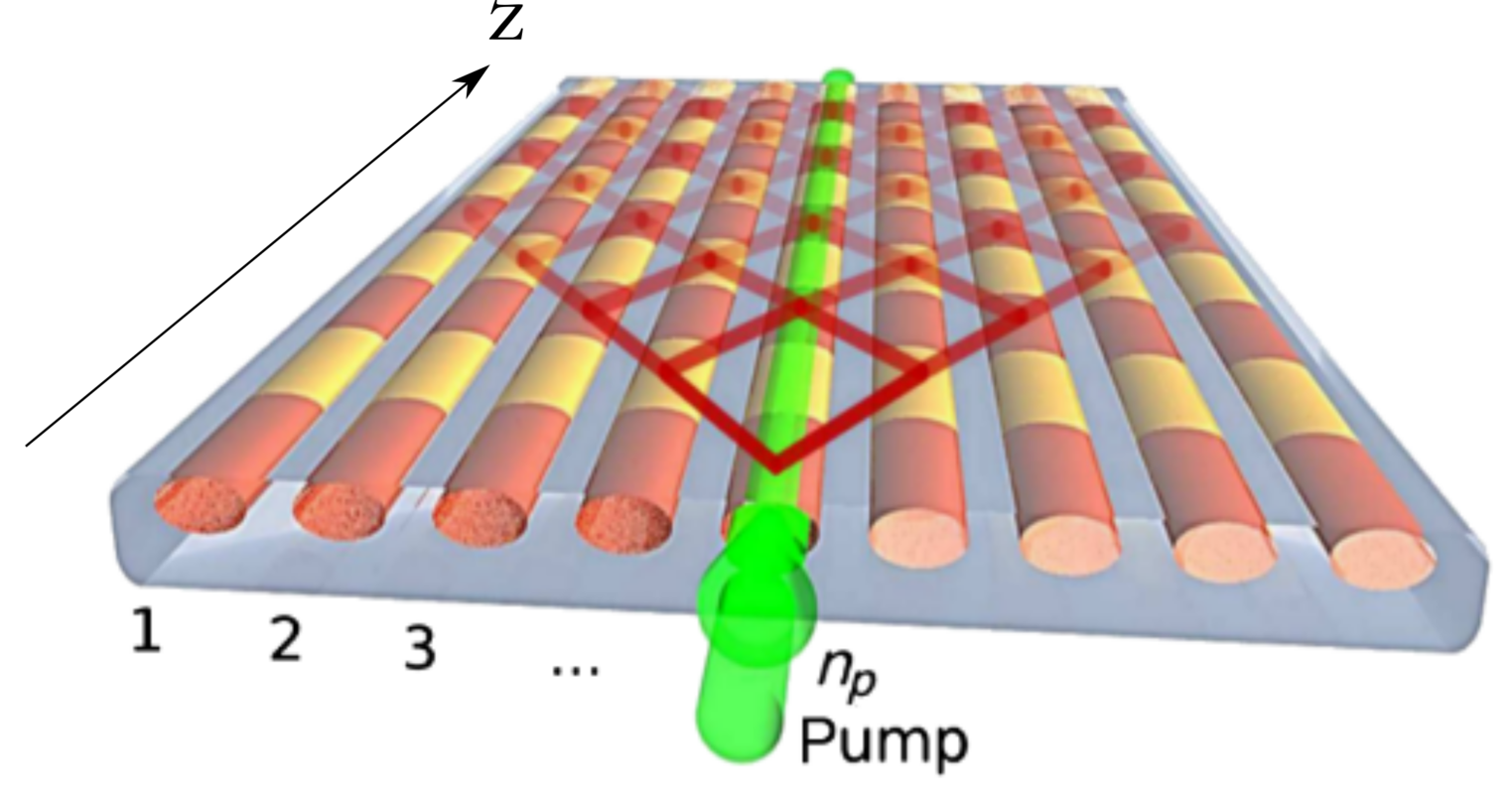}
\caption{Schematic illustration of a quadratic nonlinear waveguide array. The pump (green) is coupled to the waveguide number $n_p$. The photon pairs (red) are generated in the pump waveguide and propagate in the regime of quantum walks. As the pump continuously generates photon pairs along the propagation axis (z), signal and idler photons walk back and forth between all the waveguide (red arrows). Figure inspired from Ref.~\cite{Solntsev_2014}.\label{Aussie_chip}}
\end{figure}

As depicted in \figurename~\ref{Aussie_chip}, entanglement is achieved through quantum interference of photon pairs generated by SPDC in a $\chi^{(2)}$ nonlinear waveguide array (WGA). The WGAs are fabricated on z-cut PPLN wafers by titanium indiffusion. The WGA used for their experiment consists of 101 waveguides with a spacing of 13.5\,$\mu$m and a length of 51\,mm. The distance between each neighbour waveguides has been engineered so that the pump photons at 775\,nm never leaves the central waveguide, whereas the generated photon pairs at 1550\,nm undergo a quantum walk as schematically illustrated in \figurename~\ref{Aussie_chip}. Typical CW pump powers are below 0.5\,mW.

Starting from the pumped waveguide, photon pairs evolve over time by hopping to the left or right neighbouring waveguides. Here, the pump beam (green) can initialize photon-pair quantum walks in the central waveguide at all possible distances $z$ from the input. Basically, it implies that, at each position $z$, one has to consider the coherent superposition of created pairs hopping back and forth from the central waveguide but also the contribution of pairs created earlier. This effect leads to the excitation of a cascade of interfering 2D quantum walks, visualized by the red arrows in \figurename~\ref{Aussie_chip}. Actually, each of these quantum walks is a continuous-time nonlinear quantum walk with propagation distance $z$ providing much richer output states compared to passive quantum random walks. 

The cascaded quantum walks can be controlled by tuning the pump wavelength or changing the sample structure, providing a flexible tool for the preparation of states with a variable degree of nonclassicality in the spatial domain. For demonstration purpose, the authors have changed the phase mismatch by tuning the pump wavelength and also spanned over all possible input waveguides to show the reconfigurability of the quantum states. Interestingly, an additional theoretical investigation exploits spatial engineering of the PPLN duty cycle so as to obtain four coupled nonlinear waveguides whose efficiency is longitudinally modulated in order to produce, on demand, any desired Bell-state~\cite{sukhorukov_2015}.

\subsubsection{On-chip generation of photon-triplet states in integrated waveguide structures}

Photon-triplet states are a long standing goal in quantum optics since they are essential for fundamental tests of quantum mechanics and optical quantum technologies~\cite{GHZ_1990}. Their production has been proposed 20 years ago based on cascaded SPDC processes~\cite{GHZ_1990}. However, the direct generation of such a state suffers from an extremely low efficiency. The performances of PPLN waveguides have allowed the direct generation of photon triplets from SPDC of a single photon into a photon pair~\cite{Shalm2013}.

\begin{figure}[h!]
\includegraphics[width=0.5\textwidth]{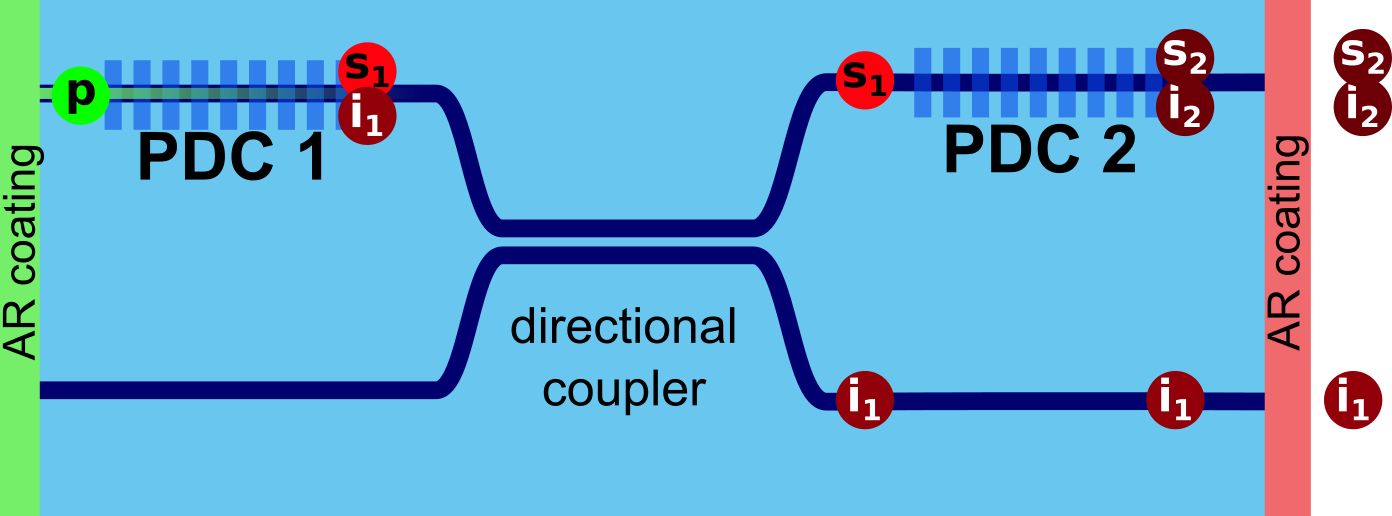}
\caption{Schematics of the device implemented for generating photon-triplets using cascaded SPDC processes. AR amounts for anti-reflection coatings at 532\,nm (input facte) and telecom wavelength (output facet). Figure inspired from Ref.~\cite{Krapick_15}.\label{triplet_chip}}
\end{figure}

As shown in \figurename~\ref{triplet_chip}, we will focus here on a fully integrated device that merges, on a single chip, two nonlinear zones and a wavelength-dependent directional coupler for such a purpose~\cite{Krapick_15}. Compared to free-space realizations, the integrated design allows maintaining the overall losses at a reduced level thanks to the ultra-low intrinsic scattering losses of waveguides and the lack of lossy interface like fibre-to-waveguide coupling. The chip indeed requires only 10\,$\mu$W of pump power for producing 11 detected photon triplets per hour and a coincidence signature of more than 27 standard deviations above the noise level. This compares very positively to free space approaches whose production rate was limited to 5 detected photon triplets per hour for several mW of pump power~\cite{Shalm2013}.

The chip is pumped with a laser beam at 532\,nm (green p-labelled photon) and the first periodically poled area efficiently generates paired photons respectively at 790\,nm (red s1-labelled photon) and 1625\,nm (brown i1-labelled photon). The on-chip directional coupler structure allows to demultiplex the paired photons as a function of their wavelengths. While the long-wavelength idler photons are transferred to the adjacent waveguide, the short-wavelength signal photons remain in the original arm. The signal photons can act as a pump for the second SPDC stage and decay to a pair of granddaughter photons at around 1580 nm (red s2 and i2 photons) within the other PPLN/W structure. Eventually, three telecom photons are produced out of one green pump photon and the tricky part of the experiment lies in the mean number of pairs (s1 and i1) to be maintained below 0.1 to ensure that only three photons are generated at a time. In other words, the triplets overall generation probability cannot be artificially compensated by increasing the pump power. Here the only free parameter is the internal conversion efficiency of SPDC which is directly related to the waveguiding structures.

Finally note that generation of tripartite telecom photons has recently been demonstrated at the University of Hefei (China) using hybrid-cascaded processes, \textit{i.e.} combining an atomic ensemble and a nonlinear PPLN waveguide~\cite{Ding_TripletOptica_2015}.

\subsubsection{Single photon nonlinear interaction}

Although not completely based on a fully integrated device as the previously discussed realizations, note that the Geneva group also reported the demonstration of photon triplet generation. The very interesting part of their approach lies in the sum-frequency generation (SFG) using two single photons as input states. Such an interaction at the single photon level has been a cornerstone in the community~\cite{Zhu:11,Friedler_04,Auffeves_07,Venkataraman_13,Immamoglu_97} for its central role in entanglement creation between independent photons. Some demonstrations have been made in artificial materials such as quantum dot in high-finesse cavities~\cite{Hu_09,Auffeves_07,Javadi2015}, or through electromagnetically induced transparency in atomic ensembles. Nowadays, a major challenge is to realise photon-photon interactions in materials that are less restrictive in terms of bandwidth and wavelength~\cite{Guerreiro_2014}. The Geneva team has taken an approach that exploits a parametric process in a nonlinear waveguide and shows for the first time a nonlinear interaction between two independent single photons via SFG.

\begin{figure}[h!]
\includegraphics[width=0.5\textwidth]{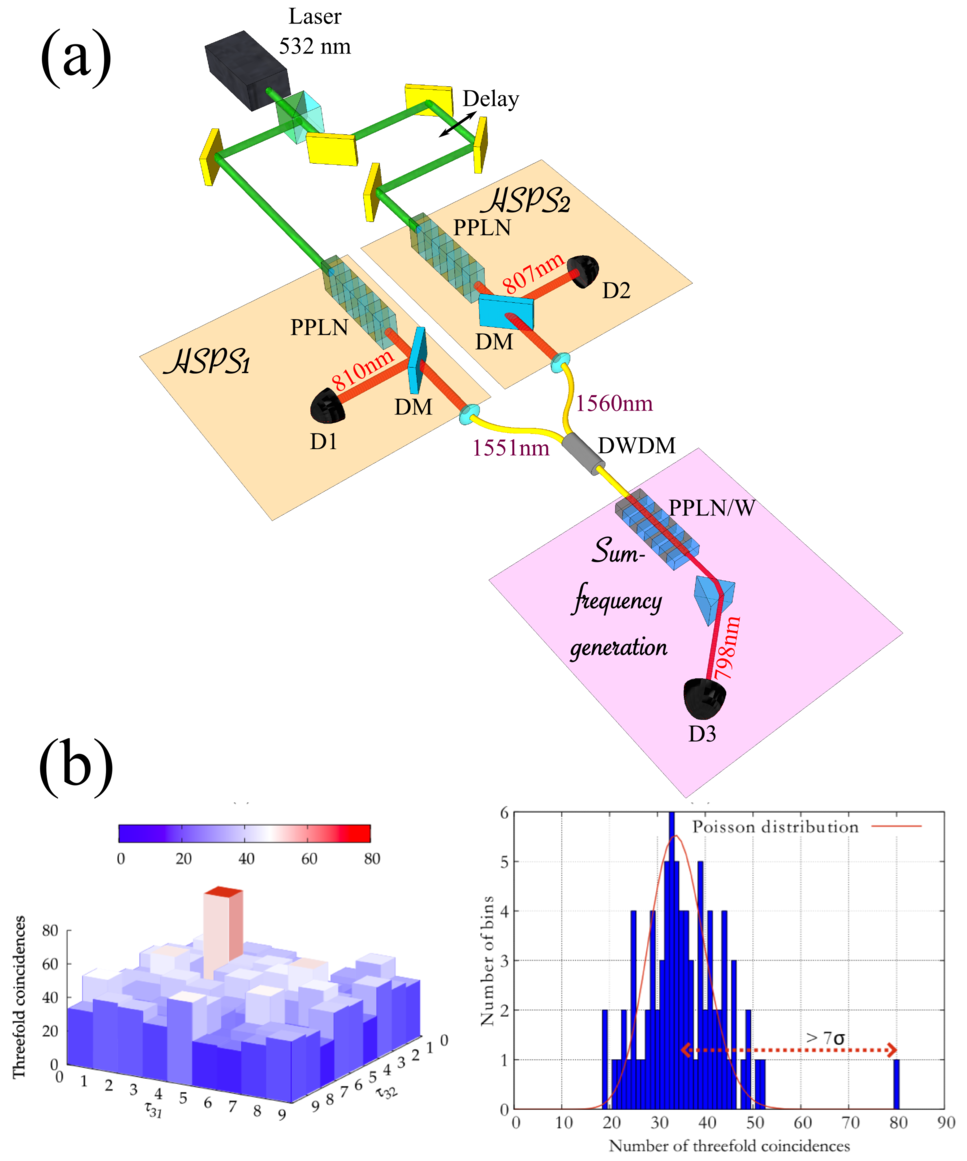}
\caption{a) Experimental setup. A mode-locked laser, emitting 10\,ps pulses at 532\,nm with a repetition rate of 430\,MHz, is used to pump two heralded single-photon sources (HSPS1 \& HSPS2) based on PPLN/Ws. The generated photons are deterministically separated by dichroic mirrors (DM), collimated, and then collected using single-mode fibres. Diffraction gratings (not shown) are employed to filter the heralding photons (810, 807\,nm) down to $\sim$0.3\,nm. In this configuration, the heralded telecom photons are projected onto a spectral mode that is matched to the acceptance bandwidth of the SFG process, which was measured to be 0.27\,nm. Those two telecom photons are combined via a dense wavelength division multiplexer (DWDM) and directed to a 4.5\,cm-long fibre-pigtailed type-0 PPLN/W. The unconverted photons are deterministically separated from the SFG photons by a prism (not shown), the latter being sent to a single-photon detector (D3). Threefold coincidences between detectors D1, D2, and D3 are recorded using a time-to-digital converter TDC.
b) A clear signature of the photon-photon interaction emerges from the background noise in the time-of-arrival three-fold coincidence histogram. The axis labeled $\tau_{31}$ shows the delay between detectors D1 and D3, while $\tau_{32}$ shows the delay between D2 and D3. Each pixel is composed of 2.3\,ns bins, defined by the laser repetition rate. The histogram counts shows a Poissonian distribution for the noise with a mean value of 35. The single peak at 80 corresponds to the true three-fold coincidences and exceeds the mean value by 7 standard deviations. Figure inspired from Ref.~\cite{Guerreiro_2014}.\label{SP_SFG}}
\end{figure}

As illustrated in \figurename~\ref{SP_SFG}(a), the authors have used two independent heralded single photon sources, HSPS1 and HSPS2, generating photons at 1560\,nm and 1551\,nm, respectively. Having two different wavelengths is of prime importance to distinguish SHG from SFG, \textit{i.e.}, to avoid the pollution of independent single photon interaction by two-photon state self interaction. The two HSPS are fibre-coupled to a PPLN/W that is quasi-phase matched to perform the SFG process 1560\,nm + 1551\,nm $\mapsto$ 778\,nm. Thanks to a single photon interaction efficiency of $1.56\times 10^{-8}$ (fibre pigtail losses included), which is directly related to the SPDC efficiency of the device, the authors have registered 80 three-fold coincidences over 260 hours as shown in \figurename~\ref{SP_SFG}(b). This result has a statistical significance of over 7 standard deviations with respect to the background (Poissonian distribution centred on 35).

As a conclusion, quantum photonics experiments have now reached such a level of complexity that only integrated optics can fit current and future requirements. Integrated photonics stands now as the first-choice approach for its reduced interface losses when the source is monolithically integrated to complex quantum circuits. The case of LN devices is particularly interesting since it exhibits all the technological requirements for future controllable photon-photon production, interaction, and manipulation.

\section{Generation \& manipulation of Squeezed states}
\label{Sec_Squeezing}

The last application of integrated photonics to the field of quantum information science, that we will discuss, concerns squeezed state of light. A key difference with the above discussed quantum functions is that, while one often consider quantum photonics to involve discrete observable (path, polarization or time-bin), squeezed state of light involves continuous variable (CV) such as the two quadrature of the electrical field of light. Among the fundamental applications, an attractive application feature is that entanglement on continuous variable (CV) of light can be generated in a deterministic way via non-linear optics and that, although propagation losses reduce le level of squeezing (or correlation), it is never completely destroyed~\cite{Braunstein2005, Solimeno2012}. Based on this principle, different realizations of squeezing at telecom wavelengths have been demonstrated~\cite{Schnabel_12v3dB, Peng_2013_TelecomSqueezing}.\\
In this context, PPLN/W can be conveniently exploited for the generation of CV non-classical light~\cite{Gupta_1997_SqueezingWg}. In particular, thanks to strong mode confinement and non-linear properties, SPDC in PPLN/Ws allows the efficient generation of high quality squeezed states without recurring to cavity-based conversion enhancements~\cite{Furusawa_2007_TwoModesSqueezing}. On one hand, compared to bulk implementations, single pass arrangement in waveguides offers better compactness and stability~\cite{Tanz_Genesis_2012}. On the other hand, in the absence of an optical resonator, the squeezing emission bandwidth is limited only by the SPDC phase-matching condition and can cover tens of THz. This last feature opens to the possibilities of wavelength demultiplexing and high-speed quantum communications.\\
In what follows, we will describe some interesting examples of applications of lithium waveguide technology to CV quantum experiments. In particular, we will report on the realization of a LN chip for the generation of squeezing at telecom wavelength in a compact and partially integrated setup pumped by a pulsed telecom laser. This will open to the presentation of a work demonstrating the possibility of enhancing pulsed homodyne detection performances thanks to a dedicated phase-matching stage implemented in a PPLN/W. Eventually, we will present a recent work where PPLN/Ws compatibility with fibred components is exploited to implement a fully guided wave squeezing experiment.\\

\begin{figure}[h]
\begin{center}
\includegraphics[width=0.5\textwidth]{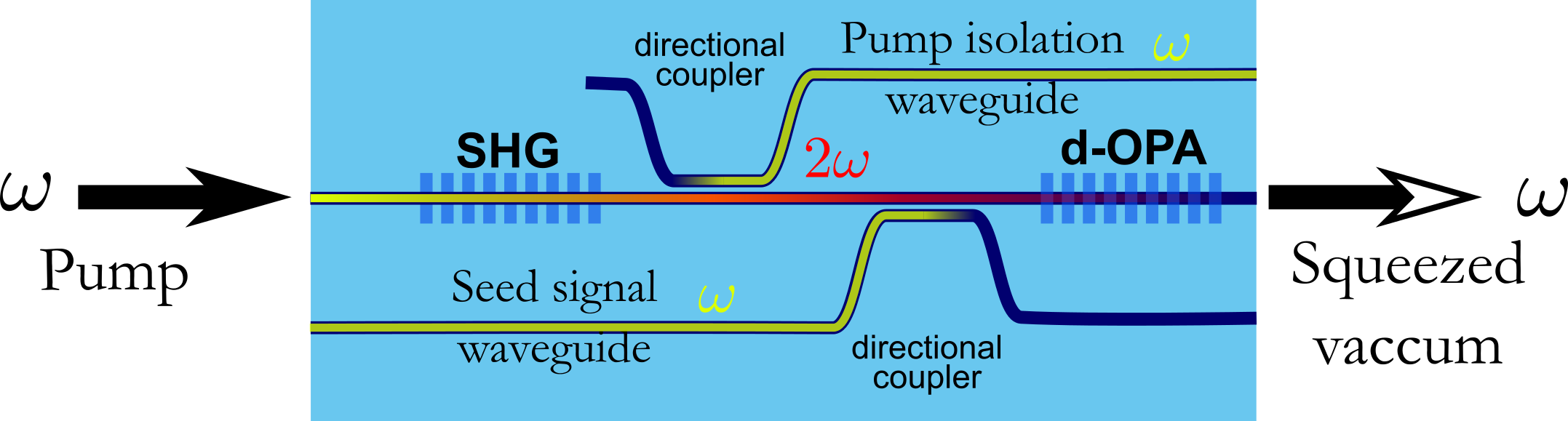}
\caption{Schematic of the integrated squeezing circuit inspired from Ref.~\cite{Fejer_2002_ChipSqueezing}. The chip is based on APE waveguide structures and includes a second harmonic generation stage, waveguide couplers, spatial mode filters, and a degenerate optical parametric amplification (D-OPA).\label{FejerSqueezing}}
\end{center}
\end{figure}

PPLN/Ws have been employed in both single mode and two mode squeezing experiments. For example, continuous wave (CW) broadband entanglement at 946\,nm has been demonstrated in Tokyo in 2007~\cite{Furusawa_2007_TwoModesSqueezing}, by combining on a 50:50 bulk beam splitter two independent squeezed states generated each by a PPLN/Ws. Correlated beams are then simultaneous measured thanks to a double homodyne detection scheme and entanglement is confirmed with a verification of Duan's inequality over a bandwidth of more than 30 MHz. As pointed out by the authors, the observed squeezing bandwidth is basically limited by the detector performances and could be improved with faster electronics. \\
In 2002, the squeezing generation in a LiNbO$_3$ integrated optical circuit is demonstrated in pulsed pump regime~\cite{Fejer_2002_ChipSqueezing}. The chip is based on APE waveguide structures and includes a second harmonic generation stage, waveguide couplers, spatial mode filters, and a degenerate optical parametric amplification (d-OPA) used to generate squeezed light at the telecom wavelength of $\approx$1538\,nm (see Fig.~\ref{FejerSqueezing}). On the component, the SHG and the d-OPA (\emph{i.e} the stimulated PDC) stages correspond to suitably periodical poled areas. The system is pumped with 20\,ps-long pulses at 1538\,nm (100\,MHz repetition rate), whose wavelength is frequency doubled by the SHG stage so as to provide a pump beam for the PDC. Another (non-converted) fraction of the same laser is sent towards a different input of the component and used as a seed for the d-OPA. The seed and the pump are combined thanks to an integrated waveguide coupler and reach together the d-OPA region. In addition to miniaturizing a part of the experimental setup, the integration reduces the amount of frequency doubled light lost in coupling to the d-OPA stage: in this sense, it increases the overall power efficiency of the device. At the chip output the squeezed light is analyzed in bulk homodyne detector. Experimentally, for 6\,W of incoming peak pump power at 1538\,nm, single mode squeezing with -1\,dB of measured noise reduction is observed.\\
As discussed by the authors of~\cite{Fejer_2002_ChipSqueezing}, a strong limitation in pulsed squeezing homodyning arises from poor detection efficiencies due to inadequate temporal mode matching between the beam under investigation and local oscillator (LO). As a matter of fact, a main difficulty in single-pass squeezers pumped in pulsed regime lies in gain induced diffraction, that distorts the squeezed light phase front, when working in high parametric gain (\emph{i.e.} high squeezing regime)~\cite{Fejer_2002_ChipSqueezing}. A valuable strategy to comply with this limitation has been demonstrated in 2008 in Tokyo~\cite{Hirano2008}. In this work, the LO pulses are suitably temporally shaped by sending them though a dedicated optical parametric amplifier realized in a periodically poled MgO:LiNbO3 waveguide. The experiment exploits the fact that parametric gain in the center portion of the amplified beam is larger than in its tails and that it can be conveniently adjusted by acting on the OPA pump beam: based on this principle, the LO temporal profile can be conveniently modified. The authors test their approach with the detection of squeezed light at 1535\,nm produced by SPDC in another periodically poled MgO:LiNbO3 waveguide. The squeezer is pumped by a frequency doubled Q-switched laser, emitting optical pulses at 1535\,nm with a duration of 3.7\,ns. A systematic study of detected squeezing as a function of the SPDC pump power is provided. Shaped LO pulses allow to achieve near-unit mode-matching efficiency and to measure, in the high gain regime, a squeezing of -4.1\,dB below the shot noise level. This value corresponds to an improvement of $\approx$1\,dB with respect to the case of non-shaped LO and proves the validity of the adopted strategy.\\

\begin{figure}[h]
\begin{center}
\includegraphics[width=0.5\textwidth]{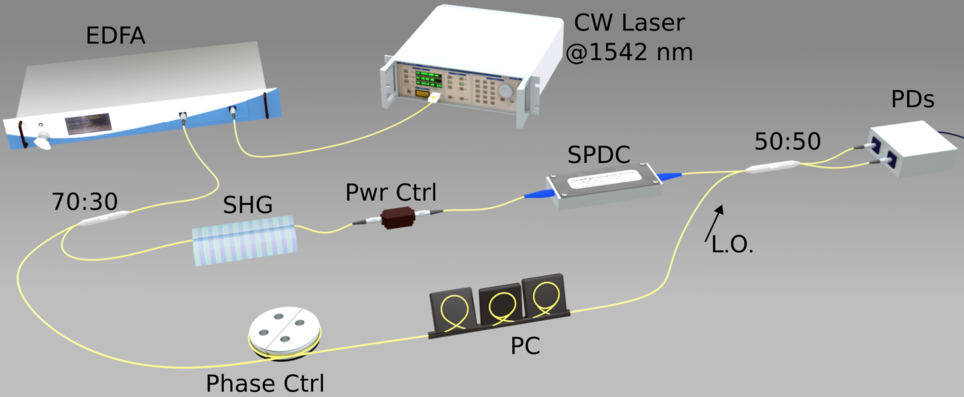}
\caption{Schematic of the experimental setup inspired from Ref.~\cite{Nice_2016_Squeezing}. A fiber coupled CW telecom laser at 1542\,nm is amplified (EDFA) and split into two beams by means of to a 70:30 fibre beam splitter. The less intense beam serves as the local oscillator (LO) while the brighter one is frequency doubled via SHG in a PPLN/W and used to pump a ridge waveguide (SPDC, PPLN/RW). The power of the beam at 771\,nm is controlled with an in-line fibred attenuator (Pwr Ctrl). At the output of the SPDC stage, the squeezed vacuum state at 1542\,nm is sent towards a fibre homodyne detector where it is optically mixed with the LO based on a balanced fibre beam splitter (50:50) followed by InGaAs photodiodes (PDs). The LO phase is scanned thanks to a fibre-stretcher module (Phase Ctrl) while a fibre polarization controller (PC) allows the polarization mode matching at the homodyne detector.\label{setupCVnice}}
\end{center}
\end{figure}

To conclude this section, we observe that in all presented implementations, squeezed states generated in PPLN/W are collected free-space to be characterized in bulk homodyne detectors. Very recently, in Nice, a entirely guided-wave realization of a squeezing experiment at telecom wavelength has been demonstrated, for the first time~\cite{Nice_2016_Squeezing} (see Fig.~\ref{setupCVnice}). In this scheme, single-mode squeezing at 1542\,nm is generated in CW pumping regime via degenerate SPDC in a periodically poled lithium niobate ridge-waveguide (PPLN/RW). At the output of the PPLN/RW, the non-classical beam is measured with a fibre homodyne detector. As for previous implementations, the SPDC pump beam is obtained by frequency doubling a CW laser operating at 1542\,nm. The SHG stage implemented in a SPE:PPLN/W. A fraction of the laser beams is used a LO for the homodyne detection. CW regime completely bypasses the LO temporal matching issues. In addition, besides the miniaturization of the setup, a major advantage of guided-wave optics lies in the achievement of a high degree of spatial mode matching between the LO and the signal without optical adjustment~\cite{Furusawa2015}. This huge benefit extremely simplifies the homodyne detector implementation. With this setup, the authors observe up to $-1.83$ dB of squeezing emitted at 1542 nm in CW pumping regime for an input pump power of 28\,mW. The proposed configuration allows implementing a plug-and-play and extremely easy experiment, entirely based on commercially available components and fully compatible with existing fibre networks.\\
In conclusion, as for discrete variable experiments in single photon regime, PPLN/Ws can be key ingredients for efficient CV experiments. More in details, as discussed, compared to their bulk counterparts, PPLN/W-based squeezing experiments offer the advantage of extremely simplifying the experimental setup, as demanded for out-of-the-lab realizations, while ensuring versatility, reliability and high non-linear optical performances.

\section{Conclusion \& Perspectives}
\label{Sec_CCL}

The impact of integrated optical devices in modern quantum optics has already been considerable. As have been outlined all along this manuscript, pertinent, and otherwise infeasible with bulk optics approaches, realizations have been reported over the last two decades, with outstanding advances in quantum sources, interfaces, relays, memories, as well as in linear optical quantum computing.
Among all the available technological quantum photonics platforms, lithium niobate has been playing a major role, due to mature fabrication processes and remarkable inherent properties, that are barely met with other materials.

Thanks to their compactness, efficiencies, and interconnects capabilities, many of the demonstrated individual devices based on lithium niobate can clearly serve as building blocks for more complex quantum systems. Those devices have also been shown compatible with other guided-wave technologies, such as standard fibre optics components. Elementary implementations of on-chip integration has begun and can be expected to be greatly expanded in the near future. While the input and output coupling continues to remain a difficult problem for all integrated optical devices, the continuing evolution of these developments, as well as the incorporation of new technologies, such as nano-wire waveguides and photonic crystal circuits, could add considerable impetus to this rapidly evolving field. There is no fundamental reason preventing a solution to this problem.

The road towards achieving exploitable quantum computers and networks might be very long and challenging. However, there exists a large space for important advances using quantum systems enabled by integrated photonics technologies, with tremendous impacts on how information will be processed and communicated. Up to now, we have only seen the premises of quantum enabled technologies. The future of quantum information science very likely lies in the development of new technologies as well as in the coherent merging of those technologies for assessing hybrid devices, for pushing the potential of dense optical integration to its maximum and exploiting the best features of all of them.

Increased density and functional integration clearly appears as one of the major directions for future deployment of quantum integrated optical devices.
A great potential might lie in integrated waveguide arrays, also referred to as quantum guidonics, where dense integration plays a crucial role, with or without nonlinear optical features. In coupled-waveguide arrays, light propagates as discrete beams, with both guided and free characteristics: these arrays are can therefore be considered as versatile two-dimensional meta-materials with, on lithium niobate, the capabilities of being (re-)configured on demand using electro-optical effect. The guidonic patterning, \textit{i.e.}, patterning the coupling coefficient between neighbour waveguides which controls the propagation, allows novel manipulation of light, constituting a complete discrete photonics corpus~\cite{Moison_09}. There is no doubt that the ``coherent'' interfacing of nanotechnologies, ferroelectric domain structuring, and integrated optics on lithium niobate should enable great advances in quantum optical systems in the near future~\cite{Gallo_15}.

\section*{Acknowledgment}
\label{Sec_ACK}

The authors acknowledge financial support from the ``Agence Nationale de la Recherche'' for the \textsc{e-Quanet} (ANR-09-BLAN-0333-01), \textsc{Conneqt} (ANR-2011-EMMA-0002), \textsc{Spocq} (ANR-14-CE32-0019), and \textsc{Inqca} (ANR-14-CE26-0038-01) projects, the European Commission for the FP7-FET \textsc{Quantip} (grant agreement N$^{\circ}$ 244026) and FP7-ITN \textsc{Picque} (grant agreement N$^{\circ}$ 608062) projects, the iXCore Research Foundation, the Simone \& Cino Del Duca Research Foundation (Institut de France), the Universit\'e Nice Sophia Antipolis, and the CNRS. The authors would like to thank D. Aktas, B. Fedrici, F. Mazeas, L. A. Ngah, and P. Vergyris for constant and fruitful help and discussions. 
ST would like to warmly thank ECT for constant support over the years.

\clearpage


\begin{thebibliography}{100}
\expandafter\ifx\csname url\endcsname\relax
  \def\url#1{{\tt #1}}\fi
\expandafter\ifx\csname urlprefix\endcsname\relax\def\urlprefix{URL }\fi
\providecommand{\eprint}[2][]{\url{#2}}

\bibitem{OBrien2009}
O'Brien J~L, Furusawa A and Vuckovic J 2009 {\em Nat. Photon.\/} {\bf 3}
  687--695

\bibitem{Tanz_Genesis_2012}
Tanzilli S, Martin A, Kaiser F, De~Micheli M~P, Alibart O and Ostrowsky D~B
  2012 {\em Laser \& Photon. Rev.\/} {\bf 6} 115--143

\bibitem{Brunner_BellRMP_2014}
Brunner N, Cavalcanti D, Pironio S, Scarani V and Wehner S 2014 {\em Rev. Mod.
  Phys.\/} {\bf 86} 419--478

\bibitem{Hensen_15}
Hensen B, Bernien H, Dreau A~E, Reiserer A, Kalb N, Blok M~S, Ruitenberg J,
  Vermeulen R~F~L, Schouten R~N, Abellan C, Amaya W, Pruneri V, Mitchell M~W,
  Markham M, Twitchen D~J, Elkouss D, Wehner S, Taminiau T~H and Hanson R 2015
  {\em Nature\/} {\bf 526} 682--686

\bibitem{Putz_LDLPRL_2016}
P\"utz G, Martin A, Gisin N, Aktas D, Fedrici B and Tanzilli S 2016 {\em Phys.
  Rev. Lett.\/} {\bf 116} 010401

\bibitem{Peruzzo_ScienceQDC_2012}
Peruzzo A, Shadbolt P, Brunner N, Popescu S and O'Brien J~L 2012 {\em
  Science\/} {\bf 338} 634--637

\bibitem{Kaiser_QDC_2012}
Kaiser F, Coudreau T, Milman P, Ostrowsky D~B and Tanzilli S 2012 {\em
  Science\/} {\bf 338} 637--640

\bibitem{Peruzzo_2010}
Peruzzo A, Lobino M, Matthews J~C~F, Matsuda N, Politi A, Poulios K, Zhou X~Q,
  Lahini Y, Ismail N, W√∂rhoff K, Bromberg Y, Silberberg Y, Thompson M~G and
  OBrien J~L 2010 {\em Science\/} {\bf 329} 1500--1503

\bibitem{Crespi_2014}
Crespi A, Osellame R, Ramponi R, Giovannetti V, Fazio R, Sansoni L, De~Nicola
  F, Sciarrino F and Mataloni P 2013 {\em Nat. Photon.\/} {\bf 7} 322--328

\bibitem{gisin_QKD_2002}
Gisin N, Ribordy G, Tittel W and Zbinden H 2002 {\em Rev. Mod. Phys.\/} {\bf
  74} 145--195

\bibitem{Scarani_QKDRMP_2009}
Scarani V, Bechmann-Pasquinucci H, Cerf N~J, Du\ifmmode~\check{s}\else
  \v{s}\fi{}ek M, L\"utkenhaus N and Peev M 2009 {\em Rev. Mod. Phys.\/} {\bf
  81} 1301--1350

\bibitem{Aktas_DWDM_2016}
Aktas D, Fedrici B, Kaiser F, Lunghi T, Labont\'e L and Tanzilli S 2016 {\em Laser \& Photon. Rev.\/} {\bf 10} 451--457

\bibitem{Korzh_2014}
Korzh B, Lim C~C~W, Houlmann R, Gisin N, Li M~J, Nolan D, Sanguinetti B, Thew R
  and Zbinden H 2014 {\em Nat. Photon.\/} {\bf 9} 163--168

\bibitem{Martin_2012}
Martin A, Alibart O, Micheli M~P~D, Ostrowsky D~B and Tanzilli S 2012 {\em New
  J. Phys.\/} {\bf 14} 025002

\bibitem{Metcalf2014}
Metcalf B~J, Spring J~B, Humphreys P~C, Thomas-Peter N, Barbieri M, Kolthammer
  W~S, Jin X~M, Langford N~K, Kundys D, Gates J~C, Smith B~J, Smith P~G~R and
  Walmsley I~A 2014 {\em Nat. Photon.\/} {\bf 8} 770--774

\bibitem{Dowling_Qmet_2008}
Dowling J~P 2008 {\em Contemp. Phys.\/} {\bf 49} 125

\bibitem{Giovannetti2011}
Giovannetti V, Lloyd S and Maccone L 2011 {\em Nat. Photon.\/} {\bf 5} 222--229

\bibitem{Deutsch85}
Deutsch D 1985 {\em Proceedings of the Royal Society of London, Series A\/}
  {\bf 400} 97--117

\bibitem{DiVincenzo95}
DiVincenzo D~P 1995 {\em Science\/} {\bf 270} 255--261

\bibitem{Nielsen00}
Nielsen M~A and Chuang I~L 2000 {\em Quantum computation and quantum
  information\/} (New York, NY, USA: Cambridge University Press) ISBN
  0-521-63503-9

\bibitem{Lanyon_2010}
Lanyon B~P, Whitfield J~D, Gillett G~G, Goggin M~E, Almeida M~P, Kassal I,
  Biamonte J~D, Mohseni M, Powell B~J, Barbieri M, Aspuru-Guzik A and White A~G
  2010 {\em Nat. Chem.\/} {\bf 2} 106--111

\bibitem{Tillman_2013}
Tillmann M, Dakic B, Heilmann R, Nolte S, Szameit A and Walther P 2013 {\em
  Nat. Photon.\/} {\bf 7} 540--544

\bibitem{Metcalf2013}
Metcalf B~J, Thomas-Peter N, Spring J~B, Kundys D, Broome M~A, Humphreys P~C,
  Jin X~M, Barbieri M, Steven~Kolthammer W, Gates J~C, Smith B~J, Langford N~K,
  Smith P~G and Walmsley I~A 2013 {\em Nat. Commun.\/} {\bf 4} 1356

\bibitem{kempe_03}
Kempe J 2003 {\em Contemp. Phys.\/} {\bf 44} 307--327

\bibitem{Solntsev_2014}
Solntsev A~S, Setzpfandt F, Clark A~S, Wu C~W, Collins M~J, Xiong C, Schreiber
  A, Katzschmann F, Eilenberger F, Schiek R, Sohler W, Mitchell A, Silberhorn
  C, Eggleton B~J, Pertsch T, Sukhorukov A~A, Neshev D~N and Kivshar Y~S 2014
  {\em Phys. Rev. X\/} {\bf 4} 031007

\bibitem{sanaka_new_2001}
Sanaka K, Kawahara K and Kuga T 2001 {\em Phys. Rev. Lett.\/} {\bf 86} 5620

\bibitem{tanzilli_ppln_2002}
Tanzilli S, Tittel W, de~Riedmatten H, Zbinden H, Baldi P, {De Micheli} M,
  Ostrowsky D and Gisin N 2002 {\em Eur. Phys. J. D\/} {\bf 18} 155--160

\bibitem{Bogaerts_LPRSilicon_2012}
Bogaerts W, De~Heyn P, Van~Vaerenbergh T, De~Vos K, Kumar~Selvaraja S, Claes T,
  Dumon P, Bienstman P, Van~Thourhout D and Baets R 2012 {\em Laser \& Photon.
  Rev.\/} {\bf 6} 47--73

\bibitem{silverstone_2014}
Silverstone J~W, Bonneau D, Ohira K, Suzuki N, Yoshida H, Iizuka N, Ezaki M,
  Natarajan C, Tanner M~G, Hadfield R~H, Zwiller V, Marshall G~D, Rarity J~G,
  O'Brien J~L and Thompson M~G 2014 {\em Nat. Photon.\/} {\bf 8} 104--108

\bibitem{Meany_2015}
Meany T, Gr√§fe M, Heilmann R, Perez-Leija A, Gross S, Steel M~J, Withford M~J
  and Szameit A 2015 {\em Laser \& Photon. Rev.\/} {\bf 9} 363--384

\bibitem{pruneri_1998}
Pruneri V, Bonfrate G, Kazansky P~G, Simonneau C, Vidakovic P and Levenson J~A
  1998 {\em Applied Physics Letters\/} {\bf 72} 1007--1009

\bibitem{angell_1994}
Angell M~J, Emerson R~M, Hoyt J~L, Gibbons J~F, Eyres L~A, Bortz M~L and Fejer
  M~M 1994 {\em Applied Physics Letters\/} {\bf 64} 3107--3109

\bibitem{deglinnocenti_2008}
Degl'Innocenti R, Majkic A, Sulser F, Mutter L, Poberaj G and G\"{u}nter P 2008
  {\em Opt. Express\/} {\bf 16} 11660--11669

\bibitem{choy_1976}
Choy M~M and Byer R~L 1976 {\em Phys. Rev. B\/} {\bf 14}(4) 1693--1706

\bibitem{shoji_1997}
Shoji I, Kondo T, Kitamoto A, Shirane M and Ito R 1997 {\em JOSA B\/} {\bf 14}
  2268--2294

\bibitem{vanherzeele_1992}
Vanherzeele H and Bierlein J~D 1992 {\em Optics letters\/} {\bf 17} 982--984

\bibitem{ballman_1965}
Ballman A~A 1965 {\em Journal of the American Ceramic Society\/} {\bf 48}
  112--113 ISSN 1551-2916

\bibitem{newnham_1975}
Newnham R~E, Miller C~S, Cross L~E and Cline T~W 1975 {\em physica status
  solidi (a)\/} {\bf 32} 69--78 ISSN 1521-396X

\bibitem{armstrong_1962}
Armstrong J~A, Bloembergen N, Ducuing J and Pershan P~S 1962 {\em Phys. Rev.\/}
  {\bf 127}(6) 1918--1939

\bibitem{miller_1965}
Miller R~C, Boyd G~D and Savage A 1965 {\em Applied Physics Letters\/} {\bf 6}
  77--79

\bibitem{fejer_1986}
Fejer M, Digonnet M and Byer R 1986 {\em Optics letters\/} {\bf 11} 230--232

\bibitem{shur_2015}
Shur V~Y, Akhmatkhanov A~R and Baturin I~S 2015 {\em Applied Physics Reviews\/}
  {\bf 2} 040604

\bibitem{yamada_1993}
Yamada M, Nada N, Saitoh M and Watanabe K 1993 {\em Applied Physics Letters\/}
  {\bf 62} 435--436

\bibitem{webjorn_1994}
Webj{\"o}rn J, Pruneri V, Russell P~S~J, Barr J and Hanna D 1994 {\em
  Electronics Letters\/} {\bf 30} 894--895

\bibitem{sugii_1978}
Sugii K, Fukuma M and Iwasaki H 1978 {\em Journal of Materials Science\/} {\bf
  13} 523--533 ISSN 0022-2461

\bibitem{sohler_1989}
Hoffmann H and Sohler W 1989 {\em Thin Solid Films\/} {\bf 175} 191 -- 200 ISSN
  0040-6090

\bibitem{Hofmann_99}
Hoffmann D, Schreiber G, Haase C, Herrmann H, Grundk\"{o}tter W, Ricken R and
  Sohler W 1999 {\em Opt. Lett.\/} {\bf 24} 896--898

\bibitem{nouroozi10}
Nouroozi R 2010 {\em All optical wavelength conversion and parametric
  amplification in Ti:PPLN channel waveguides for telecommunication
  applications\/} Ph.D. thesis Department Physik der Universit√§t Paderborn,
  Paderborn, Germany

\bibitem{Roussev06}
Roussev R~V 2006 {\em Optical-frequency mixers in periodically poled lithium
  niobate: materials, modelling and characterization\/} Ph.D. thesis Stanford
  University, United States

\bibitem{bortz93}
Bortz M~L, Eyres L~A and Fejer M~M 1993 {\em Applied Physics Letters\/} {\bf
  62} 2012--2014

\bibitem{Lenzini15}
Lenzini F, Kasture S, Haylock B and Lobino M 2015 {\em Opt. Express\/} {\bf 23}
  1748--1756

\bibitem{Parameswaran02}
Parameswaran K~R, Route R~K, Kurz J~R, Roussev R~V, Fejer M~M and Fujimura M
  2002 {\em Opt. Lett.\/} {\bf 27} 179--181

\bibitem{chanvillard_2000}
Chanvillard L, Aschi√©ri P, Baldi P, Ostrowsky D~B, de~Micheli M, Huang L and
  Bamford D~J 2000 {\em Applied Physics Letters\/} {\bf 76} 1089--1091

\bibitem{Korkishko_01}
Korkishko Y~N, Fedorov V~A, Baranov E~A, Proyaeva M~V, Morozova T~V, Caccavale
  F, Segato F, Sada C and Kostritskii S~M 2001 {\em J. Opt. Soc. Am. A\/} {\bf
  18} 1186--1191

\bibitem{Tien_71}
Tien P~K 1971 {\em Appl. Opt.\/} {\bf 10} 2395--2413

\bibitem{Ulrich_73}
Ulrich R and Torge R 1973 {\em Appl. Opt.\/} {\bf 12} 2901--2908

\bibitem{Geiss:10}
Geiss R, Diziain S, Iliew R, Etrich C, Hartung H, Janunts N, Schrempel F,
  Lederer F, Pertsch T and Kley E~B 2010 {\em Applied Physics Letters\/} {\bf
  97} 131109

\bibitem{Lu:12}
Lu H, Sadani B, Courjal N, Ulliac G, Smith N, Stenger V, Collet M, Baida F~I
  and Bernal M~P 2012 {\em Opt. Express\/} {\bf 20} 2974--2981

\bibitem{Wang:14}
Wang C, Burek M~J, Lin Z, Atikian H~A, Venkataraman V, Huang I~C, Stark P and
  Lon\v{c}ar M 2014 {\em Opt. Express\/} {\bf 22} 30924--30933

\bibitem{Geiss:15}
Geiss R, Saravi S, Sergeyev A, Diziain S, Setzpfandt F, Schrempel F, Grange R,
  Kley E~B, T\"{u}nnermann A and Pertsch T 2015 {\em Opt. Lett.\/} {\bf 40}
  2715--2718

\bibitem{Chang:16}
Chang L, Li Y, Volet N, Wang L, Peters J and Bowers J~E 2016 {\em Optica\/}
  {\bf 3} 531--535

\bibitem{Regener85}
Regener R and Sohler W 1985 {\em Applied Physics B\/} {\bf 36} 143--147 ISSN
  1432-0649

\bibitem{Migdall2013}
Migdall A, Polyakov S, Fan J and Bienfang J (eds) 2013 {\em Single-Photon
  Generation and Detection: Physics and Applications\/} ({\em Experimental
  Methods in the Physical Sciences\/} vol~45) (Academic Press) ISBN
  978-0-12-387695-9

\bibitem{Brecht_13}
Brecht B and Silberhorn C 2013 {\em Phys. Rev. A\/} {\bf 87}(5) 053810

\bibitem{Harder_13}
Harder G, Ansari V, Brecht B, Dirmeier T, Marquardt C and Silberhorn C 2013
  {\em Opt. Express\/} {\bf 21} 13975--13985

\bibitem{Kim_05}
Kim Y~H and Grice W~P 2005 {\em Opt. Lett.\/} {\bf 30} 908--910

\bibitem{Jin_13}
Jin R~B, Shimizu R, Wakui K, Benichi H and Sasaki M 2013 {\em Opt. Express\/}
  {\bf 21} 10659--10666

\bibitem{Bruno_14}
Bruno N, Martin A, Guerreiro T, Sanguinetti B and Thew R~T 2014 {\em Opt.
  Express\/} {\bf 22} 17246--17253

\bibitem{Avenhaus_09}
Avenhaus M, Eckstein A, Mosley P~J and Silberhorn C 2009 {\em Opt. Lett.\/}
  {\bf 34} 2873--2875

\bibitem{Liscidini_13}
Liscidini M and Sipe J~E 2013 {\em Phys. Rev. Lett.\/} {\bf 111}(19) 193602

\bibitem{Helt_12}
Helt L~G, Liscidini M and Sipe J~E 2012 {\em J. Opt. Soc. Am. B\/} {\bf 29}
  2199--2212

\bibitem{Eckstein_15}
Eckstein A, Boucher G, Lema√Ætre A, Filloux P, Favero I, Leo G, Sipe J~E,
  Liscidini M and Ducci S 2014 {\em Laser \& Photonics Reviews\/} {\bf 8}
  L76--L80 ISSN 1863-8899

\bibitem{Jizan_15}
Jizan I, Helt L~G, Xiong C, Collins M~J, Choi D~Y, Joon~Chae C~J, Liscidini
  Marco~Steel M~J, Eggleton B~J and Clark A~S 2015 {\em Sci. Rep.\/} {\bf 5} 12557

\bibitem{Fang_14}
Fang B, Cohen O, Liscidini M, Sipe J~E and Lorenz V~O 2014 {\em Optica\/} {\bf
  1} 281--284

\bibitem{Rozema_15}
Rozema L~A, Wang C, Mahler D~H, Hayat A, Steinberg A~M, Sipe J~E and Liscidini
  M 2015 {\em Optica\/} {\bf 2} 430--433

\bibitem{Hall_Quantum_Challenge}
Hall N 1999 {\em The British Journal for the Philosophy of Science\/} {\bf 50}
  313--315

\bibitem{Bennett_Qtele_1993}
Bennett C~H, Brassard G, Cr\'epeau C, Jozsa R, Peres A and Wootters W~K 1993
  {\em Phys. Rev. Lett.\/} {\bf 70}(13) 1895--1899

\bibitem{Bouwmeester_Qtele_1997}
Bouwmeester D, Pan J~W, Mattle K, Eibl M, Weinfurter H and Zeilinger A 1997
  {\em Nature\/} {\bf 390} 575--579

\bibitem{Boschi_Qtele_1998}
Boschi D, Branca S, de~Martini F, Hardy L and Popescu S 1998 {\em Phys. Rev.
  Lett.\/} {\bf 80} 1121--1124

\bibitem{kim_Qtele_2001}
Kim Y, Kulik S~P and Shih Y 2001 {\em Phys. Rev. Lett.\/} {\bf 86} 1370

\bibitem{Marcikic_Qtele_2003}
Marcikic I, de~Riedmatten H, Tittel W, Zbinden H and Gisin N 2003 {\em
  Nature\/} {\bf 421} 509--513

\bibitem{Ursin_Tele_Danube_2004}
Ursin R, Jennewein T, Aspelmeyer M, Kaltenbaek R, Lindenthal M and Zeilinger A
  2004 {\em Nature\/} {\bf 430} 849--849, brief com.

\bibitem{Landry_Qtele_PlainPalais_2007}
Landry O, van Houwelingen J~A~W, Beveratos A, Zbinden H and Gisin N 2007 {\em
  JOSA B\/} {\bf 24} 398--403

\bibitem{Kaiser_cw_tele_2015}
Kaiser F, Issautier A, Ngah L, Aktas D, Delord T and Tanzilli S 2015 {\em
  Selected Topics in Quantum Electronics, IEEE Journal of\/} {\bf 21} 69--77

\bibitem{Pan_Swap_1998}
Pan J~W, Bouwmeester D and Weinfurter H 1998 {\em Phys. Rev. Lett.\/} {\bf 80}
  3891--3894

\bibitem{Kaltenbaek_Inter_Indep_2009}
Kaltenbaek R, Prevedel R, Aspelmeyer M and Zeilinger A 2009 {\em Phys. Rev.
  A\/} {\bf 79} 040302(R)

\bibitem{Aboussouan_dipps_2010}
Aboussouan P, Alibart O, Ostrowsky D~B, Baldi P and Tanzilli S 2010 {\em Phys.
  Rev. A\/} {\bf 81} 021801

\bibitem{McMillan_SR2photI_2013}
McMillan A~R, Labont{\'e} L, Clark A~S, Bell B, Alibart O, Martin A, Wadsworth
  W~J, Tanzilli S and Rarity J~G 2013 {\em Sci. Rep.\/} {\bf 3} 2032

\bibitem{Collins_QRelays_2005}
Collins D, Gisin N and de~Riedmatten H 2005 {\em J. Mod. Opt.\/} {\bf 52}
  735--753

\bibitem{briegel_quantum_1998}
Briegel H, D{\"u}r W, Cirac J~I and Zoller P 1998 {\em Phys. Rev. Lett.\/} {\bf
  81} 5932

\bibitem{duan_LDQcom_2001}
Duan L, Lukin M~D, Cirac J~I and Zoller P 2001 {\em Nature\/} {\bf 414}
  413--418

\bibitem{simon_quantum_2007}
Simon C, de~Riedmatten H, Afzelius M, Sangouard N, Zbinden H and Gisin N 2007
  {\em Phys. Rev. Lett.\/} {\bf 98} 190503

\bibitem{Sangouard_DLCZRMP_2011}
Sangouard N, Simon C and de~Riedmatten H 2011 {\em Rev. Mod. Phys.\/} {\bf 83}
  33--73

\bibitem{BB84}
Bennett C~H and Brassard G 1984 {\em Proceedings of the {IEEE} {I}nternational
  {C}onference on {C}omputers, {S}ystems and {S}ignal {P}rocessing,
  {B}angalore, {I}ndia\/}  175

\bibitem{Ekert_Crypto_1991}
Ekert A~K 1991 {\em Phys. Rev. Lett.\/} {\bf 97} 661--664

\bibitem{QKD_commercial}
 {\em \url{www.idquantique.com}\/}

\bibitem{Crespi_protein_concentration_2012}
Crespi A, Lobino M, Matthews J~C~F, Politi A, Neal C~R, Ramponi R, Osellame R
  and O’Brien J~L 2012 {\em Applied Physics Letters\/} {\bf 100} 233704

\bibitem{O'brien03}
O'Brien J~L, Pryde G~J, White A~G, Ralph T~C and Branning D 2003 {\em Nature\/}
  {\bf 426} 264--267

\bibitem{Pittman03}
Pittman T~B, Fitch M~J, Jacobs B~C and Franson J~D 2003 {\em Phys. Rev. A\/}
  {\bf 68} 032316

\bibitem{Okamoto_cNOT_2011}
Okamoto R, O’Brien J~L, Hofmann H~F and Takeuchi S 2011 {\em Proceedings of the
  National Academy of Sciences\/} {\bf 108} 10067--10071

\bibitem{Politi08}
Politi A, Cryan M~J, Rarity J~G, Yu S and O'Brien J~L 2008 {\em Science\/} {\bf
  320} 646--649

\bibitem{Marshall09}
Marshall G~D, Politi A, Matthews J~C~F, Dekker P, Ams M, Withford M~J and
  O'Brien J~L 2009 {\em Opt. Express\/} {\bf 17} 12546--12554

\bibitem{Smith09}
Smith B~J, Kundys D, Thomas-Peter N, Smith P~G~R and Walmsley I~A 2009 {\em
  Opt. Express\/} {\bf 17} 13516--13525

\bibitem{Politi09}
Politi A, Matthews J~C~F and O'Brien J~L 2009 {\em Science\/} {\bf 325} 1221

\bibitem{Matthews09}
Matthews J~C~F, Politi A, Stefanov A and O'Brien J~L 2009 {\em Nat.
  Photon.\/} {\bf 3} 346--350

\bibitem{Sansoni_EntangChip_2010}
Sansoni L, Sciarrino F, Vallone G, Mataloni P, Crespi A, Ramponi R and Osellame
  R 2010 {\em Phys. Rev. Lett.\/} {\bf 105} 200503

\bibitem{Laing10}
Laing A, Peruzzo A, Politi A, Verde M~R, Halder M, Ralph T~C, Thompson M~G and
  O'Brien J~L 2010 {\em Appl. Phys. Lett.\/} {\bf 97} 211109

\bibitem{Matthews10}
Matthews J~C~F, Politi A, Bonneau D and O'Brien J~L 2011 {\em Phys. Rev.
  Lett.\/} {\bf 107}(16) 163602

\bibitem{Franson_Bell_1989}
Franson J~D 1989 {\em Phys. Rev. Lett.\/} {\bf 62} 2205--2208

\bibitem{tanzilli_ppln_2001}
Tanzilli S, de~Riedmatten H, Tittel W, Zbinden H, Baldi P, de~Micheli M,
  Ostrowsky D and Gisin N 2001 {\em Electron. Lett.\/} {\bf 37} 26--28

\bibitem{Clauser_consequences_1974}
Clauser J~F and Horne M~A 1974 {\em Phys. Rev. D\/} {\bf 10} 526--535

\bibitem{Thew_nonmax_2002}
Thew R~T, Tanzilli S, Tittel W, Zbinden H and Gisin N 2002 {\em Phys. Rev. A\/}
  {\bf 66} 062304

\bibitem{Martin_TB_2013}
Martin A, Kaiser F, Vernier A, Beveratos A, Scarani V and Tanzilli S 2013 {\em Phys.
  Rev. A\/} {\bf \textbf{87}} 020301(R)

\bibitem{suhara_generation_2007}
Suhara T, Okabe H and Fujimura M 2007 {\em {IEEE} Photon. Technol. Lett.\/}
  {\bf 19} 1093--1095

\bibitem{martin_polar_2010}
Martin A, Issautier A, Herrmann H, Sohler W, Ostrowsky D~B, Alibart O and
  Tanzilli S 2010 {\em New J. Phys.\/} {\bf 12} 103005

\bibitem{Kaiser_TypeII_2012}
Kaiser F, Issautier A, Ngah L~A, D\u{a}nil\u{a} O, Herrmann H, Sohler W, Martin
  A and Tanzilli S 2012 {\em New J. Phys.\/} {\bf \textbf{14}} 085015

\bibitem{Thya_type-II_proposal}
Thyagarajan K, Lugani J, Ghosh S, Sinha K, Martin A, Ostrowsky D~B, Alibart O
  and Tanzilli S 2009 {\em Phys. Rev. A\/} {\bf 80}(5) 052321

\bibitem{Herrmann_postselection_free_2013}
Herrmann H, Yang X, Thomas A, Poppe A, Sohler W and Silberhorn C 2013 {\em Opt.
  Express\/} {\bf 21} 27981--27991

\bibitem{Suhara_2QPM_for_polar_2009}
Suhara T, Nakaya G, Kawashima J and Fujimura M 2009 {\em Photonics Technology
  Letters, IEEE\/} {\bf 21} 1096--1098

\bibitem{Suhara_PPLNW_review}
Suhara T 2009 {\em Laser \& Photonics Reviews\/} {\bf 3} 370--393

\bibitem{yoshizawa_generation_2003}
Yoshizawa A, Kaji R and Tsuchida H 2003 {\em Electron. Lett.\/} {\bf 39} 621

\bibitem{Yoshi_polar_2004}
Yoshizawa A and Tsuchida H 2004 {\em Applied Physics Letters\/} {\bf 85}
  2457--2459

\bibitem{Herbauts_active_routing_2013}
Herbauts I, Blauensteiner B, Poppe A, Jennewein T and H\"{u}bel H 2013 {\em
  Opt. Express\/} {\bf 21} 29013--29024

\bibitem{Lim_polar1_2008}
Lim H~C, Yoshizawa A, Tsuchida H and Kikuchi K 2008 {\em Opt. Express\/} {\bf
  16} 12460--12468

\bibitem{Lim_polar2_2008}
Lim H~C, Yoshizawa A, Tsuchida H and Kikuchi K 2008 {\em Opt. Express\/} {\bf
  16} 16052--16057

\bibitem{Lim_polar3_2008}
Lim H~C, Yoshizawa A, Tsuchida H and Kikuchi K 2008 {\em Opt. Express\/} {\bf
  16} 22099--22104

\bibitem{Lim_polar4_2008}
Lim H~C, Yoshizawa A, Tsuchida H and Kikuchi K 2008 {\em Opt. Express\/} {\bf
  16} 14512--14523

\bibitem{Lim_polar_2010}
Lim H~C, Yoshizawa A, Tsuchida H and Kikuchi K 2010 {\em Optical Fiber
  Technology\/} {\bf 16} 225--235

\bibitem{Arahira_ridge_2011}
Arahira S, Namekata N, Kishimoto T, Yaegashi H and Inoue S 2011 {\em Opt.
  Express\/} {\bf 19} 16032--16043

\bibitem{jiang_generation_2007}
Jiang Y and Tomita A 2007 {\em J. Phys. B: At. Mol. Opt. Phys.\/} {\bf 40}
  437--443

\bibitem{takesue_generation_2005}
Takesue H, Inoue K, Tadanaga O, Nishida Y and Asobe M 2005 {\em Opt. Lett.\/}
  {\bf 30} 293--295

\bibitem{Kaiser_polar_transc_2013}
Kaiser F, Issautier A, Ngah L~A, Alibart O, Martin A and Tanzilli S 2013 {\em
  Laser Phys. Lett.\/} {\bf 10} 045202

\bibitem{Kaiser_SourceLong_2014}
Kaiser F, Ngah L~A, Issautier A, Delord T, Aktas D, Labont\'e L, D'Auria V,
  De~Micheli M~P, Kastberg A, Alibart O, Martin A and Tanzilli S 2014 {\em Opt.
  Com.\/} {\bf \textbf{327}} 7--16

\bibitem{Olislager_frequencybin_2010}
Olislager L, Cussey J, Nguyen A~T, Emplit P, Massar S, Merolla J~M and Huy K~P
  2010 {\em Phys. Rev. A\/} {\bf 82} 013804

\bibitem{Olislager_frequencybin_2014}
Olislager L, Woodhead E, Phan~Huy K, Merolla J~M, Emplit P and Massar S 2014
  {\em Phys. Rev. A\/} {\bf 89} 052323

\bibitem{Boyd_Litho_2011}
Boyd R~W and Dowling J~P 2011 {\em Quantum Information Processing\/} {\bf 11}
  891--901

\bibitem{Jin_2014}
Jin H, Liu F~M, Xu P, Xia J~L, Zhong M~L, Yuan Y, Zhou J~W, Gong Y~X, Wang W
  and Zhu S~N 2014 {\em Phys. Rev. Lett.\/} {\bf 113} 103601

\bibitem{Kruse_dualpath_2015}
Kruse R, Sansoni L, Brauner S, Ricken R, Hamilton C~S, Jex I and Silberhorn C
  2015 {\em Phys. Rev. A\/} {\bf 92} 053841

\bibitem{Setzpfand_nonlinear_coupler_2016}
Setzpfandt F, Solntsev A~S, Titchener J, Wu C~W, Xiong C, Schiek R, Pertsch T,
  Neshev D~N and Sukhorukov A~A 2016 {\em Laser \& Photonics Reviews\/} {\bf
  10} 131--136

\bibitem{pomarico_waveguide_2009}
Pomarico E, Sanguinetti B, Gisin N, Thew R~T, Zbinden H, Schreiber G, Thomas A
  and Sohler W 2009 {\em New J. Phys.\/} {\bf 11} 113042

\bibitem{Luo_type-II_OPO_2015}
Luo K~H, Herrmann H, Krapick S, Brecht B, Ricken R, Quiring V, Suche H, Sohler
  W and Silberhorn C 2015 {\em New Journal of Physics\/} {\bf 17} 073039

\bibitem{halder_high_2008}
Halder M, Beveratos A, Thew R~T, Jorel C, Zbinden H and Gisin N 2008 {\em New
  J. Phys.\/} {\bf 10} 023027

\bibitem{Halder_Ent_Indep_2007}
Halder M, Beveratos A, Gisin N, Scarani V, Simon C and Zbinden H 2007 {\em
  Nature Phys.\/} {\bf 3} 692--695

\bibitem{Brunner_Bell_2014}
Brunner N, Cavalcanti D, Pironio S, Scarani V and Wehner S 2014 {\em Rev. Mod.
  Phys.\/} {\bf 86}(2) 419--478

\bibitem{Acin_From_Bell_2006}
Ac\'{\i}n A, Gisin N and Masanes L 2006 {\em Phys. Rev. Lett.\/} {\bf 97}(12)
  120405

\bibitem{Acin_DIQ_2007}
Ac\'{\i}n A, Brunner N, Gisin N, Massar S, Pironio S and Scarani V 2007 {\em
  Phys. Rev. Lett.\/} {\bf 98}(23) 230501

\bibitem{Pironio_random_numbers_2010}
Pironio S, Acín A, Massar S, de~la Giroday A~B, Matsukevich D~N, Maunz P,
  Olmschenk S, Hayes D, Luo L, Manning T~A and Monroe C 2010 {\em Nature\/}
  {\bf 464} 1021--1024

\bibitem{Pironio_DIQ_2009}
Pironio S, Acín A, Brunner N, Gisin N, Massar S and Scarani V 2009 {\em New
  Journal of Physics\/} {\bf 11} 045021

\bibitem{Bancal_DIQ_2011}
Bancal J~D, Gisin N, Liang Y~C and Pironio S 2011 {\em Phys. Rev. Lett.\/} {\bf
  106}(25) 250404

\bibitem{Barreiro_DIQ_2013}
Barreiro J~T, Bancal J~D, Schindler P, Nigg D, Hennrich M, Monz T, Gisin N and
  Blatt R 2013 {\em Nat Phys\/} {\bf 9} 559--562

\bibitem{Xavier:09}
Xavier G~B, Walenta N, de~Faria G~V, Tempor„o G~P, Gisin N, Zbinden H and
  von~der Weid J~P 2009 {\em New Journal of Physics\/} {\bf 11} 045015

\bibitem{Bonneau:12}
Bonneau D, Lobino M, Jiang P, Natarajan C~M, Tanner M~G, Hadfield R~H, Dorenbos
  S~N, Zwiller V, Thompson M~G and O'Brien J~L 2012 {\em Phys. Rev. Lett.\/}
  {\bf 108}(5) 053601

\bibitem{Zhang:14}
Zhang P, Aungskunsiri K, Mart\'{\i}n-L\'opez E, Wabnig J, Lobino M, Nock R~W,
  Munns J, Bonneau D, Jiang P, Li H~W, Laing A, Rarity J~G, Niskanen A~O,
  Thompson M~G and O'Brien J~L 2014 {\em Phys. Rev. Lett.\/} {\bf 112}(13)
  130501

\bibitem{Kaiser_ultra_broadband_2016}
Kaiser F, Aktas D, Fedrici B, Lunghi T, Labont\'{e} L and Tanzilli S 2016 {\em Appl. Phys. Lett.\/} {\bf 108} 231108

\bibitem{Polyakov_2011_ReviewSinglePhoton}
Eisaman M~D, Fan J, Migdall A and Polyakov S~V 2011 {\em Rev. Sci. Instrum.\/}
  {\bf 82} 071101

\bibitem{Fasel_HSPS_2004}
Fasel S, Alibart O, Tanzilli S, Baldi P, Beveratos A, Gisin N and Zbinden H
  2004 {\em New J. Phys.\/} {\bf 6} 163

\bibitem{Walmsley_2008_HSPS}
Mosley P~J, Lundeen J~S, Smith B~J and Walmsley I~A 2008 {\em New J. Phys.\/}
  {\bf 10} 093011

\bibitem{Laurat_2011_EPJD}
D'Auria V, Morin O, Fabre C and Laurat J 2012 {\em Eur. Phys. J. D\/} {\bf 66}
  249

\bibitem{alibart_HSPS_2005}
Alibart O, Ostrowsky D~B, Baldi P and Tanzilli S 2005 {\em Opt. Lett.\/} {\bf
  30} 1539--1541

\bibitem{Krapick_HSPStwocolor_2013}
Krapick S, Herrmann H, Quiring V, Brecht B, Suche H and Silberhorn C 2013 {\em
  New J. Phys.\/} {\bf \textbf{15}} 033010

\bibitem{Ngah_HSPS_2014}
Ngah L~A, Alibart O, Labont\'e L, D'Auria V and Tanzilli S 2014 {\em Laser \&
  Photon. Rev.\/} {\bf \textbf{9}} L1--L5

\bibitem{Meany_HybridQcircuits_2014}
Meany T, Ngah L~A, Collins M~J, Clark A~S, Williams R~J, Eggleton B~J, Steel
  M~J, Withford M~J, Alibart O and Tanzilli S 2014 {\em Laser \& Photon.
  Rev.\/} {\bf \textbf{8}} L42--L46

\bibitem{Castelletto_2008_ReviewHSPS}
Castelletto S~A and Scholten R~E 2008 {\em Eur. Phys. J. Appl. Phys.\/} {\bf
  41} 181--194

\bibitem{Brida_2012_LowNoiseHSPS}
Brida G, Degiovanni I~P, Genovese M, Piacentini F, Traina P, {Della Frera} A,
  Tosi A, {Bahgat Shehata} A, Scarcella C, Gulinatti A, Ghioni M, Polyakov S~V,
  Migdall A and Giudice A 2012 {\em Appl. Phys. Lett.\/} {\bf 101} 221112

\bibitem{Yamamoto_2007_10GHz}
Zhang Q, Xie X, Takesue H, Nam S~W, Langrock C, Fejer M~M and Yamamoto Y 2007
  {\em Opt. Express\/} {\bf 15} 10288

\bibitem{Sasaki_2014_SciRep}
Jin R~B, Shimizu R, Morohashi I, Wakui K, Takeoka M, Izumi S, Sakamoto T,
  Fujiwara M, Yamashita T, Miki S, Terai H, Wang Z and Sasaki M 2014 {\em Sci.
  Rep.\/} {\bf 4} 7468

\bibitem{Migdall_HSPSPRA_2002}
Migdall A~L, Branning D and Castelletto S 2002 {\em Phys. Rev. A\/} {\bf 66}
  053805

\bibitem{weihs_photonic_2001}
Tittel W and Weihs G 2001 {\em Quant. Inf. Comp.\/} {\bf 1} 3--56

\bibitem{Julsgard_QMem_2004}
Julsgard B, Sherson J, Cirac J~I, Fiur{\`a}sek J and Polzik E~S 2004 {\em
  Nature\/} {\bf 432} 482--486

\bibitem{Langer_LongLivedQM_2005}
Langer C, Ozeri R, Jost J~D, Chiaverini J, de~Marco B, Ben-Kish A, Blakestad
  R~B, Britton J, Hume D~B, Itano W~M, Leibfried D, Reichle R, Rosenband T,
  Schaetz T, Schmidt P~O and Wineland D~J 2005 {\em Phys. Rev. Lett.\/} {\bf
  95} 060502

\bibitem{Giorgi_FrequencyHopping_2003}
Giorgi G, Mataloni P and de~Martini F 2003 {\em Phys. Rev. Lett.\/} {\bf 90}
  027902

\bibitem{Tanz_Interface_2005}
Tanzilli S, Halder M, Tittel W, Alibart O, Baldi P, Gisin N and Zbinden H 2005
  {\em Nature\/} {\bf 437} 116--120

\bibitem{curtz_CFDC_2010}
Curtz N, Thew R, Simon C, Gisin N and Zbinden H 2010 {\em Opt. Express\/} {\bf
  18} 22099

\bibitem{takesue_SPFDC_2010}
Takesue H 2010 {\em Phys. Rev. A\/} {\bf 82} 013833

\bibitem{Ikuta_InterfaceNatCom_2011}
Ikuta R, Kusaka Y, Kitano T, Kato H, Yamamoto T, Koashi M and Imoto N 2011, {\em Nat. Commun.\/} {\bf 2} 1544

\bibitem{Chaneliere_RemoteAtomEnt_06}
Matsukevich D~N, Chaneli\`ere T, Jenkins S~D, Lan S~Y, Kennedy T~A~B and
  Kuzmich A 2006 {\em Phys. Rev. Lett.\/} {\bf 96} 030405

\bibitem{Pan_QMRb_2008}
Chen Y~A, Chen S, Yuan Z~S, Zhao B, Chuu C~S, Schmiedmayer J and Pan J~W 2008
  {\em Nature Phys.\/} {\bf 4} 103--107

\bibitem{Pan_QMRb_2009}
Zhao B, Chen Y~A, Bao X~H, Strassel T, Chuu C~S, Jin X~M, Schmiedmayer J, Yuan
  Z~S, Chen S and Pan J~W 2009 {\em Nature Phys.\/} {\bf 5} 95--99

\bibitem{Shapiro_LongDistTele_2001}
Lloyd S, Shahriar M~S, Shapiro J~H and Hemmer P~R 2001 {\em Phys. Rev. Lett.\/}
  {\bf 87} 167903

\bibitem{simon_quantum_2010}
Simon C, Afzelius M, Appel J, de~la Giroday A~B, Dewhurst S~J, Gisin N, Hu C~Y,
  Jelezko F, Kr\"oll S, M{\"u}ller J~H, Nunn J, Polzik E~S, Rarity J~G,
  de~Riedmatten H, Rosenfeld W, Shields A~J, Sk\"old N, Stevenson R~M, Thew R,
  Walmsley I~A, Weber M~C, Weinfurter H, Wrachtrup J and Young R~J 2010 {\em
  Eur. Phys. J. D\/} {\bf 58} 1--22

\bibitem{arXiv:1607.07314}
Ikuta R, Nozaki S, Yamamoto T, Koashi M and Imoto N 2016, {\em e-print arXiv:1607.07314\/}

\bibitem{arXiv:1607.01350}
Farrera P, Maring N, Albrecht B, Heinze G and de Riedmatten H 2016, {\em e-print arXiv:1607.01350\/}

\bibitem{Diamanti_PRA_05}
Diamanti E, Takesue H, Honjo T, Inoue K and Yamamoto Y 2005 {\em Phys. Rev.
  A\/} {\bf 72} 052311

\bibitem{Thew_GHzQKD_2006}
Thew R~T, Tanzilli S, Krainer L, Zeller S~C, Rochas A, Rech I, Cova S, Zbinden
  H and Gisin N 2006 {\em New J. Phys.\/} {\bf 8} 32

\bibitem{Matsukevich_DSPS_2006}
Matsukevich D~N, Chaneli\`ere T, Jenkins S~D, Lan S~Y, Kennedy T~A~B and
  Kuzmich A 2006 {\em Phys. Rev. Lett.\/} {\bf 97} 013601

\bibitem{Chen_DSPS_2006}
Chen S, Chen Y~A, Strassel T, Yuan Z~S, Zhao B, Schmiedmayer J and Pan J~W 2006
  {\em Phys. Rev. Lett.\/} {\bf 97} 173004

\bibitem{Simon_QR_2007}
Simon C, de~Riedmatten H, Afzelius M, Sangouard N, Zbinden H and Gisin N 2007
  {\em Phys. Rev. Lett.\/} {\bf 98} 190503

\bibitem{Simon_multiQR_2010}
Simon C, de~Riedmatten H and Afzelius M 2010 {\em Phys. Rev. A\/} {\bf 82}
  010304

\bibitem{Shor_DecoQC_1996}
Shor P~W 1995 {\em Phys. Rev. A\/} {\bf 52} R2493--R2496

\bibitem{Knill_QECC_1997}
Knill E and Laflamme R 1997 {\em Phys. Rev. A\/} {\bf 55} 900--911

\bibitem{Kok_QC_2007}
Kok P, Munro W~J, Nemoto K, Ralph T~C, Dowling J~P and Milburn G~J 2007 {\em
  Rev. Mod. Phys.\/} {\bf 79} 135--174

\bibitem{knill_QC_2010}
Knill E 2010 {\em Nature\/} {\bf 463} 441--443

\bibitem{tittel_PEQMSS_2009}
Tittel W, Afzelius M, Chaneli\`ere T, Cone R~L, Kr\"oll S, Moiseev S and
  Sellars M 2009 {\em Laser \& Photon. Rev.\/} {\bf 4} 244--267

\bibitem{chaneliere_storage_2005}
Chaneli\`ere T, Matsukevich D~N, Jenkins S~D, Lan S~Y, Kennedy T~A~B and
  Kuzmich A 2005 {\em Nature\/} {\bf 438} 833--836

\bibitem{reim_towards_2010}
Reim K~F, Nunn J, Lorenz V~O, Sussman B~J, Lee K~C, Langford N~K, Jaksch D and
  Walmsley I~A 2010 {\em Nat. Photon.\/} {\bf 4} 218--221

\bibitem{de_riedmatten_ss_2008}
de~Riedmatten H, Afzelius M, Staudt M~U, Simon C and Gisin N 2008 {\em
  Nature\/} {\bf 456} 773--777

\bibitem{chaneliere_light_2010}
Chaneli\`ere T, Bonarota M, Damon V, Lauro R, Ruggiero J, Lorger\'e I and
  Gou\"et J~L 2010 {\em J. Lumin.\/} {\bf 130} 1572--1578

\bibitem{lvovsky_optical_2009}
Lvovsky A~I, Sanders B~C and Tittel W 2009 {\em Nat. Photon.\/} {\bf 3}
  706--714

\bibitem{Hammerer_interface_10}
Hammerer K, S\o{}rensen A~S and Polzik E~S 2010 {\em Rev. Mod. Phys.\/} {\bf
  82} 1041--1093

\bibitem{staudt_interference_2007}
Staudt M~U, Afzelius M, de~Riedmatten H, {Hastings-Simon} S~R, Simon C, Ricken
  R, Suche H, Sohler W and Gisin N 2007 {\em Phys. Rev. Lett.\/} {\bf 99}
  173602

\bibitem{staudt_fidelity_2007}
Staudt M~U, {Hastings-Simon} S~R, Nilsson M, Afzelius M, Scarani V, Ricken R,
  Suche H, Sohler W, Tittel W and Gisin N 2007 {\em Phys. Rev. Lett.\/} {\bf
  98} 113601

\bibitem{Kraus_CRIB_2006}
Kraus B, Tittel W, Gisin N, Nilsson M, Kr\"oll S and Cirac J~I 2006 {\em Phys.
  Rev. A\/} {\bf 73} 020302

\bibitem{Gisin_PEQM_2007}
Gisin N, Moiseev S~A and Simon C 2007 {\em Phys. Rev. A\/} {\bf 76} 014302

\bibitem{Lauritzen_SSM_2010}
Lauritzen B, Min\'a\v{r} J, de~Riedmatten H, Afzelius M, Sangouard N, Simon C
  and Gisin N 2010 {\em Phys. Rev. Lett.\/} {\bf 104} 080502

\bibitem{mitsunaga_time-domain_1992}
Mitsunaga M 1992 {\em Opt. Quantum Electron.\/} {\bf 24} 1137--1150

\bibitem{sinclair_spectroscopic_2010}
Sinclair N, Saglamyurek E, George M, Ricken R, Mela C~L, Sohler W and Tittel W
  2010 {\em J. Lumin.\/} {\bf 130} 1586--1593

\bibitem{Saglamyurek_BroadbandWQM_2011}
Saglamyurek E, Sinclair N, Jin J, Slater J~A, Oblak D, Bussi\`{e}res F, George
  M, Ricken R, Sohler W and Tittel W 2011 {\em Nature\/} {\bf 469} 512--515

\bibitem{Walmsley_BroadMapping_2007}
Nunn J, Walmsley I~A, Raymer M~G, Surmacz K, Waldermann F~C, Wang Z and Jaksch
  D 2007 {\em Phys. Rev. A\/} {\bf 75} 011401(R)

\bibitem{Bentivegnae1400255}
Bentivegna M, Spagnolo N, Vitelli C, Flamini F, Viggianiello N, Latmiral L,
  Mataloni P, Brod D~J, Galv{\~a}o E~F, Crespi A, Ramponi R, Osellame R and
  Sciarrino F 2015 {\em Science Adv.\/} {\bf 1} e1400255

\bibitem{Martin_IQR_2012}
Martin A, Alibart O, De~Micheli M~P, Ostrowsky D~B and Tanzilli S 2012 {\em New
  J. Phys.\/} {\bf \textbf{14}} 025002

\bibitem{Shalm_2013}
Shalm L~K, Hamel D~R, Yan Z, Simon C, Resch K~J and Jennewein T 2013 {\em Nat.
  Phys.\/} {\bf 9} 19--22

\bibitem{Hubel_2010}
Hubel H, Hamel D~R, Fedrizzi A, Ramelow S, Resch K~J and Jennewein T 2010 {\em
  Nature\/} {\bf 466} 601--603

\bibitem{Hamel_2014}
Hamel D~R, Shalm L~K, H{\"u}bel H, Miller A~J, Marsili F, Verma V~B, Mirin R~P,
  Nam S~W, Resch K~J and Jennewein T 2014 {\em Nat. Photon.\/} {\bf 8} 801--807

\bibitem{Guerreiro_2014}
Guerreiro T, Martin A, Sanguinetti B, Pelc J~S, Langrock C, Fejer M~M, Gisin N,
  Zbinden H, Sangouard N and Thew R~T 2014 {\em Phys. Rev. Lett.\/} {\bf
  113}(17) 173601

\bibitem{Knill01}
Knill E, Laflamme R and Milburn G~J 2001 {\em Nature\/} {\bf 409} 46--52

\bibitem{Dowling_book}
Dowling J~P 2013 {\em Schr\"odinger's Killer App: Race to Build the World's
  First Quantum Computer\/} (CRC Press)

\bibitem{sukhorukov_2015}
Titchener J~G, Solntsev A~S and Sukhorukov A~A 2015 {\em Phys. Rev. A\/} {\bf
  92} 033819

\bibitem{GHZ_1990}
Greenberger D~M, Horne M~A, Shimony A and Zeilinger A 1990 {\em Am. J. Phys.\/}
  {\bf 58} 1131--1143

\bibitem{Shalm2013}
Shalm L~K, Hamel D~R, Yan Z, Simon C, Resch K~J and Jennewein T 2013 {\em Nat.
  Phys.\/} {\bf 9} 19--22

\bibitem{Krapick_15}
Krapick S, Brecht B, Quiring V, Ricken R, Herrmann H and Silberhorn C 2015
  On-chip generation of photon-triplet states in integrated waveguide
  structures {\em CLEO: 2015\/} p FM2E.3

\bibitem{Zhu:11}
Zhu C and Huang G 2011 {\em Opt. Express\/} {\bf 19} 23364--23376

\bibitem{Ding_TripletOptica_2015}
Ding D-S, Zhang W, Shi S, Zhou Z-Y, Li Y, Shi B-S and Guo G-C 2015 {\em Optica\/} {\bf 2} 642--645

\bibitem{Friedler_04}
Friedler I, Kurizki G and Petrosyan D 2004 {\em EPL (Europhys. Lett.)\/} {\bf
  68} 625

\bibitem{Auffeves_07}
Auff\`eves-Garnier A, Simon C, G\'erard J~M and Poizat J~P 2007 {\em Phys. Rev.
  A\/} {\bf 75} 053823

\bibitem{Venkataraman_13}
Venkataraman V, Saha K and Gaeta A~L 2013 {\em Nat. Photon.\/} {\bf 7} 138--141

\bibitem{Immamoglu_97}
Imamo\ifmmode~\bar{g}\else \={g}\fi{}lu A, Schmidt H, Woods G and Deutsch M
  1997 {\em Phys. Rev. Lett.\/} {\bf 79}(8) 1467--1470

\bibitem{Hu_09}
Hu C~Y, Munro W~J, O'Brien J~L and Rarity J~G 2009 {\em Phys. Rev. B\/} {\bf 80} 205326

\bibitem{Javadi2015}
Javadi A, Sollner I, Arcari M, Hansen S~L, Midolo L, Mahmoodian S, Kirsanske G,
  Pregnolato T, Lee E~H, Song J~D, Stobbe S and Lodahl P 2015 {\em Nat. Commun.\/} {\bf 6} 8655

\bibitem{Braunstein2005}
Braunstein S~L and van Loock P 2005 {\em Rev. Mod. Phys.\/} {\bf 77} 513--577

\bibitem{Solimeno2012}
Buono D, Nocerino G, Porzio A and Solimeno S 2012 {\em Phys. Rev. A\/} {\bf 86}
  042308

\bibitem{Schnabel_12v3dB}
Mehmet M, Ast S, Eberle T, Steinlechner S, Vahlbruch H and Schnabel R 2011 {\em
  Opt. Express\/} {\bf 19} 25763--25772

\bibitem{Peng_2013_TelecomSqueezing}
Jun-Jun Z, Xiao-Min G, Xu-Yang W, Ning W, Yong-Min L and Kun-Chi P 2013 {\em
  Chinese Phys. Lett.\/} {\bf 30} 060302

\bibitem{Gupta_1997_SqueezingWg}
Anderson M~E, McAlister D~F, Raymer M~G and Gupta M~C 1997 {\em J. Opt. Soc.
  Am. B\/} {\bf 14} 3180

\bibitem{Furusawa_2007_TwoModesSqueezing}
Yoshino K~i, Aoki T and Furusawa A 2007 {\em Appl. Phys. Lett.\/} {\bf 90}
  041111

\bibitem{Fejer_2002_ChipSqueezing}
Kanter G, Kumar P, Roussev R, Kurz J, Parameswaran K and Fejer M 2002 {\em Opt.
  Express\/} {\bf 10} 177

\bibitem{Hirano2008}
Eto Y, Tajima T, Zhang Y and Hirano T 2008 {\em Opt. Express\/} {\bf 16} 10650

\bibitem{Nice_2016_Squeezing}
Kaiser F, Fedrici B, Zavatta A, D'Auria V and Tanzilli S 2016 {\em Optica\/} {\bf 3} 362--365

\bibitem{Furusawa2015}
Masada G, Miyata K, Politi A, Hashimoto T, O'Brien J~L and Furusawa A 2015 {\em Nat. Photonics\/} {\bf 9} 316--319

\bibitem{Moison_09}
Moison J~M, Belabas N, Minot C and Levenson J~A 2009 {\em Opt. Lett.\/} {\bf 34} 2462--2464

\bibitem{Gallo_15}
Gallo K and Baghban M~A 2015 Recent developments on the lithium niobate
  material platform: The silicon of nonlinear optics? {\em Advanced Solid State
  Lasers\/} (OSA) p ATu3A.2

\end{thebibliography}

\end{document}